\documentclass[journal]{IEEEtran}

\usepackage{amsmath}
\usepackage{amssymb}
\usepackage{amsfonts}
\usepackage{amsthm}
\usepackage{mathtools}
\usepackage{bm}
\usepackage{mathrsfs}
\usepackage{algorithm}
\usepackage{algpseudocode}

\usepackage{caption}
\usepackage{subcaption}
\usepackage{graphicx}
\usepackage{booktabs}
\usepackage{array}
\usepackage{multirow}

\usepackage{cite}
\usepackage{url}

\usepackage{xcolor}
\usepackage[
    colorlinks=true,
    linkcolor=blue,
    citecolor=blue,
    urlcolor=blue
]{hyperref}

\usepackage{graphicx}
\usepackage{float}

\newtheorem{theorem}{Theorem}
\newtheorem{lemma}{Lemma}
\newtheorem{assumption}{Assumption}
\newtheorem{proposition}{Proposition}

\newtheorem{definition}{Definition}
\newtheorem{remark}{Remark}

\newcommand{\bx}{\bm{x}}
\newcommand{\bu}{\bm{u}}
\newcommand{\bfv}{\bm{f}}
\newcommand{\bg}{\bm{g}}
\newcommand{\bW}{\bm{W}}
\newcommand{\bphi}{\bm{\phi}}
\newcommand{\bsigma}{\bm{\sigma}}
\newcommand{\bR}{\bm{R}}
\newcommand{\bP}{\bm{P}}
\newcommand{\bI}{\bm{I}}

\usepackage{amsmath}
\usepackage{amssymb}
\usepackage{amsfonts}
\usepackage{bm}
\usepackage{mathtools}
\providecommand{\bc}{\bm{c}}

\providecommand{\be}{\bm{e}}
\providecommand{\bfv}{\bm{f}}
\providecommand{\bg}{\bm{g}}

\providecommand{\bs}{\bm{s}}

\providecommand{\bu}{\bm{u}}

\providecommand{\bw}{\bm{w}}
\providecommand{\bx}{\bm{x}}

\providecommand{\bI}{\bm{I}}

\providecommand{\bP}{\bm{P}}

\providecommand{\bR}{\bm{R}}

\providecommand{\bU}{\bm{U}}

\providecommand{\bW}{\bm{W}}

\providecommand{\bphi}{\bm{\phi}}

\providecommand{\bsigma}{\bm{\sigma}}

\begin{document}

% ============================================================
% TITLE
% ============================================================

\title{
Safety-Critical Koopman-MPC-Initialized Practical Fixed-Time Integral Reinforcement Learning for Unknown Nonlinear Systems
}

% ============================================================
% AUTHORS
% ============================================================

\author{
Tien Dat Vu
\thanks{
T. D. Vu is with the Faculty of Mechanical Engineering,
Ho Chi Minh City University of Technology (HCMUT),
Vietnam National University Ho Chi Minh City (VNU-HCM),
Ho Chi Minh City, Vietnam
(e-mails: dat.vuv@hcmut.edu.vn).
}%
\thanks{
Corresponding author: Tien Dat Vu
(e-mail: dat.vuv@hcmut.edu.vn).
}
}

\maketitle

\begin{abstract}

This paper addresses a challenging problem in
cyber--physical systems: reinforcement learning
(RL)--based control of nonlinear systems with
unknown drift dynamics, while simultaneously
achieving fixed-time convergence, optimality,
ensuring safety during operation, and satisfying
input constraints. Although numerous studies have
addressed individual aspects of this
problem~\cite{Cohen2023SafeExploration,Rabiee2025BackupCBF,Wang2024FixedTime,Gong2025,Kokolakis2025Predefined,Wang2026SafeFixedTime,10970061},
no existing work has, to date, addressed all
these challenges within a unified framework.
The present work addresses these challenges in turn.
First, integral reinforcement learning (IRL) is
employed to handle the unknown drift dynamics.
Unlike existing IRL approaches, the proposed
framework constructs the cost function through
inverse optimal control to enforce fixed-time
stability throughout both learning and actual
closed-loop operation. In contrast
to~\cite{Kokolakis2025Predefined,Wang2026SafeFixedTime},
where the reconstruction of the cost function
remains an open question, the present work
explicitly establishes this construction in
Theorem~\ref{thm:fixed_time_inverse_optimal}.
Specifically, integral concurrent learning is
used to identify the unknown component of the
dynamics, and the resulting estimate is then
used to construct the cost function. This
estimate serves to reconstruct the cost function
for the actual system, rather than to replace
the actual system dynamics with an identified
model for control. The second challenge is to
ensure that the system operates both safely
and optimally, while also achieving the desired
stability properties and respecting input constraints.
To this end, the proposed cost function is
constructed to accommodate fixed-time stability,
safety, and input constraints. A barrier function
is designed to encode the system constraints
and is embedded directly into the cost function
to be minimized, thereby addressing these
objectives jointly. Building on this formulation,
a new weight-update law is proposed to meet
these requirements. Finally, the paper develops
a principled method for selecting the initial
learning weights using Koopman model predictive
control (Koopman--MPC), as detailed in
Algorithm~\ref{alg:koopman_policy_initialization},
so that initialization does not compromise system
safety or stability. A nonlinear control example
is presented to validate the theoretical results.

\end{abstract}

% ============================================================
% KEYWORDS
% ============================================================

\begin{IEEEkeywords}
Practical fixed-time stability,
integral reinforcement learning,
inverse optimal control,
safety-critical control,
Koopman operator,
unknown nonlinear systems,
critic-only learning,
experience replay.
\end{IEEEkeywords}

% ============================================================
% INTRODUCTION
% ============================================================

\section{Introduction}
\label{sec:introduction}

Modern cyber--physical systems (CPSs) are becoming increasingly
complex in their structure, dynamics, and interaction between
physical, computational, and communication components, making
an accurate first-principles model increasingly difficult to
obtain \cite{Wan2021CPS,Lee2015CPS}.
At the same time, stricter timing, performance, and safety
requirements have become important design considerations in
nonlinear control systems \cite{Lee2015CPS,Wan2021CPS,Kokolakis2025Predefined,Wang2024FixedTimeFormation,Wang2026SafeFixedTime,11677079,10596995}. The concept of "predefined-time" has long been developed through works such as~\cite{SanchezTorres2015PredefinedTime,8961989}, but in the context of CPSs, a more profound requirement arises: the system must not only converge within a desired timeframe but also achieve optimality with respect to a specific cost function. This presents a challenge one that has been partially addressed in~\cite{Kokolakis2025Predefined} though the problem becomes significantly more complex when the system model is unknown.
Optimal control provides a systematic framework for handling
such demanding performance objectives \cite{Wan2021CPS}.
However, nonlinear optimal control is fundamentally limited
by the Hamilton--Jacobi--Bellman (HJB) equation, which
generally does not admit an explicit analytical solution,
while unknown system dynamics further prevent direct
construction of the optimal feedback law
\cite{Doya2000,Vamvoudakis2010}.
When the dynamics are unknown, guaranteeing convergence
within a uniform time bound independent of the initial
condition becomes particularly difficult
\cite{Polyakov2012,Kokolakis2025Predefined}.
Moreover, when nonlinear systems operate under state
constraints, ensuring that the closed-loop trajectories
remain within a prescribed safe set must also be considered
without sacrificing stability and control performance~\cite{Tee2009BLF,Ames2017CBFQP,Wabersich2021SafetyFilter,Malisani2014InteriorPenalty,Krstic2024InverseSafety,Wang2026SafeFormation}.
In addition to state constraints, practical control systems
are also subject to input constraints arising from actuator
limitations, which must be respected while simultaneously
achieving the desired stability, safety, and optimality
objectives.

RL and adaptive dynamic programming
(ADP) have emerged as data-driven approaches for approximating
the solution of the HJB equation without solving it
analytically \cite{Doya2000,Vrabie2009,Vamvoudakis2010}.
In particular, IRL
replaces the differential Bellman equation with an integral
relation constructed from measured trajectories, thereby
reducing the dependence of the learning process on unknown
drift dynamics \cite{Modares2014}.
In parallel, fixed-time and predefined-time optimal
stabilization have provided a systematic way to impose
convergence-time requirements that are independent of the
initial condition
\cite{Polyakov2012,Jimenez2017Predefined,Kokolakis2025Predefined}.
Kokolakis \emph{et al.} further developed fixed- and
predefined-time critic-learning laws using current and
recorded data, showing that sufficiently rich stored
information can replace the persistent-excitation
requirement
\cite{Kokolakis2023FixedTime,Kokolakis2025Predefined}.
These ideas also motivate an inverse-optimal viewpoint\cite{Sanchez2013InverseOptimal},
in which a prescribed Lyapunov decay is used to construct
a compatible running cost and optimal feedback law
\cite{Jimenez2017Predefined,Kokolakis2025Predefined}.
For an unknown system, however, such a construction first
requires the drift information; in this work, the drift
is therefore approximated by a neural basis representation
and its unknown weights are estimated from integral data
through integral concurrent learning~\cite{Cohen2023AdaptiveSafety} under a
full-row-rank condition
(see Eq.~\eqref{eq:integral_identification}).

% Several related developments address only parts of this
% problem.
% Modares \emph{et al.} combined IRL with experience replay
% for partially unknown continuous-time systems and used
% recorded data to relax the persistent-excitation requirement
% \cite{Modares2014}.
% Han \emph{et al.} developed IRL for unknown constrained
% nonlinear systems by combining system identification,
% experience replay, and actor--critic learning, but
% fixed-time critic convergence was not established
% \cite{Han2024ETIRL}.
% Hong \emph{et al.} investigated adaptive dynamic programming
% for partially unknown nonlinear systems, where the
% closed-loop and learning errors were characterized through
% robust or ultimate-bounded stability rather than fixed-time
% convergence \cite{Hong2025DETADP}.
% Zhang \emph{et al.} incorporated fixed-time prescribed
% performance into critic-only optimal tracking, but the
% neural and closed-loop signals were again established as
% uniformly ultimately bounded
% \cite{Zhang2026ETFixedTime}.
% Kokolakis \emph{et al.} achieved predefined-time critic
% learning and optimal stabilization using current and
% replayed data; however, the HJB residual used by the
% learning law depends explicitly on the system dynamics
% \cite{Kokolakis2025Predefined}.
% Meanwhile, Koopman-based representations provide a
% data-driven means of constructing approximate lifted
% dynamics and associated value-function surrogates,
% but their use as informed initializations for
% safety-critical practical fixed-time IRL remains
% insufficiently explored.

\subsection{Related Work and Motivation}

Several related developments address only parts of this
problem.
Modares \emph{et al.} combined IRL with experience replay
for partially unknown continuous-time systems and used
recorded data to relax the persistent-excitation requirement
\cite{Modares2014}.
Han \emph{et al.} developed IRL for unknown constrained
nonlinear systems by combining system identification,
experience replay, and actor--critic learning, but
fixed-time critic convergence was not established
\cite{Han2024ETIRL}.
Hong \emph{et al.} investigated adaptive dynamic programming
for partially unknown nonlinear systems, where the
closed-loop and learning errors were characterized through
robust or ultimate-bounded stability rather than fixed-time
convergence \cite{Hong2025DETADP}.
Zhang \emph{et al.} incorporated fixed-time prescribed
performance into critic-only optimal tracking, but the
neural and closed-loop signals were again established as
uniformly ultimately bounded
\cite{Zhang2026ETFixedTime}.

Kokolakis \emph{et al.} investigated predefined-time and
fixed-time learning and optimal control under known
system dynamics
\cite{Kokolakis2025Predefined,Kokolakis2023FixedTime}.
In particular, their predefined-time formulation uses
current and replayed data to achieve predefined-time
critic learning and optimal stabilization, but the HJB
residual explicitly depends on the system dynamics.
These developments do not provide an explicit,
constructive procedure for designing a running cost
that meets the desired convergence specifications.
Gong \emph{et al.} developed a fixed-time framework for
secure control \cite{Gong2025};
however, the presentation provides limited detail on
how to construct the optimal control problem to ensure
the required fixed-time property.

For unknown multi-agent dynamics, Wang \emph{et al.}
developed an ADP-based fixed-time formation control
scheme using an identifier inspired by the work of
Bhasin \emph{et al.}
\cite{Wang2024FixedTimeFormation,Bhasin2013ACI}.
Nevertheless, the identifier update remains strongly
dependent on persistent excitation, and the fixed-time
control formulation is developed using the identified
dynamics rather than directly using the true plant
dynamics.
Nancai \emph{et al.} adopted a similar identification-based
approach to predefined-time control of unknown systems
\cite{Cai2026PredefinedTime}.
Although the connection between cost-function design
and the desired convergence property is explained more
explicitly, a concrete cost-construction formula is
still not provided.
Moreover, the control formulation relies on a
least-squares identified model, and its value-iteration
procedure may entail substantial computational effort;
safety constraints are not addressed.
These identification-based approaches motivate a
closer examination of how model errors affect
convergence guarantees for the actual plant.

Safety has also been incorporated into learning-based
optimal control.
Wang \emph{et al.} proposed a barrier-based safe
fixed-time RL framework for multi-agent systems
\cite{Wang2026SafeFixedTime}.
However, the assumption of known dynamics limits its
applicability to complex CPSs with substantial model
uncertainty.
Ke Wang \emph{et al.} combined a barrier-augmented cost
with actor--critic IRL and state extrapolation to learn
avoidance policies for moving unsafe zones
\cite{10970061};
their analysis establishes uniform ultimate boundedness
of the plant state and weight errors rather than
fixed-time convergence.
Ping Wang \emph{et al.} formulated safe optimal UAV
formation tracking as a constrained zero-sum differential
game, incorporating a CBF into the cost and using
critic-only RL with experience replay
\cite{Wang2026SafeFormation}.
This formulation addresses collision avoidance,
external disturbances, and asymmetric input constraints,
with reported guarantees of stability and forward
invariance of the safe set, rather than an explicitly
prescribed fixed-time convergence guarantee. 

However, none of the aforementioned works address an important yet often overlooked factor: weight initialization. Poor weight initialization can lead to instability in the system during the learning process~\cite{Xu2024RLWeightInitialization}. According to the perspective in~\cite{Xu2024RLWeightInitialization}, if the weights are initialized randomly, the control signal will also be random; this introduces a deficiency in this approach. In study~\cite{Xu2024RLWeightInitialization}, this issue is addressed by designing a Nonlinear Model Predictive Control (NMPC) controller for a known dynamic model, and then solving an inverse problem to determine appropriate weighting parameters such that the control signal from the Adaptive Dynamic Programming (ADP) approach approximates that of the NMPC. The primary drawback of this method is that it requires an explicit model to derive the control signal. Furthermore, the study does not provide a clear discussion on the stability of the NMPC when used for policy initialization; it would be problematic if the initial control signal generated by the NMPC were inherently unstable. This, however, gives rise to an open question: how can one employ MPC to obtain a stable initial control signal when the model is unknown? One technical solution proposed in this context is KoopmanMPC—a technique that has seen significant development over the past decade~\cite{Williams2015Koopman,Brunton2016Koopman,Korda2018}. The specific details of how this technique is handled will be clarified in the section~\ref{koopman}.

These results leave a gap between time-critical optimal
learning, unknown-dynamics reconstruction, and
safety-critical control as well as finding initial weights for the learning process in a systematic way.
In particular, an inverse-optimal fixed-time construction
cannot be directly implemented when the drift required
by the optimality conditions is unavailable, whereas
existing IRL schemes for unknown nonlinear systems do
not simultaneously establish practical fixed-time critic
convergence, practical fixed-time closed-loop stability,
and forward invariance of a prescribed safe set
\cite{Han2024ETIRL,Hong2025DETADP,
Kokolakis2025Predefined,Zhang2026ETFixedTime}.
% Furthermore, conventional critic initialization does
% not explicitly exploit the information contained in
% data-driven lifted dynamics, leaving the initial
% value-function approximation potentially inaccurate
% and the subsequent learning process sensitive to the
% selected initial weights.

\subsection{Contribution of the paper
}

To address this gap, this paper contributes a
safety-critical KoopmanMPC-initialized practical fixed-time
IRL framework for unknown nonlinear systems. First
establish a unified framework for safe learning via
reinforcement learning for unknown nonlinear continuous-time
systems, aiming to simultaneously guarantee three key
properties: Fixed-time stability, safety, and optimality.
Furthermore, the results
in~\cite{Kokolakis2025Predefined,Gong2025,Wang2026SafeFixedTime}
are extended. Specifically,
Theorem~\ref{thm:fixed_time_inverse_optimal}
formulates a unique closed-form expression for selecting
a cost function such that the signals generated during
the learning process satisfy fixed-time stability for
the system. Notably, the reinforcement learning (RL)
approach employed here is Inverse Reinforcement Learning
(IRL) rather than standard RL as presented
in~\cite{Kokolakis2025Predefined,Gong2025,Wang2026SafeFixedTime},
which constitutes the core distinction from existing
works. In addition, the proposed framework explicitly
incorporates input constraints arising from actuator
limitations, ensuring that the learned control signals
remain within the admissible input set while
simultaneously satisfying the fixed-time stability,
safety, and optimality requirements. A tailored
stability analysis for the proposed method also
features specific characteristics, such as the
application of
Lemma~\ref{lem:derivative_limited_window_bound},
which represents a key contribution of this research.

Develop a new systematic method/algorithm for weight
initialization, thereby significantly improving the
probability of system stability. The proposed algorithm
is Algorithm 1, which extends the approach
in~\cite{Xu2024RLWeightInitialization,VuDoan2026WarmStart}.
Unlike~\cite{Xu2024RLWeightInitialization},
this study provides a weight initialization scheme for
nonlinear systems with unknown dynamic models, serving
as the core differentiator. Furthermore, the present
study offers a more comprehensive discussion on the
stability of the signals corresponding to the
initialized weights---an aspect that was not fully
clarified in~\cite{Xu2024RLWeightInitialization}.

\noindent\textit{Notation:}
$\mathbb R^n$ and $\mathbb R^{n\times m}$
denote real vector and matrix spaces.
Bold symbols denote vectors or matrices,
whereas nonbold symbols denote scalars.
$\|\cdot\|$ denotes the Euclidean vector norm
or induced matrix $2$-norm, while
$\|\cdot\|_F$ denotes the Frobenius norm.
The infinity norm is defined as
\(
\|\bx\|_\infty:=\max_j |x_j|
\).
$(\cdot)^\top$, $\otimes$, and $I_n$
denote transpose, Kronecker product, and
the identity matrix, respectively.
$\operatorname{tr}(\cdot)$ and
$\operatorname{diag}(\cdot)$ denote
the matrix trace and diagonal assembly,
respectively.
$\bP\succ0$ and $\bP\succeq0$ denote
positive definiteness and positive
semidefiniteness, respectively,
for symmetric matrices.
$\lambda_{\min}(\bP)$ and
$\lambda_{\max}(\bP)$ denote the minimum
and maximum eigenvalues of $\bP$,
respectively.
$\nabla V$ denotes the gradient of a
scalar function, and $C^1(\Omega)$
denotes continuously differentiable
functions on $\Omega$.
The superscript $(\cdot)^\dagger$
denotes the Moore--Penrose pseudoinverse,
The hyperbolic secant function is
defined as
$\operatorname{sech}(x)=2/(e^x+e^{-x})$.
The symbols
\(\mathbf 0_n\) and \(\mathbf 1_n\) denote
the \(n\)-dimensional all-zero and all-one
column vectors, respectively.

% ============================================================
% II. PROBLEM FORMULATION AND DATA-DRIVEN SYSTEM IDENTIFICATION
% ============================================================

\section{Problem Formulation and Data-Driven System Identification}
\label{sec:problem_identification}

\subsection{Problem Formulation}

Consider the nonlinear control-affine system
\begin{equation}
    \dot{\bx}
    =
    \bfv(\bx)
    +
    \bg(\bx)\bu,
    \label{eq:plant}
\end{equation}
where \(\bx\in\mathbb R^n\) is the system state, \(\bu\in\mathbb R^m\) is the control input, \(\bfv:\mathbb R^n\rightarrow\mathbb R^n\) is an unknown nonlinear drift, and \(\bg:\mathbb R^n\rightarrow\mathbb R^{n\times m}\) is a known input matrix. The control objective is to learn the unknown drift from measured state--input data and subsequently construct an optimal feedback policy that drives the state to the origin within a uniform finite settling time.

\begin{definition}[Admissible policy~\cite{AbuKhalaf2005}]
\label{def:admissible_secure_policy}
A continuous policy \(\bu:\Omega\rightarrow\mathcal U\) is admissible
on \(\Omega\), denoted by \(\bu\in\Psi(\Omega)\), if
\(\bu(\bm0)=\bm0\), the resulting closed-loop system is
asymptotically stable on \(\Omega\) in the absence of exogenous signals,
and the associated infinite-horizon cost is finite.
\end{definition}

\begin{assumption}
\label{ass:plant}
Let \(\Omega\subset\mathbb R^n\) be a compact set containing the origin in its interior. The unknown drift \(\bfv(\bx)\) and the known input matrix \(\bg(\bx)\) are locally Lipschitz on an open neighborhood of \(\Omega\), with \(\bfv(\bm 0)=\bm 0\). For every admissible input considered herein, the corresponding solution of \eqref{eq:plant} exists uniquely while it remains in \(\Omega\).
\end{assumption}

\begin{lemma}[Fixed-time stability~\cite{Polyakov2012}]
\label{lem:fixed_time}
Let \(V(\bx)\) be a positive-definite continuously
differentiable function satisfying
\(\dot V\le-aV^{\theta_1}-bV^{\theta_2}\),
where \(a>0\), \(b>0\), and
\(0<\theta_1<1<\theta_2\).
Then, the origin is fixed-time stable and its
settling time satisfies
\begin{equation}
    T_{\max}
    \le
    \frac{1}{a(1-\theta_1)}
    +
    \frac{1}{b(\theta_2-1)}.
    \label{eq:fixed_time_bound}
\end{equation}
\end{lemma}

\subsection{Neural Approximation of the Unknown Dynamics}

Inspired by~\cite{Bhasin2013ACI,Wang2024FixedTimeFormation} Since \(\bfv(\bx)\) is unavailable for control design, it is first reconstructed from measured data. By the universal approximation property~\cite{Kokolakis2025Predefined}, each continuous drift on the compact set \(\Omega\) can be represented as
\begin{equation}
    \bfv(\bx)
    =
    \bW_f^{*\top}\bsigma_f(\bx)
    +
    \bm{\varepsilon}_f(\bx),
    \label{eq:f_approximation}
\end{equation}
where \(\bsigma_f(\bx)\in\mathbb R^{N_f}\) is a prescribed vector of continuously differentiable basis functions, \(\bW_f^*\in\mathbb R^{N_f\times n}\) is the unknown ideal weight matrix, and \(\bm{\varepsilon}_f(\bx)\in\mathbb R^n\) is the approximation error. On \(\Omega\), \(\|\bsigma_f(\bx)\|\le\bar\sigma_f\) and \(\|\bm{\varepsilon}_f(\bx)\|\le\bar\varepsilon_f\) for some finite constants \(\bar\sigma_f>0\) and \(\bar\varepsilon_f>0\).

The identified drift is subsequently represented by
\(
\hat{\bfv}(\bx)
=
\hat{\bW}_f^\top\bsigma_f(\bx)
\),
where
\(
\hat{\bW}_f
\)
denotes the estimated weight matrix associated with the prescribed basis
vector
\(
\bsigma_f(\bx)
\).

\subsection{Integral Composite Concurrent-Learning Identtification~\cite{Cohen2023AdaptiveSafety}}
\label{subsec:integral_identification}

Let
\(
\tau_0<\tau_1<\cdots
\)
denote the sampling instants used to construct integral identification
data. Integrating \eqref{eq:plant} over
\(
[\tau_\ell,\tau_{\ell+1}]
\)
and using \eqref{eq:f_approximation} gives
\begin{equation}
\begin{aligned}
    &\bx(\tau_{\ell+1})-\bx(\tau_\ell)
    -
    \int_{\tau_\ell}^{\tau_{\ell+1}}
    \bg(\bx(\tau))\bu(\tau)\,d\tau
    \\
    &\qquad =
    \bW_f^{*\top}
    \int_{\tau_\ell}^{\tau_{\ell+1}}
    \bsigma_f(\bx(\tau))\,d\tau
    +
    \int_{\tau_\ell}^{\tau_{\ell+1}}
    \bm{\varepsilon}_f(\bx(\tau))\,d\tau .
\end{aligned}
\label{eq:integral_identification}
\end{equation}

For each interval, define
\(
\bm y_{f,\ell}
:=
\bx(\tau_{\ell+1})-\bx(\tau_\ell)
-
\int_{\tau_\ell}^{\tau_{\ell+1}}
\bg(\bx(\tau))\bu(\tau)\,d\tau
\),
\(
\bm\phi_{f,\ell}
:=
\int_{\tau_\ell}^{\tau_{\ell+1}}
\bsigma_f(\bx(\tau))\,d\tau
\),
and
\(
\bm d_{f,\ell}
:=
\int_{\tau_\ell}^{\tau_{\ell+1}}
\bm{\varepsilon}_f(\bx(\tau))\,d\tau
\).
Then
\(
\bm y_{f,\ell}
=
\bW_f^{*\top}\bm\phi_{f,\ell}
+
\bm d_{f,\ell}
\).

The informative data are stored in the history stack
\(
\mathcal H_f
=
\{(\bm y_{f,\ell},\bm\phi_{f,\ell})\}_{\ell=0}^{M_f-1}
\),
with cumulative information matrix
\(
\bm{\mathcal S}_f
:=
\sum_{\ell=0}^{M_f-1}
\bm\phi_{f,\ell}\bm\phi_{f,\ell}^{\top}
\).
Whenever a new data segment is admitted,
\(
\bm{\mathcal S}_f
\)
is updated by
\(
\bm{\mathcal S}_f^{+}
=
\bm{\mathcal S}_f
+
\bm\phi_{f,\mathrm{new}}
\bm\phi_{f,\mathrm{new}}^{\top}
\).

\begin{assumption}[Finite excitation]
\label{ass:finite_excitation}
There exist a finite \(M_f\geq N_f\) and a constant
\(\alpha_f>0\) such that the stored integral data satisfy
\(
\bm{\mathcal S}_f
=
\sum_{\ell=0}^{M_f-1}
\bm\phi_{f,\ell}\bm\phi_{f,\ell}^{\top}
\succeq
\alpha_f\bm I
\).
\end{assumption}

In addition to the stored data, define the current integral data over a
moving window of length \(T_f>0\) by
\(
\bm y_f(t)
:=
\bx(t)-\bx(t-T_f)
-
\int_{t-T_f}^{t}
\bg(\bx(\tau))\bu(\tau)\,d\tau
\)
and
\(
\bm\phi_f(t)
:=
\int_{t-T_f}^{t}
\bsigma_f(\bx(\tau))\,d\tau
\).
From \eqref{eq:plant} and \eqref{eq:f_approximation}, these quantities
satisfy
\(
\bm y_f(t)
=
\bW_f^{*\top}\bm\phi_f(t)
+
\bm d_f(t)
\),
where
\(
\bm d_f(t)
:=
\int_{t-T_f}^{t}
\bm{\varepsilon}_f(\bx(\tau))\,d\tau
\).
Similarly, every stored pair satisfies
\(
\bm y_{f,\ell}
=
\bW_f^{*\top}\bm\phi_{f,\ell}
+
\bm d_{f,\ell}
\).

Using the current estimate \(\hat{\bW}_f(t)\), define the current and
stored integral prediction errors as
\(
\bm e_f(t)
:=
\bm y_f(t)
-
\hat{\bW}_f^\top(t)\bm\phi_f(t)
\)
and
\(
\bm e_{f,\ell}(t)
:=
\bm y_{f,\ell}
-
\hat{\bW}_f^\top(t)\bm\phi_{f,\ell}
\),
respectively. Consider the composite prediction-error objective
\(
J_f
:=
\frac{1}{2}\|\bm e_f(t)\|^2
+
\frac{k_f}{2}
\sum_{\ell=0}^{M_f-1}
\|\bm e_{f,\ell}(t)\|^2
\),
where \(k_f>0\). Since
\(
\nabla_{\hat{\bW}_f}J_f
=
-\bm\phi_f(t)\bm e_f^\top(t)
-
k_f
\sum_{\ell=0}^{M_f-1}
\bm\phi_{f,\ell}\bm e_{f,\ell}^\top(t)
\),
the gradient update
\(
\dot{\hat{\bW}}_f
=
-\bm{\Gamma}_f
\nabla_{\hat{\bW}_f}J_f
\)
gives
\begin{equation}
\dot{\hat{\bW}}_f
=
\bm{\Gamma}_f
\left[
    \bm\phi_f(t)\bm e_f^\top(t)
    +
    k_f
    \sum_{\ell=0}^{M_f-1}
    \bm\phi_{f,\ell}\bm e_{f,\ell}^\top(t)
\right].
\label{eq:icl_identifier}
\end{equation}
where
\(
\bm{\Gamma}_f=\bm{\Gamma}_f^\top\succ0
\).
Hence, the identifier combines the current moving-window information
with the informative historical data retained in the history stack.

Let
\(
\tilde{\bW}_f
:=
\hat{\bW}_f-\bW_f^*
\).
The above regression relations immediately yield
\(
\bm e_f
=
-\tilde{\bW}_f^\top\bm\phi_f+\bm d_f
\)
and
\(
\bm e_{f,\ell}
=
-\tilde{\bW}_f^\top\bm\phi_{f,\ell}
+
\bm d_{f,\ell}
\).
Accordingly, the identification dynamics are analyzed with
\(
V_f
:=
\frac{1}{2}
\operatorname{tr}
\left(
\tilde{\bW}_f^\top
\bm{\Gamma}_f^{-1}
\tilde{\bW}_f
\right)
\).

\begin{lemma}[Composite identification convergence]
\label{lem:identification_error}
Suppose that Assumptions~\ref{ass:plant}--\ref{ass:finite_excitation}
hold and that the integral approximation residuals are bounded.
Then the weight-estimation error generated by
\eqref{eq:icl_identifier} is uniformly ultimately bounded. In
particular, there exist constants \(c_f>0\) and \(\Delta_f\geq0\)
such that
\begin{equation}
    \dot V_f
    \leq
    -c_fV_f+\Delta_f .
    \label{eq:icl_weight_bound}
\end{equation}
Consequently, the reconstructed drift
\(
\hat{\bfv}(\bx,t)
=
\hat{\bW}_f^\top(t)\bsigma_f(\bx)
\)
satisfies
\(
\|\bfv(\bx)-\hat{\bfv}(\bx,t)\|
\leq
\bar\sigma_f\|\tilde{\bW}_f(t)\|_F
+
\bar\varepsilon_f
\)
for every
\(
\bx\in\Omega
\).
If
\(
\bm{\varepsilon}_f\equiv\bm0
\),
then
\(
\Delta_f=0
\)
and
\(
\tilde{\bW}_f(t)\to\bm0
\).
\end{lemma}

% \begin{proof}
% Differentiating \(V_f\) along \eqref{eq:icl_identifier} gives
% \(
% \dot V_f
% =
% \operatorname{tr}
% [
% \tilde{\bW}_f^\top
% \bm\phi_f\bm e_f^\top
% ]
% +
% k_f
% \sum_{\ell=0}^{M_f-1}
% \operatorname{tr}
% [
% \tilde{\bW}_f^\top
% \bm\phi_{f,\ell}
% \bm e_{f,\ell}^{\top}
% ]
% \).
% Substituting the current and stored prediction errors yields
% \(
% \dot V_f
% =
% -\|\tilde{\bW}_f^\top\bm\phi_f\|^2
% -
% k_f
% \sum_{\ell=0}^{M_f-1}
% \|\tilde{\bW}_f^\top\bm\phi_{f,\ell}\|^2
% +
% \Delta_{f,r}
% \),
% where \(\Delta_{f,r}\) collects the bounded terms induced by
% \(\bm d_f\) and \(\bm d_{f,\ell}\).
% By Assumption~\ref{ass:finite_excitation},
% \(
% \sum_{\ell=0}^{M_f-1}
% \|\tilde{\bW}_f^\top\bm\phi_{f,\ell}\|^2
% =
% \operatorname{tr}
% [
% \tilde{\bW}_f^\top
% \bm{\mathcal S}_f
% \tilde{\bW}_f
% ]
% \geq
% \alpha_f\|\tilde{\bW}_f\|_F^2
% \).
% Using Young's inequality and the bounded integral residuals therefore
% gives
% \(
% \dot V_f
% \leq
% -\bar c_f
% \|\tilde{\bW}_f\|_F^2
% +
% \Delta_f
% \)
% for some
% \(
% \bar c_f>0
% \)
% and
% \(
% \Delta_f\geq0
% \).
% Since
% \(
% V_f
% \leq
% [2\lambda_{\min}(\bm{\Gamma}_f)]^{-1}
% \|\tilde{\bW}_f\|_F^2
% \),
% \eqref{eq:icl_weight_bound} follows for a suitable \(c_f>0\).

% % Finally,
% % \(
% % \bfv(\bx)-\hat{\bfv}(\bx,t)
% % =
% % -\tilde{\bW}_f^\top(t)\bsigma_f(\bx)
% % +
% % \bm{\varepsilon}_f(\bx)
% % \),
% % which gives
% % \(
% % \|\bfv(\bx)-\hat{\bfv}(\bx,t)\|
% % \leq
% % \bar\sigma_f\|\tilde{\bW}_f(t)\|_F
% % +
% % \bar\varepsilon_f
% % \).
% \end{proof}

\begin{proof}
Differentiating \(V_f\) along \eqref{eq:icl_identifier}, and noting that the ideal weight matrix is constant so that
\(\dot{\tilde{\bW}}_f=\dot{\hat{\bW}}_f\), gives
\begin{equation}
\dot V_f
=
\operatorname{tr}
[
\tilde{\bW}_f^\top
\bm\phi_f\bm e_f^\top
]
+
k_f
\sum_{\ell=0}^{M_f-1}
\operatorname{tr}
[
\tilde{\bW}_f^\top
\bm\phi_{f,\ell}
\bm e_{f,\ell}^{\top}
].
\label{eq:Vf_dot_initial}
\end{equation}
From the current regression relation and its stored counterparts, the corresponding prediction errors can be written as
\(
\bm e_f
=
-\tilde{\bW}_f^\top\bm\phi_f+\bm d_f
\)
and
\(
\bm e_{f,\ell}
=
-\tilde{\bW}_f^\top\bm\phi_{f,\ell}
+
\bm d_{f,\ell}
\),
respectively. Substituting these expressions into \eqref{eq:Vf_dot_initial} yields
\begin{equation}
\begin{aligned}
\dot V_f
={}&
-\operatorname{tr}
[
\tilde{\bW}_f^\top
\bm\phi_f
\bm\phi_f^\top
\tilde{\bW}_f
]
-
k_f
\sum_{\ell=0}^{M_f-1}
\operatorname{tr}
[
\tilde{\bW}_f^\top
\bm\phi_{f,\ell}
\bm\phi_{f,\ell}^\top
\tilde{\bW}_f
]
\\
&+
\operatorname{tr}
[
\tilde{\bW}_f^\top
\bm\phi_f
\bm d_f^\top
]
+
k_f
\sum_{\ell=0}^{M_f-1}
\operatorname{tr}
[
\tilde{\bW}_f^\top
\bm\phi_{f,\ell}
\bm d_{f,\ell}^\top
].
\end{aligned}
\label{eq:Vf_dot_expanded}
\end{equation}
Using
\(
\operatorname{tr}
[
\tilde{\bW}_f^\top
\bm\phi_f
\bm\phi_f^\top
\tilde{\bW}_f
]
=
\|\tilde{\bW}_f^\top\bm\phi_f\|^2
\)
and, similarly,
\(
\operatorname{tr}
[
\tilde{\bW}_f^\top
\bm\phi_{f,\ell}
\bm\phi_{f,\ell}^\top
\tilde{\bW}_f
]
=
\|\tilde{\bW}_f^\top\bm\phi_{f,\ell}\|^2
\),
one obtains
\begin{equation}
\begin{aligned}
\dot V_f
={}&
-\|\tilde{\bW}_f^\top\bm\phi_f\|^2
-
k_f
\sum_{\ell=0}^{M_f-1}
\|\tilde{\bW}_f^\top\bm\phi_{f,\ell}\|^2
\\
&+
\operatorname{tr}
[
\tilde{\bW}_f^\top
\bm\phi_f
\bm d_f^\top
]
+
k_f
\sum_{\ell=0}^{M_f-1}
\operatorname{tr}
[
\tilde{\bW}_f^\top
\bm\phi_{f,\ell}
\bm d_{f,\ell}^\top
].
\end{aligned}
\label{eq:Vf_dot_residual}
\end{equation}

The last two terms in \eqref{eq:Vf_dot_residual} arise from the integral approximation residuals and therefore must be bounded explicitly. By the Cauchy--Schwarz inequality for the Frobenius inner product,
\(
\left|
\operatorname{tr}
[
\tilde{\bW}_f^\top
\bm\phi_f
\bm d_f^\top
]
\right|
\leq
\|\tilde{\bW}_f^\top\bm\phi_f\|
\|\bm d_f\|
\)
and
\(
\left|
\operatorname{tr}
[
\tilde{\bW}_f^\top
\bm\phi_{f,\ell}
\bm d_{f,\ell}^\top
]
\right|
\leq
\|\tilde{\bW}_f^\top\bm\phi_{f,\ell}\|
\|\bm d_{f,\ell}\|
\).
For arbitrary constants
\(
\epsilon_{f,0}\in(0,2)
\)
and
\(
\epsilon_{f,\ell}\in(0,2)
\),
Young's inequality gives
\begin{equation}
\|\tilde{\bW}_f^\top\bm\phi_f\|
\|\bm d_f\|
\leq
\frac{\epsilon_{f,0}}{2}
\|\tilde{\bW}_f^\top\bm\phi_f\|^2
+
\frac{1}{2\epsilon_{f,0}}
\|\bm d_f\|^2,
\label{eq:young_current}
\end{equation}
and
\begin{equation}
\|\tilde{\bW}_f^\top\bm\phi_{f,\ell}\|
\|\bm d_{f,\ell}\|
\leq
\frac{\epsilon_{f,\ell}}{2}
\|\tilde{\bW}_f^\top\bm\phi_{f,\ell}\|^2
+
\frac{1}{2\epsilon_{f,\ell}}
\|\bm d_{f,\ell}\|^2.
\label{eq:young_stored}
\end{equation}
Substituting \eqref{eq:young_current} and \eqref{eq:young_stored} into \eqref{eq:Vf_dot_residual} gives
\begin{equation}
\begin{aligned}
\dot V_f
\leq{}&
-
\left(
1-\frac{\epsilon_{f,0}}{2}
\right)
\|\tilde{\bW}_f^\top\bm\phi_f\|^2
\\
&-
k_f
\sum_{\ell=0}^{M_f-1}
\left(
1-\frac{\epsilon_{f,\ell}}{2}
\right)
\|\tilde{\bW}_f^\top\bm\phi_{f,\ell}\|^2
\\
&+
\frac{1}{2\epsilon_{f,0}}
\|\bm d_f\|^2
+
\frac{k_f}{2}
\sum_{\ell=0}^{M_f-1}
\frac{1}{\epsilon_{f,\ell}}
\|\bm d_{f,\ell}\|^2.
\end{aligned}
\label{eq:Vf_dot_young}
\end{equation}

Since the integral approximation residuals are bounded, there exist finite constants
\(
\bar d_f\geq0
\)
and
\(
\bar d_{f,\ell}\geq0
\)
such that
\(
\|\bm d_f\|\leq\bar d_f
\)
and
\(
\|\bm d_{f,\ell}\|\leq\bar d_{f,\ell}
\).
Define
\(
\underline\epsilon_f
:=
\min_{\ell=0,\ldots,M_f-1}
\left(
1-\frac{\epsilon_{f,\ell}}{2}
\right)>0
\).
Then \eqref{eq:Vf_dot_young} implies
\begin{equation}
\begin{aligned}
\dot V_f
\leq{}&
-
\left(
1-\frac{\epsilon_{f,0}}{2}
\right)
\|\tilde{\bW}_f^\top\bm\phi_f\|^2
\\
&-
k_f\underline\epsilon_f
\sum_{\ell=0}^{M_f-1}
\|\tilde{\bW}_f^\top\bm\phi_{f,\ell}\|^2
+
\Delta_{f,r},
\end{aligned}
\label{eq:Vf_dot_bounded_residual}
\end{equation}
where
\(
\Delta_{f,r}
:=
\frac{\bar d_f^2}{2\epsilon_{f,0}}
+
\frac{k_f}{2}
\sum_{\ell=0}^{M_f-1}
\frac{\bar d_{f,\ell}^2}{\epsilon_{f,\ell}}
\geq0
\).

By Assumption~\ref{ass:finite_excitation},
\(
\bm{\mathcal S}_f
=
\sum_{\ell=0}^{M_f-1}
\bm\phi_{f,\ell}\bm\phi_{f,\ell}^\top
\succeq
\alpha_f I
\).
Hence,
\begin{equation}
\begin{aligned}
\sum_{\ell=0}^{M_f-1}
\|\tilde{\bW}_f^\top\bm\phi_{f,\ell}\|^2
&=
\sum_{\ell=0}^{M_f-1}
\operatorname{tr}
[
\tilde{\bW}_f^\top
\bm\phi_{f,\ell}
\bm\phi_{f,\ell}^\top
\tilde{\bW}_f
]
\\
&=
\operatorname{tr}
[
\tilde{\bW}_f^\top
\bm{\mathcal S}_f
\tilde{\bW}_f
]
\\
&\geq
\alpha_f
\operatorname{tr}
[
\tilde{\bW}_f^\top
\tilde{\bW}_f
]
\\
&=
\alpha_f
\|\tilde{\bW}_f\|_F^2.
\end{aligned}
\label{eq:stored_data_lower_bound}
\end{equation}
Using \eqref{eq:stored_data_lower_bound} in \eqref{eq:Vf_dot_bounded_residual}, and discarding the additional nonpositive current-data term, gives
\begin{equation}
\dot V_f
\leq
-
k_f\underline\epsilon_f\alpha_f
\|\tilde{\bW}_f\|_F^2
+
\Delta_{f,r}.
\label{eq:Vf_dot_weight_norm}
\end{equation}
Thus, defining
\(
\bar c_f
:=
k_f\underline\epsilon_f\alpha_f>0
\)
and
\(
\Delta_f:=\Delta_{f,r}\geq0
\),
one obtains
\begin{equation}
\dot V_f
\leq
-\bar c_f
\|\tilde{\bW}_f\|_F^2
+
\Delta_f.
\label{eq:Vf_dot_intermediate}
\end{equation}

Finally, from the definition
\(
V_f
=
\frac{1}{2}
\operatorname{tr}
[
\tilde{\bW}_f^\top
\bm{\Gamma}_f^{-1}
\tilde{\bW}_f
]
\)
and
\(
\bm{\Gamma}_f\succ0
\),
it follows that
\(
V_f
\leq
[2\lambda_{\min}(\bm{\Gamma}_f)]^{-1}
\|\tilde{\bW}_f\|_F^2
\),
or equivalently
\(
\|\tilde{\bW}_f\|_F^2
\geq
2\lambda_{\min}(\bm{\Gamma}_f)V_f
\).
Substituting this inequality into \eqref{eq:Vf_dot_intermediate} yields
\begin{equation}
\dot V_f
\leq
-
2\bar c_f
\lambda_{\min}(\bm{\Gamma}_f)
V_f
+
\Delta_f.
\label{eq:Vf_dot_final_step}
\end{equation}
Therefore, by defining
\(
c_f
:=
2\bar c_f
\lambda_{\min}(\bm{\Gamma}_f)>0
\),
one obtains
\(
\dot V_f
\leq
-c_fV_f+\Delta_f
\),
which proves \eqref{eq:icl_weight_bound}.
\end{proof}

The online reconstructed model
\(
\hat{\bfv}(\bx,t)
=
\hat{\bW}_f^\top(t)\bsigma_f(\bx)
\)
is subsequently used in the inverse-optimal construction.

\begin{lemma}[Bounded identification error]
\label{lem:identification_error}
Under Assumptions~\ref{ass:plant}--\ref{ass:finite_excitation},
the identification error is uniformly bounded on \(\Omega\). In particular,
there exists a finite constant \(\bar\Delta_f>0\) such that
\begin{equation}
    \left\|
        \bfv(\bx)-\hat{\bfv}(\bx,t)
    \right\|
    \leq
    \bar\Delta_f,
    \qquad
    \forall\,\bx\in\Omega .
    \label{eq:identification_error_bound}
\end{equation}
\end{lemma}

\begin{proof}
From
\(
\bfv(\bx)
=
\bW_f^{*\top}\bsigma_f(\bx)
+
\bm{\varepsilon}_f(\bx)
\)
and
\(
\hat{\bfv}(\bx,t)
=
\hat{\bW}_f^\top(t)\bsigma_f(\bx)
\),
one has
\(
\bfv(\bx)-\hat{\bfv}(\bx,t)
=
-\tilde{\bW}_f^\top(t)\bsigma_f(\bx)
+
\bm{\varepsilon}_f(\bx)
\).
Hence,
\(
\|\bfv(\bx)-\hat{\bfv}(\bx,t)\|
\leq
\bar\sigma_f\|\tilde{\bW}_f(t)\|_F
+
\bar\varepsilon_f
\).
By Lemma~\ref{lem:identification_error},
\(
\tilde{\bW}_f(t)
\)
is uniformly ultimately bounded. Therefore, there exists a finite
constant \(\bar W_f>0\) such that
\(
\|\tilde{\bW}_f(t)\|_F\leq\bar W_f
\),
and thus \eqref{eq:identification_error_bound} holds with
\(
\bar\Delta_f
:=
\bar\sigma_f\bar W_f
+
\bar\varepsilon_f
\).
\end{proof}

\section{Practical Fixed-Time Integral Reinforcement Learning}
\label{sec:koopman_fixed_time_irl}

% ============================================================
% III-A. CONSTRAINED FIXED-TIME OPTIMAL-CONTROL FORMULATION
% ============================================================

% =========================================================
% BASIC DEFINITIONS FOR CONSTRAINED FIXED-TIME IRL
% =========================================================

Motivated by the problem formulations in
\cite{Cohen2023SafeExploration,Wang2026SafeFixedTime},
we first introduce several definitions
that lay the groundwork for the subsequent
formulation of the constrained fixed-time
optimal control problem.

% \begin{definition}[Safe set\cite{Cohen2023AdaptiveSafety}]
% \label{def:safe_set}
% Let
% \(
% h_\ell:\mathbb R^n\rightarrow\mathbb R
% \),
% \(
% \ell=1,\ldots,n_h
% \), where $n_h \in \mathbb{N}_{>0}$ denotes the number of safety constraints
% be continuously differentiable constraint functions.
% The admissible state set and its interior are defined as
% \(
% \mathcal C
% :=
% \{
% \bx\in\Omega:
% h_\ell(\bx)\geq0,\;
% \ell=1,\ldots,n_h
% \}
% \)
% and
% \(
% \operatorname{Int}\mathcal C
% :=
% \{
% \bx\in\Omega:
% h_\ell(\bx)>0,\;
% \ell=1,\ldots,n_h
% \}
% \),
% respectively.
% A closed-loop trajectory is called safe if
% \(
% \bx(0)\in\operatorname{Int}\mathcal C
% \)
% implies
% \(
% \bx(t)\in\operatorname{Int}\mathcal C
% \)
% for all
% \(
% t\geq0
% \).
% \end{definition}

% \begin{definition}[Safe set\cite{Cohen2023AdaptiveSafety}]
% \label{def:safe_set}
% Let
% \(
% h_\ell:\mathbb R^n\rightarrow\mathbb R
% \),
% \(
% \ell=1,\ldots,n_h
% \),
% be continuously differentiable constraint functions,
% where
% \(
% n_h\in\mathbb N_{>0}
% \)
% denotes the number of safety constraints.
% The admissible state set and its interior are
% defined, respectively, as
% \(
% \mathcal C
% :=
% \{
% \bx\in\Omega:
% h_\ell(\bx)\geq0,\;
% \ell=1,\ldots,n_h
% \}
% \)
% and
% \(
% \operatorname{Int}\mathcal C
% :=
% \{
% \bx\in\Omega:
% h_\ell(\bx)>0,\;
% \ell=1,\ldots,n_h
% \}
% \).
% A closed-loop trajectory is called safe if
% \(
% \bx(0)\in\operatorname{Int}\mathcal C
% \)
% implies
% \(
% \bx(t)\in\operatorname{Int}\mathcal C
% \)
% for all
% \(
% t\geq0
% \).
% \end{definition}

\begin{definition}[Safe set\cite{Cohen2023AdaptiveSafety,Cohen2023SafeExploration,Tee2009BLF,Ames2017CBFQP}]
\label{def:safe_set}
Let
\(
h_\ell:\mathbb R^n\rightarrow\mathbb R
\),
\(
\ell=1,\ldots,n_h
\),
be continuously differentiable constraint functions,
where
\(
n_h\in\mathbb N_{>0}
\)
denotes the number of safety constraints.
The admissible state set, its interior, and its
boundary are defined, respectively, as
\(
\mathcal C
:=
\{
\bx\in\Omega:
h_\ell(\bx)\geq0,\;
\ell=1,\ldots,n_h
\}
\),
\(
\operatorname{Int}\mathcal C
:=
\{
\bx\in\Omega:
h_\ell(\bx)>0,\;
\ell=1,\ldots,n_h
\}
\),
and
\(
\partial\mathcal C
:=
\mathcal C\setminus\operatorname{Int}\mathcal C
\).
A closed-loop trajectory is called safe if
\(
\bx(0)\in\operatorname{Int}\mathcal C
\)
implies
\(
\bx(t)\in\operatorname{Int}\mathcal C
\)
for all
\(
t\geq0
\).
\end{definition}

\begin{definition}[Input-admissible policy~\cite{Modares2014}]
\label{def:input_admissible_policy}
Let
\(
\bm U
:=
\operatorname{diag}
\{
\bar u_1,\ldots,\bar u_m
\}
\succ0
\)
and define
\(
\mathbb U
:=
\{
\bu\in\mathbb R^m:
|u_j|<\bar u_j,\;
j=1,\ldots,m
\}
\).
A feedback policy
\(
\boldsymbol\mu:\operatorname{Int}\mathcal C\rightarrow\mathbb U
\)
is called input admissible if
\(
\boldsymbol\mu(\bx)\in\mathbb U
\)
for every
\(
\bx\in\operatorname{Int}\mathcal C
\).
In particular, any policy of the form
\(
\boldsymbol\mu(\bx)
=
-\bm U\tanh(\bm\zeta(\bx))
\)
with finite
\(
\bm\zeta(\bx)
\)
is input admissible since
\(
|\tanh(\zeta_j)|<1
\)
implies
\(
|\mu_j(\bx)|<\bar u_j
\). \noindent\textit{Note:}
Input admissibility only guarantees
satisfaction of the prescribed input
constraints and does not, by itself,
imply admissibility in the sense of
Definition~\ref{def:admissible_secure_policy}, which additionally
requires closed-loop stability and
finite infinite-horizon cost
\end{definition}

% \begin{definition}[Barrier-compatible value function~\cite{Cohen2023AdaptiveSafety}]
% \label{def:barrier_compatible_value}
% Let
% \(
% V_0:\operatorname{Int}\mathcal C\rightarrow\mathbb R_{\geq0}
% \)
% be positive definite and let
% \(
% \mathcal B:\operatorname{Int}\mathcal C
% \rightarrow
% \mathbb R_{\geq0}
% \)
% satisfy
% \(
% \mathcal B(\bm0)=0
% \)
% and
% \(
% \mathcal B(\bx)\rightarrow\infty
% \)
% as
% \(
% \bx\rightarrow\partial\mathcal C
% \).
% For
% \(
% \kappa_B>0
% \),
% the function
% \(
% V_d(\bx)
% :=
% V_0(\bx)
% +
% \kappa_B\mathcal B(\bx)
% \)
% is called a barrier-compatible desired value function.
% Hence,
% \(
% V_d(\bx)\rightarrow\infty
% \)
% as
% \(
% \bx\rightarrow\partial\mathcal C
% \),
% so boundedness of
% \(
% V_d(\bx(t))
% \)
% precludes finite-time escape through the state-constraint boundary.
% \end{definition}

\begin{definition}[Barrier-compatible value function~\cite{Cohen2023AdaptiveSafety,Tee2009BLF}]
\label{def:barrier_compatible_value}
Let
\(
\bm0\in\operatorname{Int}\mathcal C
\)
and
\(
V_0,\mathcal B\in C^1(\operatorname{Int}\mathcal C)
\).
Suppose that
\(
V_0:\operatorname{Int}\mathcal C
\rightarrow\mathbb R_{\geq0}
\)
is positive definite, satisfying
\(
V_0(\bm0)=0
\)
and
\(
V_0(\bx)>0
\)
for all
\(
\bx\neq\bm0
\),
and is radially unbounded whenever
\(
\operatorname{Int}\mathcal C
\)
is unbounded.
Let
\(
\mathcal B:\operatorname{Int}\mathcal C
\rightarrow\mathbb R_{\geq0}
\)
satisfy
\(
\mathcal B(\bm0)=0
\)
and
\(
\mathcal B(\bx)\rightarrow\infty
\)
as
\(
\bx\rightarrow\partial\mathcal C
\).
For
\(
\kappa_B>0
\),
the function
\(
V_d(\bx)
:=
V_0(\bx)
+
\kappa_B\mathcal B(\bx)
\)
is called a barrier-compatible desired
value function.
Consequently,
\(
V_d(\bx)\rightarrow\infty
\)
as
\(
\bx\rightarrow\partial\mathcal C
\),
and its sublevel sets are compact
in
\(
\operatorname{Int}\mathcal C
\).
Hence, boundedness of
\(
V_d(\bx(t))
\)
precludes finite-time escape through
the state-constraint boundary.
\end{definition}

\subsection{Constrained Fixed-Time Optimal-Control Formulation}

With the unknown nonlinear system~\eqref{eq:plant}, according to
Definition~\ref{def:safe_set},
the state constraints are represented by
\(
h_\ell(\bx)>0
\),
\(
\ell=1,\ldots,n_h
\),
with
\begin{equation}
    \mathcal C
    :=
    \{
        \bx\in\Omega:
        h_\ell(\bx)>0,\;
        \ell=1,\ldots,n_h
    \}.
\end{equation}
Each
\(
h_\ell:\Omega\rightarrow\mathbb R
\)
is continuously differentiable and satisfies
\(
h_\ell(\bm 0)>0
\).

To encode the state constraints in the value function, define
\begin{equation}
    b_\ell(\bx)
    :=
    \log
    \left(
        1+\frac{1}{\gamma_\ell h_\ell(\bx)}
    \right),
\end{equation}
where
\(
\gamma_\ell>0
\),
and the recentered barrier
\begin{equation}
    \mathcal B(\bx)
    :=
    \sum_{\ell=1}^{n_h}
    \beta_\ell
    [
        b_\ell(\bx)-b_\ell(\bm 0)
    ]^2,
\end{equation}
where
\(
\beta_\ell>0
\).
Then
\(
\mathcal B(\bm 0)=0
\),
\(
\mathcal B(\bx)\geq0
\),
and
\begin{equation}
    \mathcal B(\bx)\rightarrow\infty
\end{equation}
whenever
\(
h_\ell(\bx)\rightarrow0^+
\).

Following
Definition~\ref{def:barrier_compatible_value},
select
\begin{equation}
    V_d(\bx)
    :=
    V_0(\bx)
    +
    \kappa_B\mathcal B(\bx)
\end{equation}
is the desird vulue fuction of~\eqref{eq:plant}, 
where
\(
\kappa_B>0
\).
Hence,
\(
V_d(\bm 0)=0
\),
\(
V_d(\bx)>0
\)
for
\(
\bx\neq\bm 0
\),
and \(V_d(\bx)\rightarrow\infty\) as
\(
\bx\rightarrow\partial\mathcal C
\).

The desired fixed-time evolution is prescribed as
\begin{equation}
\dot V_d(\bx)
=
-aV_d^{\gamma_1}(\bx)
-bV_d^\nu(\bx),
\label{eq:desired_constrained_fixed_time_decay}
\end{equation}
where
\(
a,b>0
\)
and
\(
\frac12<\gamma_1<1<\nu
\)
are prescribed design parameters.
For the subsequent integral reinforcement-learning
and composite stability analysis, define
\(
\mu:=(\gamma_1+1)/2
\)
and
\(
\gamma_2:=2\nu-1
\),
so that
\(
\frac34<\mu<1<\nu
\)
and
\(
0<\gamma_1<1<\gamma_2
\).

By Lemma~\ref{lem:fixed_time},
any feedback satisfying
\eqref{eq:desired_constrained_fixed_time_decay}
renders the origin fixed-time stable with
\begin{equation}
T_{\max}
\leq
[a(1-\gamma_1)]^{-1}
+
[b(\nu-1)]^{-1},
\end{equation}
independently of the initial condition.
Together with
Definitions~\ref{def:safe_set}, \ref{def:input_admissible_policy}
and
\ref{def:barrier_compatible_value},
this specifies the constrained fixed-time target used in the subsequent
inverse-optimal and integral reinforcement-learning developments.

The actuator constraints are incorporated through the nonquadratic
control penalty~\cite{Modares2014}.

\begin{equation}
    \begin{aligned}
    \mathcal U_c(\bu)
    :={}&
    2\sum_{j=1}^{m}
    r_j\bar u_j
    \int_0^{u_j}
    \tanh^{-1}
    (s/\bar u_j)\,ds
    \\
    ={}&
    2\int_{\bm 0}^{\bu}
    \big[
    \bR\bm U
    \tanh^{-1}(\bm U^{-1}\bs)
    \big]^\top
    d\bs.
    \end{aligned}
\label{U}
\end{equation}

where
\( \bR
    =
    \operatorname{diag}(r_1,\ldots,r_m)
    \succ0\), and
\(
\tanh^{-1}(\cdot)
\)
be applied componentwise. Inspired by the works in \cite{Xu2024RLWeightInitialization,Kokolakis2023FixedTime,Kim2022SafeModelBasedRL,Cohen2020ApproximateOptimalCBF,10970061}, we consider the following constrained performance index:
\begin{equation}
\begin{aligned}
    J(\bx_0,\bu)
    =
    \int_0^\infty
    \Big[
        \bm L_1(\bx)\bu
        +
        L_2(\bx)
        +
        \mathcal B(\bx)
        +
        \mathcal U_c(\bu)
    \Big]dt ,
\end{aligned}
\label{eq:constrained_fixed_time_cost}
\end{equation}

where
\(\bm L_1:\Omega\to\mathbb R^{1\times m}\)
and \(L_2:\Omega\to\mathbb R\)
are to be constructed such that
\(V_d(\bx)\) satisfies the associated
HJB equation and the resulting policy
achieves the prescribed fixed-time decay
\eqref{eq:desired_constrained_fixed_time_decay}
for the identified dynamics.
Here, \(\mathcal B(\bx)\) penalizes
proximity to the state-constraint boundary,
while \(\mathcal U_c(\bu)\) penalizes
control effort through a non-quadratic
structure accounting for actuator limits.
The term \(\bm L_1(\bx)\bu\) introduces
state-control coupling, whereas
\(L_2(\bx)\) provides the state-dependent
cost component required by the
inverse-optimal construction. The running cost need not be pointwise
positive definite for every admissible
state-input pair. As discussed in
\cite[Sec.~8.4, pp.~524--525]{Haddad2008NonlinearControl},
state-control cross-weighting may render
the running cost negative for certain
inputs without necessarily precluding
inverse optimality. The latter requires
appropriate HJB verification,
admissibility, and transversality
conditions. The constructed coefficients
satisfy \(\bm L_1(\bm0)=\bm0\) and
\(L_2(\bm0)=0\), as established in
Lemma~\ref{lem:origin_extension}.

\begin{remark}
The running cost in
\eqref{eq:constrained_fixed_time_cost}
differs from that in
\cite{Kokolakis2025Predefined},
which focuses on prescribed stability
properties without explicitly incorporating
safety or state constraints.
Furthermore, unlike the discrete-time
cost formulation in
\cite{Xu2024RLWeightInitialization},
the present work constructs a
continuous-time barrier-augmented
inverse-optimal running cost that
accounts for state constraints while
enforcing the prescribed fixed-time
Lyapunov decay.
\end{remark}

% \begin{remark}[Safety and Optimality]
% Barrier-based safety control has been
% investigated through barrier Lyapunov
% functions \cite{Tee2009BLF},
% CBF-based safety filters \cite{Ames2017CBFQP},
% predictive safety filters
% \cite{Wabersich2021SafetyFilter},
% interior-penalty optimal control
% \cite{Malisani2014InteriorPenalty},
% and inverse-optimal safety filters
% \cite{Krstic2024InverseSafety}.
% These approaches incorporate safety
% through stability certificates,
% control constraints, or optimality
% criteria, with distinct performance
% guarantees. Since the present objective
% requires both prescribed fixed-time
% convergence and infinite-horizon
% optimality, we adopt a barrier-compatible
% inverse-optimal formulation, embedding
% state constraints into the desired
% value function and constructing the
% associated running cost through the
% HJB equation. This unifies safety,
% convergence, and inverse optimality
% under the stated feasibility and
% stability conditions.
% \end{remark}

\begin{remark}[Safety and Optimality]
Barrier-based safety control has been
investigated through barrier Lyapunov
functions \cite{Tee2009BLF},
CBF-based safety filters \cite{Ames2017CBFQP},
predictive safety filters
\cite{Wabersich2021SafetyFilter},
interior-penalty optimal control
\cite{Malisani2014InteriorPenalty},
and inverse-optimal safety filters
\cite{Krstic2024InverseSafety}.
These approaches incorporate safety
through stability certificates,
control constraints, or optimality
criteria, with distinct performance
guarantees. The present framework
lies at the intersection of CBF-based
safety filtering and interior-penalty
optimal control, combining explicit
safety enforcement with the integration
of barrier functions into the optimal
control formulation. Since the present
objective requires both prescribed
fixed-time convergence and infinite-horizon
optimality, we adopt a barrier-compatible
inverse-optimal formulation, embedding
state constraints into the desired
value function and constructing the
associated running cost through the
HJB equation. This unifies safety,
convergence, and inverse optimality
under the stated feasibility and
stability conditions, as formulated
in \eqref{eq:constrained_fixed_time_cost}.
\end{remark}

For the identified dynamics
\(
\dot{\bx}
=
\hat{\bfv}(\bx)
+
\bg(\bx)\bu
\),
the Hamiltonian associated with
\eqref{eq:constrained_fixed_time_cost} is
\(
H
=
\bm L_1(\bx)\bu
+
L_2(\bx)
+
\mathcal B(\bx)
+
\mathcal U_c(\bu)
+
\nabla V_d^\top(\bx)\hat{\bfv}(\bx)
+
\nabla V_d^\top(\bx)\bg(\bx)\bu
\).
The corresponding HJB equation is
\(
\min_{\bu\in\mathbb U}
H(\bx,\bu,\nabla V_d)
=
0
\).
Solving for
\(
\bu
\)
gives
\begin{equation}
\begin{aligned}
    \bu^*(\bx)
    =
    -\bm U
    \tanh
    \Bigg[
        \frac{1}{2}
        \bm U^{-1}\bR^{-1}
        \Big(
            \bm L_1^\top(\bx)
            +
            \bg^\top(\bx)\nabla V_d(\bx)
        \Big)
    \Bigg].
\end{aligned}
\label{eq:general_constrained_optimal_policy}
\end{equation}

\begin{remark}[Compact-Domain Invariance]
\label{rmk3}
Compact operating domains are commonly
employed in neural ADP and IRL for
identification, approximation, and
stability analysis
\cite{Vamvoudakis2010,Modares2014,
Han2024ETIRL,Chang2024IRLTracking,
Tan2024ETIRL,Liang2024DETIRL,
Jia2025ETIRL,Hong2025DETADP}.
However, guaranteeing that trajectories
remain within such domains is nontrivial
and may require stabilizing or backup
controllers
\cite{Cohen2023SafeExploration,
Rabiee2025BackupCBF}.
Unlike approaches that rely solely on
an assumed operating domain, the present
work embeds a barrier into the prescribed
value function. Under the established
closed-loop bounds, the resulting
boundedness of $V_d(\bx)$ prevents
trajectories from reaching the
state-constraint boundary, thereby
establishing forward invariance of
a compact sublevel set contained in
the admissible region.
This provides a safety-based justification
for the compact-domain assumption
used in the subsequent learning analysis.
\end{remark}

Remark~\ref{rmk3} also highlights a trade-off involved in using a barrier function to define a compact set for the system's operation. It restricts the system's state and operating region; consequently, the learning process—or the effectiveness of the control signals generated by the algorithm—is only meaningful within this domain. In return, however, we obtain an explicit bound that facilitates computation.

\subsection{Reverse Construction of the Fixed-Time Optimal Cost}
\label{subsec:reverse_fixed_time_cost}

The following theorem establishes an
inverse-optimal construction in which the
weighted minimum-effort control satisfying the
prescribed fixed-time decay is recovered as the
minimizing policy~\cite{Petersen1987ControlEffort,Primbs1999CLFRecedingHorizon}. The associated
cost components are constructed explicitly
from the desired value function and the
identified system dynamics.

\begin{theorem}[Constrained Fixed-Time Inverse-Optimal Construction]
\label{thm:fixed_time_inverse_optimal}
Consider the identified nonlinear system
$\dot{\bx}=\hat{\bfv}(\bx)+\bg(\bx)\bu$
and the prescribed value function $V_d(\bx)$
satisfying the desired fixed-time decay
\eqref{eq:desired_constrained_fixed_time_decay}.
Define
\[
\bc_d(\bx)=\bg^\top(\bx)\nabla V_d(\bx),
\]
\[
\sigma_d(\bx)=
\nabla V_d^\top(\bx)\hat{\bfv}(\bx)
+aV_d^{\gamma_1}(\bx)+bV_d^\nu(\bx).
\]

Suppose that, for all
$\bx\in\Omega\setminus\{\bm0\}$,
\[
\bc_d(\bx)\neq\bm0,\qquad
|\sigma_d(\bx)|
<\sum_{j=1}^{m}\bar u_j|c_{d,j}(\bx)|,
\]
where $c_{d,j}(\bx)$ denotes the $j$-th
component of $\bc_d(\bx)$.
Let $\lambda^*(\bx)$ be the unique solution of
\begin{equation}
\bc_d^\top(\bx)\bU
\tanh\left(
\frac{\lambda^*(\bx)}{2}
\bU^{-1}\bR^{-1}\bc_d(\bx)
\right)
=\sigma_d(\bx),
\label{eq:constrained_lagrange_multiplier}
\end{equation}

and define
\begin{equation}
\bu_c(\bx)=
-\bU\tanh\left(
\frac{\lambda^*(\bx)}{2}
\bU^{-1}\bR^{-1}\bc_d(\bx)
\right).
\label{eq:continuous_fixed_time_controller}
\end{equation}
If the running-cost components are chosen as
\begin{equation}
\bm L_1^\top(\bx)
=
[\lambda^*(\bx)-1]\bc_d(\bx),
\label{eq:L1_design}
\end{equation}
and
\begin{equation}
\begin{aligned}
L_2(\bx)
={}&-\nabla V_d^\top(\bx)\hat{\bfv}(\bx)
-B(\bx)\\
&-\mathcal U_c(\bu_c(\bx))
+\lambda^*(\bx)\sigma_d(\bx),
\end{aligned}
\label{eq:L2_design}
\end{equation}
then $\bu_c(\bx)$ uniquely minimizes
$\mathcal U_c(\bu)$ subject to
$\bc_d^\top(\bx)\bu=-\sigma_d(\bx)$
and uniquely minimizes the Hamiltonian
associated with
\[
\ell_c(\bx,\bu)=
\bm L_1(\bx)\bu+L_2(\bx)
+B(\bx)+\mathcal U_c(\bu).
\]
Consequently,
\[
\min_{\bu\in\mathcal U}
H(\bx,\bu,\nabla V_d(\bx))=0,
\]
where
$\mathcal U=\{\bu\in\mathbb R^m:
|u_j|<\bar u_j,\ j=1,\ldots,m\}$.
Moreover, the nominal closed-loop system
satisfies
$\dot V_d=-aV_d^{\gamma_1}-bV_d^\nu$.
\end{theorem}

\begin{proof}
Along the identified dynamics,
\[
\dot V_d
=\nabla V_d^\top\hat{\bfv}
+\bc_d^\top\bu.
\]
Imposing
\eqref{eq:desired_constrained_fixed_time_decay}
yields the constraint
\[
\bc_d^\top(\bx)\bu=-\sigma_d(\bx).
\]
We therefore consider
\[
\min_{\bu\in\mathcal U}\mathcal U_c(\bu)
\quad\text{s.t.}\quad
\bc_d^\top(\bx)\bu+\sigma_d(\bx)=0.
\]
The associated Lagrangian is
\[
\mathcal L_c(\bu,\lambda_c)
=\mathcal U_c(\bu)
+\lambda_c(\bc_d^\top\bu+\sigma_d).
\]
Since
\[
\nabla_{\bu}\mathcal U_c(\bu)
=2\bR\bU\tanh^{-1}(\bU^{-1}\bu),
\]
the stationarity condition
$\nabla_{\bu}\mathcal L_c=\bm0$
gives
\[
\bu(\lambda_c)
=-\bU\tanh\left(
\frac{\lambda_c}{2}
\bU^{-1}\bR^{-1}\bc_d
\right).
\]
Substitution into the equality constraint
gives
\eqref{eq:constrained_lagrange_multiplier}.

To establish uniqueness, define
\[
G_{\bx}(\lambda_c)
=\sum_{j=1}^{m}
c_{d,j}\bar u_j
\tanh\left(
\frac{\lambda_c c_{d,j}}
{2r_j\bar u_j}
\right).
\]
Its derivative satisfies
\[
G_{\bx}'(\lambda_c)
=\frac12\sum_{j=1}^{m}
\frac{c_{d,j}^2}{r_j}
\operatorname{sech}^2\left(
\frac{\lambda_c c_{d,j}}
{2r_j\bar u_j}
\right)>0
\]
whenever $\bc_d\neq\bm0$, and
\[
\lim_{\lambda_c\to\pm\infty}
G_{\bx}(\lambda_c)
=\pm\sum_{j=1}^{m}
\bar u_j|c_{d,j}|.
\]
The strict input-feasibility condition
therefore guarantees a unique finite
solution $\lambda^*(\bx)$.
Moreover,
\[
\nabla_{\bu}^2\mathcal U_c(\bu)
=
\operatorname{diag}\left(
\frac{2r_j}
{1-(u_j/\bar u_j)^2}
\right)\succ\bm0.
\]
Hence, \eqref{eq:continuous_fixed_time_controller}
is the unique constrained minimum-effort
solution, with
$|u_{c,j}|<\bar u_j$.
Since $\bc_d^\top\bu_c=-\sigma_d$,
\[
\dot V_d
=\nabla V_d^\top\hat{\bfv}
-\sigma_d
=-aV_d^{\gamma_1}-bV_d^\nu,
\]
establishing the prescribed nominal
fixed-time decay.

Next, the Hamiltonian associated with
the identified model is
\[
\begin{aligned}
H(\bx,\bu,\nabla V_d)
={}&(\bm L_1^\top+\bc_d)^\top\bu
+\mathcal U_c(\bu)\\
&+L_2+B+\nabla V_d^\top\hat{\bfv}.
\end{aligned}
\]
From
\eqref{eq:general_constrained_optimal_policy},
its minimizing policy is
\[
\bu^*
=-\bU\tanh\left[
\frac12\bU^{-1}\bR^{-1}
(\bm L_1^\top+\bc_d)
\right].
\]
Requiring $\bu^*=\bu_c$ and using
the injectivity of $\tanh(\cdot)$
gives
\[
\bm L_1^\top+\bc_d
=\lambda^*\bc_d,
\]
which establishes
\eqref{eq:L1_design}.

With this choice,
\[
H
=\lambda^*\bc_d^\top\bu
+\mathcal U_c(\bu)
+L_2+B+\nabla V_d^\top\hat{\bfv}.
\]
The stationarity condition
\[
\nabla_{\bu}H
=\lambda^*\bc_d
+2\bR\bU\tanh^{-1}(\bU^{-1}\bu)
=\bm0
\]
holds at $\bu=\bu_c$.
Since
$\nabla_{\bu}^2H
=\nabla_{\bu}^2\mathcal U_c\succ\bm0$,
$\bu_c$ is the unique Hamiltonian
minimizer. Evaluating at $\bu_c$,
\[
\begin{aligned}
H(\bx,\bu_c,\nabla V_d)
={}&-\lambda^*\sigma_d
+\mathcal U_c(\bu_c)\\
&+L_2+B+\nabla V_d^\top\hat{\bfv}.
\end{aligned}
\]
Imposing
$H(\bx,\bu_c,\nabla V_d)=0$
yields \eqref{eq:L2_design}.
Consequently,
\[
\min_{\bu\in\mathcal U}
H(\bx,\bu,\nabla V_d)=0
\]
on $\Omega\setminus\{\bm0\}$,
completing the proof.
\end{proof}

\begin{remark}[Alternative computational formulations]
The scalar multiplier $\lambda_c$ can be
computed using symbolic or numerical
root-finding tools in MATLAB.
Alternatively, one may consider a
quadratic control objective
$\bu^\top\bR\bu$ subject to the prescribed
Lyapunov-decay condition and additional
input constraints. Under this formulation,
the Lagrange multiplier can be determined
more conveniently and even admits an
explicit expression when only the
Lyapunov-decay equality constraint is
considered. However, incorporating
additional input constraints requires
reformulating the optimization problem
and revisiting the inverse-optimal
construction in
Theorem~\ref{thm:fixed_time_inverse_optimal}.
To retain a general formulation
that incorporates input constraints
directly into the control penalty,
the present work adopts the constrained
optimization approach developed in
the proof of
Theorem~\ref{thm:fixed_time_inverse_optimal}.
\end{remark}

\begin{remark}

It is worth emphasizing that
$V_d(\bx)$ is a single prescribed
value function shared by both the
true and identified systems, rather
than two independently defined value
functions. The inverse-optimal
construction is performed using the
identified dynamics to enforce the
prescribed fixed-time decay of this
common function. When the resulting
controller is applied to the true
system, the same $V_d(\bx)$ is retained,
while the identification error introduces
an additional term in its time derivative.
This residual is explicitly accounted
for in the subsequent stability analysis.

\end{remark}

\begin{lemma}[Continuous Extension at the Origin]
\label{lem:origin_extension}
Suppose that the conditions of
Theorem~\ref{thm:fixed_time_inverse_optimal}
hold in a punctured neighborhood of
$\bm0$, and that
$V_d$ is continuously differentiable
with $V_d(\bm0)=0$.
Assume that $\bg$ and $\hat{\bfv}$
are continuous at the origin and
\(\lim_{\bx\to\bm0}
|\sigma_d(\bx)|/\|\bc_d(\bx)\|=0\).
Then
\(\lim_{\bx\to\bm0}\bu_c(\bx)=\bm0\),
\(\lim_{\bx\to\bm0}\bm L_1(\bx)=\bm0\),
and
\(\lim_{\bx\to\bm0}L_2(\bx)=0\).
Consequently, defining
\(\bu_c(\bm0)=\bm0\),
\(\bm L_1(\bm0)=\bm0\), and
\(L_2(\bm0)=0\)
provides continuous extensions,
and the HJB equation remains valid
at the origin.
\end{lemma}

\begin{proof}
For $\bx\neq\bm0$, consider the
minimum-Euclidean-norm solution of
$\bc_d^\top\bu=-\sigma_d$, given by
\(\bu_0(\bx)=
-\sigma_d(\bx)\bc_d(\bx)/
\|\bc_d(\bx)\|^2\).
By construction,
\(\bc_d^\top\bu_0=-\sigma_d\)
and
\(\|\bu_0\|=
|\sigma_d|/\|\bc_d\|\to0\).
Hence, $\bu_0\in\mathcal U$
sufficiently near the origin.
Since $\bu_c$ is the unique
constrained minimum-effort solution,
\(0\leq\mathcal U_c(\bu_c)
\leq\mathcal U_c(\bu_0)\to0\).

To establish convergence of $\bu_c$,
recall from \eqref{U} that
\(\nabla_{\bu}\mathcal U_c(\bu)
=2\bR\bU\tanh^{-1}(\bU^{-1}\bu)\).
Consequently,
\(\nabla_{\bu}^2\mathcal U_c(\bu)
=\operatorname{diag}
\left(
2r_j/[1-(u_j/\bar u_j)^2]
\right)\).
Since $|u_j|<\bar u_j$,
\(\nabla_{\bu}^2\mathcal U_c(\bu)
\succeq2r_{\min}\bI\succ\bm0\),
where \(r_{\min}=\min_jr_j>0\).
Thus, $\mathcal U_c$ is strongly
convex on $\mathcal U$.
Using
\(\mathcal U_c(\bm0)=0\) and
\(\nabla_{\bu}\mathcal U_c(\bm0)=\bm0\),
strong convexity yields
\(\mathcal U_c(\bu)\geq
r_{\min}\|\bu\|^2\).
Therefore,
\(0\leq r_{\min}\|\bu_c\|^2
\leq\mathcal U_c(\bu_c)
\leq\mathcal U_c(\bu_0)\to0\),
which implies
\(\bu_c\to\bm0\).

The stationarity condition gives
\(\lambda^*\bc_d
=-\nabla_{\bu}\mathcal U_c(\bu_c)
\to\bm0\).
Since
\(\bc_d=\bg^\top\nabla V_d\to\bm0\),
we obtain
\(\bm L_1^\top
=(\lambda^*-1)\bc_d\to\bm0\).
Moreover, using
\(\bc_d^\top\bu_c=-\sigma_d\),
we have
\(\lambda^*\sigma_d
=-(\lambda^*\bc_d)^\top\bu_c
=\nabla_{\bu}\mathcal U_c(\bu_c)^\top
\bu_c\to0\).
Together with
\(\nabla V_d^\top\hat{\bfv}\to0\),
\(\mathcal B\to0\), and
\(\mathcal U_c(\bu_c)\to0\),
equation~\eqref{eq:L2_design}
yields \(L_2\to0\).

Finally, since $V_d$ is
continuously differentiable and
positive definite with
$V_d(\bm0)=0$, one has
$\nabla V_d(\bm0)=\bm0$.
With the continuous extensions
$\bu_c(\bm0)=\bm0$,
$\bm L_1(\bm0)=\bm0$,
$L_2(\bm0)=0$, and
$\mathcal B(\bm0)=0$, the Hamiltonian
at the origin reduces to
\(H(\bm0,\bu,\nabla V_d(\bm0))
=\mathcal U_c(\bu)\geq0\),
with equality uniquely at
$\bu=\bm0$.
Hence,
\(\min_{\bu\in\mathcal U}
H(\bm0,\bu,\nabla V_d(\bm0))=0\),
completing the proof.
\end{proof}

\subsection{Koopman-Initialized Critic-Only Integral Reinforcement Learning}

% Define
% \(
% V_B(\bx)
% :=
% \kappa_B\mathcal B(\bx)
% \)
% and write
% \(
% V^*(\bx)
% =
% V_B(\bx)
% +
% \bW^{*\top}\bphi(\bx)
% +
% \varepsilon(\bx)
% \),
% where
% \(
% \bphi(\bx)\in\mathbb R^{N_c}
% \),
% \(
% \bW^*\in\mathbb R^{N_c}
% \),
% and
% \(
% \varepsilon(\bx)
% \)
% is the value-function approximation error. Hence,
% \(
% \nabla V^*
% =
% \nabla V_B
% +
% (\nabla\bphi)^\top\bW^*
% +
% \nabla\varepsilon
% \).

% The critic estimate is selected as
% \(
% \hat V(\bx)
% =
% V_B(\bx)
% +
% \hat{\bW}^{\top}\bphi(\bx)
% \),
% with
% \(
% \nabla\hat V(\bx)
% =
% \nabla V_B(\bx)
% +
% (\nabla\bphi(\bx))^\top\hat{\bW}
% \).

Define
\(
V_B(\bx)
:=
\kappa_B\mathcal B(\bx)
\)
and by the
higher-order Weierstrass approximation theorem~\cite{Vamvoudakis2010,Kokolakis2025Predefined,Gong2025} write
\(
V^*(\bx)
=
V_B(\bx)
+
\bW^{*\top}\bphi(\bx)
+
\varepsilon(\bx)
\),
where
\(
\bphi(\bx)\in\mathbb R^{N_c}
\),
\(
\bW^*\in\mathbb R^{N_c}
\),
and
\(
\varepsilon(\bx)
\)
is the value-function approximation error. Hence,
\(
\nabla V^*
=
\nabla V_B
+
(\nabla\bphi)^\top\bW^*
+
\nabla\varepsilon
\).

The critic estimate is selected as
\(
\hat V(\bx)
=
V_B(\bx)
+
\hat{\bW}^{\top}\bphi(\bx)
\),
with
\(
\nabla\hat V(\bx)
=
\nabla V_B(\bx)
+
(\nabla\bphi(\bx))^\top\hat{\bW}
\).

Accordingly, the critic-based constrained feedback policy
is constructed by replacing the desired value-function
gradient in the nominal inverse-optimal controller
\eqref{eq:general_constrained_optimal_policy}
with the estimated critic gradient, yielding
\begin{equation}
\begin{aligned}
\hat{\bu}(\bx,\hat{\bW})
=
-\bm U\tanh\Bigg[
\frac{1}{2}\bm U^{-1}\bR^{-1}
\Big(
\bm L_1^\top(\bx)
+
\bg^\top(\bx)\nabla\hat V(\bx)
\Big)
\Bigg].
\end{aligned}
\label{eq:constrained_critic_policy}
\end{equation}

\begin{proposition}[Critic approximation residual~\cite{Vamvoudakis2010,Kokolakis2025Predefined}]
\label{prop:critic_approximation_residual}
On the compact set
\(
\Omega
\),
suppose that the critic approximation converges
uniformly in the
\(
C^1
\)
sense as
\(
N_C\to\infty
\).
Then,
\[
\lim_{N_C\to\infty}
\sup_{\bx\in\Omega}
\left(
|\varepsilon(\bx)|
+
\|\nabla\varepsilon(\bx)\|
\right)
=0.
\]
\end{proposition}

\begin{remark}[Approximation accuracy and computational trade-off]
Proposition~\ref{prop:critic_approximation_residual}
provides one possible means of reducing the
ultimate bounds established in 
Theorems~\ref{thm:critic_practical_fixed_time_convergence}
and~\ref{thm:composite_constrained_stability}
by increasing the number of critic neurons
\(
N_C
\).
However, approximation residuals are only one
of several factors determining the UUB radii.
Increasing
\(
N_C
\)
may improve approximation accuracy at the
expense of computational complexity, whereas
a smaller network may sacrifice approximation
capability~\cite{ParkSandberg1991}.
This trade-off suggests a possible extension
using an adaptive network architecture, in which
\(
N_C
\)
is adjusted online through neuron allocation
or growing-and-pruning mechanisms, as explored
in~\cite{Platt1991RAN,Han2010SORBF}.
\end{remark}

\subsubsection{Koopman-Based Critic Initialization}
\label{koopman}

A critical issue in critic-based reinforcement learning
is the initialization of the estimated weights
\(
\hat{\bW}(0)
\)~\cite{Xu2024RLWeightInitialization},
which directly determine the initial control policy.
Poor initialization may induce large control transients
and even compromise closed-loop stability before
learning becomes effective.
This sensitivity was explicitly demonstrated in
\cite{VuDoan2026WarmStart}, where Fig.~7 shows
that adverse critic initializations can lead to
severely degraded closed-loop responses.
Although critic-weight adaptation has been extensively
studied in
\cite{Vamvoudakis2010,Modares2014,Han2024ETIRL},
systematic initialization receives comparatively
limited attention, with initial weights often
prescribed rather than constructed from data.
The difficulty is further compounded when the
system dynamics are unknown, making it challenging
to obtain an informed initial policy.

The Koopman operator, originally introduced in
\cite{Koopman1931}, provides a linear
operator-theoretic representation of nonlinear
dynamics through observable functions.
Its spectral characterization was further developed
in \cite{Mezic2005}, while data-driven approximations
enabled finite-dimensional linear predictors for
nonlinear control
\cite{Williams2015Koopman,Brunton2016Koopman,Korda2018}.
Motivated by these developments and inspired by the perspective of the~\cite{Xu2024RLWeightInitialization}, we construct
a Koopman-based surrogate value function from
finite trajectory data and use it to initialize
the critic weights, thereby providing an informed
starting point for subsequent integral
reinforcement learning.

In~\cite{Xu2024RLWeightInitialization},
Nonlinear Model Predictive Control (NMPC)
was employed to initialize the actor-network
weights through supervised learning.
Since the system model was assumed to be known,
the NMPC problem could be solved to generate
optimal state-control trajectories, which were
subsequently used to determine the initial
network weights, as described in Algorithm 1
of~\cite{Xu2024RLWeightInitialization}.

In contrast, the main difficulty in the present
work is that the system dynamics are unknown.
An alternative is to follow the Koopman-based
identification framework developed in
\cite{Su2026KoopmanSafe}, where suitable
observable dictionaries must first be selected
to construct a lifted dynamical model from
measured data. However, finite-dimensional
Koopman approximations inevitably introduce
approximation errors, since an exact Koopman
representation generally requires an
infinite-dimensional observable space.
Moreover, for control-affine nonlinear systems,
the lifted dynamics generally exhibit a
bilinear dependence on the control input,
rather than the purely linear structure
commonly adopted in finite-dimensional
approximations. Consequently, directly
designing a controller based on such an
approximate lifted model may introduce
additional model-dependent errors.

Integral reinforcement learning (IRL)
provides another data-driven alternative
by exploiting measured trajectories of
the actual system without requiring
explicit knowledge of the drift dynamics.
Nevertheless, the input matrix
$\bg(\bx)$ must generally be available
to reconstruct the control policy.
Fortunately, in many practical
control-affine systems, $\bg(\bx)$
is known, whereas the drift dynamics
remain unknown.

Motivated by these observations, the
present work extends the initialization
strategy of~\cite{Xu2024RLWeightInitialization}
to unknown nonlinear systems by employing
a data-driven Koopman lifting model.
Unlike~\cite{Xu2024RLWeightInitialization},
which requires known system dynamics
to generate reference control trajectories,
the proposed approach first identifies
a finite-dimensional Koopman model
from measured data and subsequently
uses it to construct a constrained
linear MPC problem.

\begin{figure*}[!t]
    \centering
    \includegraphics[width=\textwidth]{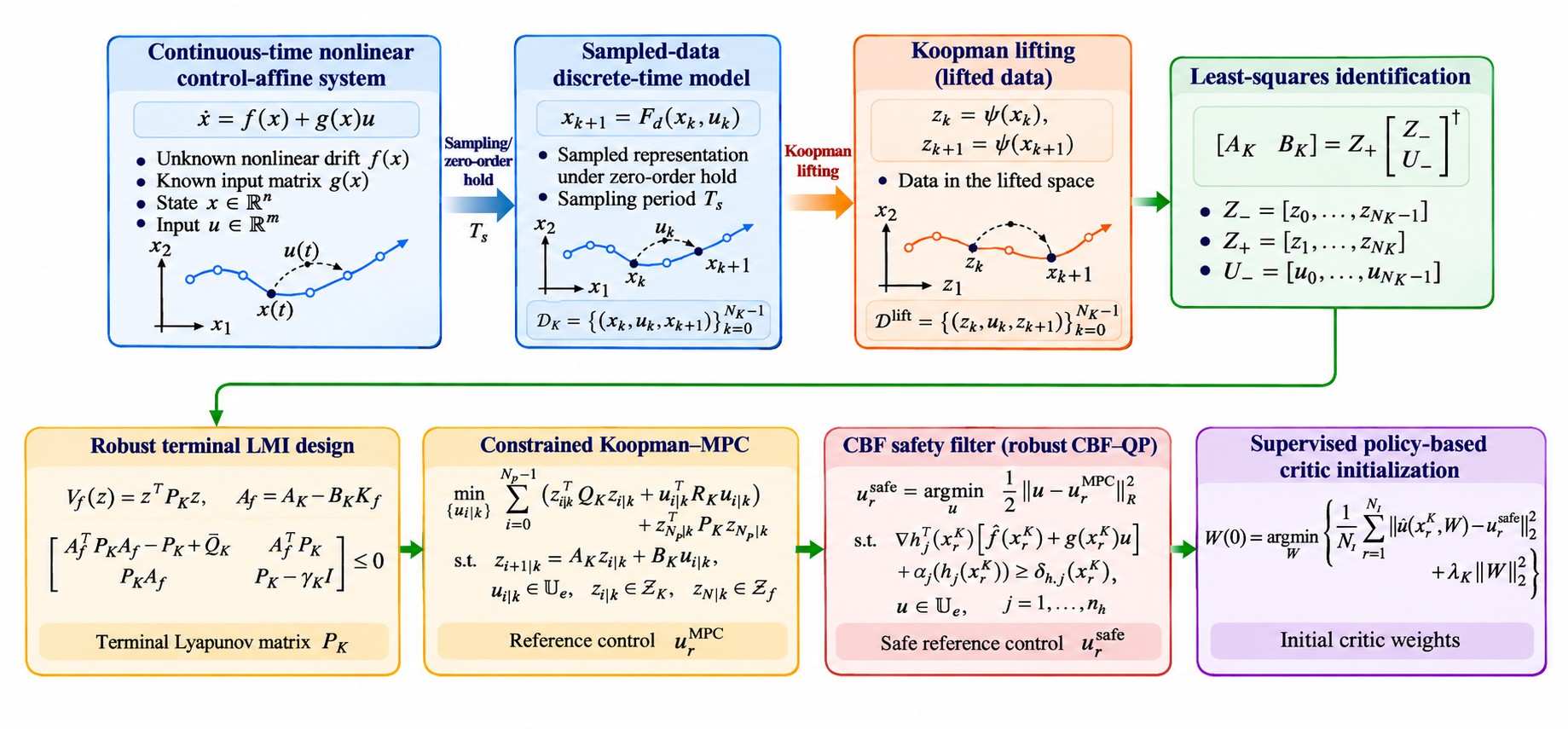}
    \caption{KoopmanMPC-based critic initialization
    through sampled-data identification,
    robust LMI terminal design, constrained
    MPC, and supervised policy matching.}
    \label{fig:koopman_lifting}
\end{figure*}

The choice of MPC is motivated by the
requirement that control actions satisfy
the prescribed constraints even during
the critic initialization stage.
By explicitly incorporating input and
state constraints into the optimization
problem, Koopman-based MPC provides
a systematic means of generating
constraint-aware reference control actions.
These actions are then used as supervised
learning targets to initialize the
critic weights by minimizing the
discrepancy between the MPC-generated
controls and the critic-induced policy.
Unlike the actor-network initialization
in~\cite{Xu2024RLWeightInitialization},
the present approach directly initializes
the critic weights within a critic-only
learning architecture (See Figure~\ref{fig:koopman_lifting}). The resulting Koopman--MPC-based
policy initialization constitutes
one of the contributions of the
proposed framework.

To initialize the critic weights, consider
the sampled dataset
\(
\mathcal D_K
=
\{(\bx_k,\bu_k,\bx_{k+1})\}_{k=0}^{N_K-1}
\).
Let
\(
\bm z_k=\boldsymbol\psi(\bx_k)
\in\mathbb R^{n_K}
\),
with
\(
\boldsymbol\psi(\bm0)=\bm0
\), where \(\boldsymbol\psi(.)\) is the observer for koopman lifting .
Defining
\(
\bm Z_-=[\bm z_0,\ldots,\bm z_{N_K-1}]
\),
\(
\bm Z_+=[\bm z_1,\ldots,\bm z_{N_K}]
\),
and
\(
\bm U_-=[\bu_0,\ldots,\bu_{N_K-1}]
\),
the Koopman matrices are identified by
\begin{equation}
\left[\bm A_K\;\bm B_K\right]
=
\bm Z_+
\begin{bmatrix}
\bm Z_-\\
\bm U_-
\end{bmatrix}^{\dagger}.
\label{eq:koopman_identification}
\end{equation}

The identified lifted dynamics are modeled as
\(
\bm z_{k+1}
=
\bm A_K\bm z_k+\bm B_K\bu_k+\bw_k
\),
where
\(
\bw_k=\bm\varepsilon_{K,k}
\)
denotes the Koopman approximation residual,
satisfying
\(
\bw_k\in\mathcal W
:=
\{\bw:\|\bw\|\leq\bar\varepsilon_K\}
\)
on the considered operating domain.

Recent studies have investigated robust
constraint satisfaction, recursive
feasibility, and stability of Koopman-based
MPC under modeling errors
\cite{Yan2026RobustKoopmanMPC,Schimperna2026KoopmanMPC,Strasser2026SafEDMD}.
Motivated by these results, a robust
terminal Lyapunov function is constructed
through an LMI-based design before
formulating the constrained Koopman--MPC
problem~\cite{Boyd1994LMI,Zhou1996RobustControl,
Dullerud2000RobustControl,Xie2026MinMaxMPC}.

Consider the auxiliary feedback
\(
\bu_k=-\bm K_f\bm z_k
\),
where
\(
\bm A_f:=\bm A_K-\bm B_K\bm K_f
\)
is Schur stable. The corresponding
disturbed dynamics become
\(
\bm z_{k+1}
=
\bm A_f\bm z_k+\bw_k
\).
Choose
\(
V_f(\bm z)=\bm z^\top\bm P_K\bm z
\),
with
\(
\bm P_K\succ0
\),
and define
\(
\bar{\bm Q}_K
=
\bm Q_K+\bm K_f^\top\bm R_K\bm K_f
\),
where
\(
\bm Q_K\succ0
\)
and
\(
\bm R_K\succ0
\).

The desired robust dissipation condition is
\begin{equation}
\Delta V_f
\leq
-\bm z_k^\top\bar{\bm Q}_K\bm z_k
+
\gamma_K^2\|\bw_k\|^2,
\label{eq:koopman_robust_dissipation}
\end{equation}
where
\(
\Delta V_f
=
V_f(\bm z_{k+1})-V_f(\bm z_k)
\).
Substituting the disturbed dynamics gives
\(
\Delta V_f
=
\bm z_k^\top
(\bm A_f^\top\bm P_K\bm A_f-\bm P_K)
\bm z_k
+
2\bm z_k^\top\bm A_f^\top\bm P_K\bw_k
+
\bw_k^\top\bm P_K\bw_k
\).
Hence,
\eqref{eq:koopman_robust_dissipation}
is ensured by
\begin{equation}
\begin{bmatrix}
\bm A_f^\top\bm P_K\bm A_f
-\bm P_K+\bar{\bm Q}_K
&
\bm A_f^\top\bm P_K
\\
\bm P_K\bm A_f
&
\bm P_K-\gamma_K^2\bm I
\end{bmatrix}
\preceq\bm0.
\label{eq:koopman_robust_lmi}
\end{equation}

To determine
\(
\bm K_f
\)
and
\(
\bm P_K
\)
simultaneously, introduce the change
of variables
\(
\bm X=\bm P_K^{-1}
\)
and
\(
\bm Y=\bm K_f\bm X
\),
and define
\(
\bm F=\bm A_K\bm X-\bm B_K\bm Y
\),
\(
\eta_K=\gamma_K^2
\),
and
\[
\bm C=
\begin{bmatrix}
\bm Q_K^{1/2}\bm X\\
\bm R_K^{1/2}\bm Y
\end{bmatrix}.
\]
Applying the Schur complement and
a congruence transformation to
\eqref{eq:koopman_robust_lmi}
yields the equivalent LMI
\begin{equation}
\begin{bmatrix}
\bm X & \bm0 & \bm F^\top & \bm C^\top\\
\bm0 & \eta_K\bm I & \bm I & \bm0\\
\bm F & \bm I & \bm X & \bm0\\
\bm C & \bm0 & \bm0 & \bm I
\end{bmatrix}
\succeq\bm0.
\label{eq:koopman_synthesis_lmi}
\end{equation}

The terminal feedback and Lyapunov
matrix are therefore jointly obtained
from the semidefinite program
\begin{equation}
\begin{aligned}
\min_{\bm X,\bm Y,\eta_K}\quad&
\eta_K\\
\mathrm{s.t.}\quad&
\bm X\succeq\epsilon_X\bm I,\\
&
\bm X\preceq\epsilon_P^{-1}\bm I,\\
&
\eqref{eq:koopman_synthesis_lmi},
\end{aligned}
\label{eq:koopman_terminal_sdp}
\end{equation}
where
\(
\epsilon_X,\epsilon_P>0
\).
The decision variables enter
\eqref{eq:koopman_terminal_sdp}
affinely, making it a convex
semidefinite program.
A feasible solution yields
\begin{equation}
\bm P_K=\bm X^{-1},
\qquad
\bm K_f=\bm Y\bm X^{-1},
\qquad
\gamma_K=\sqrt{\eta_K}.
\label{eq:koopman_terminal_recovery}
\end{equation}
Consequently,
\(
\bm P_K\succ0
\)
and the recovered terminal feedback
satisfies the robust dissipation
condition
\eqref{eq:koopman_robust_dissipation}.

Summing
\eqref{eq:koopman_robust_dissipation}
over
\(
k=0,\ldots,N-1
\)
gives
\(
V_f(\bm z_N)
+
\sum_{k=0}^{N-1}
\bm z_k^\top\bar{\bm Q}_K\bm z_k
\leq
V_f(\bm z_0)
+
\gamma_K^2
\sum_{k=0}^{N-1}\|\bw_k\|^2
\).
If
\(
\bw\in\ell_2
\),
the state energy is finite and
\(
\bm z_k\rightarrow\bm0
\)
for the auxiliary terminal closed loop.
For bounded residuals, the dissipation
inequality provides an energy-based
robustness estimate without requiring
the stronger
\(
\ell_2
\)
assumption.

\begin{remark}[Finite-Dimensional Koopman Approximation]
The Koopman operator is generally
infinite-dimensional, and exact
finite-dimensional invariant subspaces
are not guaranteed
\cite{Brunton2016Koopman,Korda2018Koopman}.
Moreover, lifted control-affine dynamics
may contain bilinear input-dependent terms
\cite{Bruder2021BilinearKoopman}.
The identified linear model is therefore
treated as an approximation with a bounded
residual, explicitly accounted for by
the robust dissipation condition
\eqref{eq:koopman_robust_lmi}.
\end{remark}

Using the LMI-derived
\(
\bm P_K
\)
as the terminal cost and
\(
\kappa_f(\bm z)=-\bm K_f\bm z
\)
as the terminal controller, the
constrained Koopman--MPC problem
for a prediction horizon
\(
N_p
\)
is formulated as
\begin{equation}
\begin{aligned}
\min_{\{\bu_{i|k}\}_{i=0}^{N_p-1}}
\quad&
\sum_{i=0}^{N_p-1}
\left(
\bm z_{i|k}^\top\bm Q_K\bm z_{i|k}
+
\bu_{i|k}^\top\bm R_K\bu_{i|k}
\right)\\
&+
\bm z_{N_p|k}^\top
\bm P_K\bm z_{N_p|k}\\
\mathrm{s.t.}\quad&
\bm z_{0|k}
=
\boldsymbol\psi(\bx_k),\\
&
\bm z_{i+1|k}
=
\bm A_K\bm z_{i|k}
+
\bm B_K\bu_{i|k},\\
&
\bu_{i|k}\in\mathbb U_e,\\
&
\bm z_{i|k}\in\mathcal Z_K,\\
&
\bm z_{N_p|k}\in\mathcal Z_f,
\end{aligned}
\label{eq:koopman_constrained_mpc}
\end{equation}
where the stage constraints apply for
\(
i=0,\ldots,N_p-1
\),
and
\(
\mathbb U_e
=
\{\bu:|u_j|\leq(1-\epsilon_u)\bar u_j\}
\),
with
\(
0<\epsilon_u<1
\).
The sets
\(
\mathcal Z_K
\)
and
\(
\mathcal Z_f
\)
denote the lifted-state and terminal
constraints, respectively~
\cite{Rawlings2026MPC,Ellis2017EMPC,Kouvaritakis2016MPC,Lee2026MPC,Guay2016RobustMPC,Zhu2026EventMPC,Song2026ComplexMPC,Camacho2026MPC,Kwon2005RHC,Rakovic2019MPC}
.

For nominal terminal admissibility,
\(
\kappa_f(\bm z)\in\mathbb U_e
\)
and
\(
\bm A_f\bm z\in\mathcal Z_f
\)
are imposed for every
\(
\bm z\in\mathcal Z_f
\).
Robust terminal invariance additionally
requires
\(
\bm A_f\mathcal Z_f\oplus\mathcal W
\subseteq\mathcal Z_f
\),
where \(\oplus\) denotes the Minkowski sum.
Together with suitable constraint
tightening, these conditions provide
a basis for extending nominal feasibility
guarantees to the disturbed lifted
dynamics.

% For nominal terminal admissibility,
% \(
% \kappa_f(\bm z)\in\mathbb U_\epsilon
% \)
% and
% \(
% \bm A_f\bm z\in\mathcal Z_f
% \)
% are imposed for every
% \(
% \bm z\in\mathcal Z_f
% \).
% Robust terminal invariance additionally
% requires
% \(
% \bm A_f\mathcal Z_f\oplus\mathcal W
% \subseteq\mathcal Z_f
% \).
% Together with suitable constraint
% tightening, these conditions are relevant
% to extending nominal feasibility
% guarantees to the disturbed lifted
% dynamics.

% \begin{figure}[t]
% \centering
% \includegraphics[width=\columnwidth]{AH/Achi.png}
% \caption{Koopman-based critic initialization
% through sampled-data identification,
% robust LMI terminal design, constrained
% MPC, and supervised policy matching.}
% \label{fig:koopman_lifting}
% \end{figure}

% For each feasible initialization state
% \(
% \bx_r^K\in\Omega_K\subseteq\Omega
% \),
% the MPC reference action is selected as
% \(
% \bu_r^{\mathrm{MPC}}=\bu_{0|r}^*
% \).
% The resulting dataset is
% \(
% \mathcal D_{\mathrm{MPC}}
% =
% \{(\bx_r^K,\bu_r^{\mathrm{MPC}})\}_{r=1}^{N_I}
% \).

For each feasible initialization state
\(
\bx_r^K\in\Omega_K\subseteq\Omega
\),
the MPC reference action is selected as
\(
\bu_r^{\mathrm{MPC}}=\bu_{0|r}^*
\).
To ensure that the initialization actions
also satisfy the prescribed safety
constraints, a robust CBF-based safety
filter is applied to the MPC ooutput~\cite{Ames2017CBFQP}.
Specifically, the safe reference action
is obtained from
\begin{equation}
\begin{aligned}
\bu_r^{\mathrm{safe}}
=\arg\min_{\bu}\quad&
\frac12
\|\bu-\bu_r^{\mathrm{MPC}}\|_{\bR}^2\\
\mathrm{s.t.}\quad&
\nabla h_j^\top(\bx_r^K)
\left[
\hat{\boldsymbol f}(\bx_r^K)
+\bg(\bx_r^K)\bu
\right]\\
&\quad+
\alpha_j(h_j(\bx_r^K))
\geq
\delta_{h,j}(\bx_r^K),\\
&\bu\in\mathbb U_\epsilon,
\quad j=1,\ldots,n_h,
\end{aligned}
\label{eq:koopman_mpc_cbf_filter}
\end{equation}
where
\(
\delta_{h,j}(\bx)
\geq
\left|
\nabla h_j^\top(\bx)
[\boldsymbol f(\bx)-\hat{\boldsymbol f}(\bx)]
\right|
\)
bounds the effect of the identification
error on the CBF derivative, and
\(
\alpha_j
\)
is an extended class-\(\mathcal K\)
function. Under CBF feasibility and
the stated error bounds, the filtered
actions satisfy the robust CBF
conditions at the initialization states.
The resulting dataset is
\(
\mathcal D_{\mathrm{safe}}
=
\{(\bx_r^K,\bu_r^{\mathrm{safe}})\}_{r=1}^{N_I}
\). 
At this point, a natural question arises:
``If the identified dynamics $\hat{\boldsymbol f}(\bx)$
are already available, why not design MPC directly
using this model instead of introducing Koopman lifting?
Furthermore, once the MPC-generated control actions
have been obtained, why not apply them directly rather
than proceeding with the more elaborate learning
framework developed in the subsequent sections?''
The answer is straightforward. Although the resulting
control actions may stabilize the system and satisfy
the prescribed safety constraints, they are not
necessarily optimal with respect to the intended
infinite-horizon performance objective. Moreover,
designing MPC directly on the identified nonlinear
dynamics is a reasonable alternative. However, this
would require solving a nonlinear optimization problem,
potentially involving substantial computational
complexity. In particular, accurate approximation
using radial basis function (RBF) neural networks
may require a large number of basis functions, whose
nonlinear structures further complicate the optimization
problem. In contrast, Koopman lifting provides an
approximately linear representation in a higher-dimensional
space, allowing a linear MPC problem to be formulated
and solved more efficiently. Under appropriate convexity
assumptions, the resulting optimization problem admits
a globally optimal solution in the lifted space, although
this does not necessarily imply global optimality for
the original nonlinear system. Nevertheless,
finite-dimensional Koopman approximations inevitably
introduce residual errors, making a claim of global
closed-loop stability considerably stronger and more
difficult to justify. Such a guarantee is not essential
in the present framework, since the Koopman--MPC
controller serves only as a data-informed surrogate
for initializing the critic weights, rather than as
the final control policy. The subsequent learning-based
control policy is ultimately responsible for the
closed-loop stability and performance of the actual
system, as established through the corresponding
stability analysis.

The initial critic weights are selected
by minimizing the discrepancy between
the MPC-generated controls and the
critic-induced policy
\eqref{eq:constrained_critic_policy}:
\begin{equation}
\begin{aligned}
\hat{\bW}(0)
=
\arg\min_{\bW}
\Bigg\{
&\frac{1}{N_I}
\sum_{r=1}^{N_I}
\left\|
\hat{\bu}(\bx_r^K,\bW)
-
\bu_r^{\mathrm{safe}}
\right\|_2^2\\
&+
\lambda_K\|\bW\|_2^2
\Bigg\},
\qquad\lambda_K>0.
\end{aligned}
\label{eq:koopman_policy_matching_loss}
\end{equation}

% Since the critic-induced policy
% contains a hyperbolic tangent,
% the optimization above is generally
% nonlinear. Applying the inverse
% hyperbolic tangent yields a
% computationally convenient transformed
% policy-matching relation.

% Using
% \(
% \hat V(\bx,\bW)
% =
% \bW^\top\bphi(\bx)
% +
% \kappa_B\mathcal B(\bx)
% \),
% the transformed relation is
% \(
% \bm A_r^K\bW\approx\bm y_r^K
% \),
% where
% \(
% \bm A_r^K
% =
% \bg^\top(\bx_r^K)
% \nabla\bphi^\top(\bx_r^K)
% \)
% and
% \begin{equation}
% \begin{aligned}
% \bm y_r^K
% ={}&
% -2\bR\bU
% \tanh^{-1}\left(
% \bU^{-1}\bu_r^{\mathrm{safe}}
% \right)\\
% &-
% \bL_1(\bx_r^K)
% -
% \kappa_B\bg^\top(\bx_r^K)
% \nabla\mathcal B(\bx_r^K).
% \end{aligned}
% \label{eq:koopman_mpc_policy_target}
% \end{equation}

% Stacking
% \(
% \bm{\mathsf A}_K
% =
% [(\bm A_1^K)^\top,\ldots,
% (\bm A_{N_I}^K)^\top]^\top
% \)
% and
% \(
% \bm{\mathsf y}_K
% =
% [(\bm y_1^K)^\top,\ldots,
% (\bm y_{N_I}^K)^\top]^\top
% \),
% the transformed regularized
% least-squares problem becomes
% \(
% \min_{\bW}
% \|\bm{\mathsf A}_K\bW-\bm{\mathsf y}_K\|_2^2
% +\lambda_K\|\bW\|_2^2
% \).
% Since
% \(
% \lambda_K>0
% \),
% the normal matrix is positive definite,
% and the unique solution is
% \begin{equation}
% \hat{\bW}(0)
% =
% \left(
% \bm{\mathsf A}_K^\top\bm{\mathsf A}_K
% +
% \lambda_K\bm I
% \right)^{-1}
% \bm{\mathsf A}_K^\top\bm{\mathsf y}_K.
% \label{eq:koopman_policy_initial_weights}
% \end{equation}

This solution minimizes the transformed
policy residual rather than the
original control-space loss in
\eqref{eq:koopman_policy_matching_loss}.
Thus, the two formulations are not
generally equivalent.
The overall Koopman--MPC-based critic
initialization procedure is summarized
in Algorithm~\ref{alg:koopman_policy_initialization}.

% Note that \eqref{eq:koopman_policy_matching_loss}
% is a standard offline supervised-learning
% problem, which can be solved using
% gradient-based optimizers such as
% SGD, AdaGrad, RMSProp, or Adam
% \cite{Duchi2011AdaGrad,Tieleman2012RMSProp,
% Kingma2015Adam}.
% This allows the critic weights to be
% initialized from MPC-generated data
% before online integral reinforcement
% learning begins.

Note that \eqref{eq:koopman_policy_matching_loss}
can also be regarded as a standard offline
supervised-learning problem, which can be
solved using gradient-based optimizers
such as SGD, AdaGrad, RMSProp, or Adam
\cite{Duchi2011AdaGrad,Tieleman2012RMSProp,
Kingma2015Adam}.
This enables the critic weights to be
initialized from MPC-generated data
before online integral reinforcement
learning begins.

\begin{remark}[Policy Matching and Sampled-Data Consistency]
Exact policy matching requires solving
a system of equations that may be
inconsistent due to differences between
the MPC reference policy and the
critic-induced policy. Regularized
least squares therefore provides a
well-defined approximate initialization.
Moreover, under zero-order hold,
identical control inputs applied at
the same state and sampling instant
produce identical instantaneous state
and Lyapunov derivatives. However,
this pointwise consistency does not
guarantee continuous-time stability,
which also depends on intersample
evolution. Accordingly, Koopman--MPC
provides constraint-aware reference
actions for critic initialization,
while stability of the subsequent
learned closed loop is established
separately.
\end{remark}

\begin{algorithm}[t]
\caption{Koopman--MPC-Based Safe Critic Initialization}
\label{alg:koopman_policy_initialization}
\begin{algorithmic}[1]
\Require Identified Koopman model,
critic policy, feasible sampling region
$\Omega_K$, $N_I$, and $\lambda_K>0$.
\Ensure Initial critic weights $\hat{\bW}(0)$.

\State Compute $\bm P_K$ and $\bm K_f$
from \eqref{eq:koopman_terminal_sdp}.

\State Initialize
$\mathcal D_{\mathrm{safe}}\gets\emptyset$.

\For{$r=1,\ldots,N_I$}
    \State Sample a feasible state
    $\bx_r^K\in\Omega_K$.

    \State Solve
    \eqref{eq:koopman_constrained_mpc}.

    \State Set
    $\bu_r^{\mathrm{MPC}}
    \gets\bu_{0|r}^*$.

    \State Solve the robust CBF-QP
    \eqref{eq:koopman_mpc_cbf_filter}
    to obtain $\bu_r^{\mathrm{safe}}$.

    \State Append
    $(\bx_r^K,\bu_r^{\mathrm{safe}})$
    to $\mathcal D_{\mathrm{safe}}$.
\EndFor

\State Compute $\hat{\bW}(0)$
using \eqref{eq:koopman_policy_matching_loss}.

\State \Return $\hat{\bW}(0)$.
\end{algorithmic}
\end{algorithm}

% \begin{algorithm}[t]
% \caption{Koopman--MPC-Based Critic Initialization}
% \label{alg:koopman_policy_initialization}
% \begin{algorithmic}[1]
% \Require Identified Koopman model,
% critic policy, feasible sampling region
% $\Omega_K$, $N_I$, and $\lambda_K>0$.
% \Ensure Initial critic weights $\hat{\bW}(0)$.

% \State Compute the robust terminal matrix
% $\bm P_K$ from \eqref{eq:koopman_terminal_sdp}.

% \State Initialize
% $\mathcal D_{\mathrm{MPC}}\gets\emptyset$.

% \For{$r=1,\ldots,N_I$}
%     \State Sample a feasible state
%     $\bx_r^K\in\Omega_K$.

%     \State Solve
%     \eqref{eq:koopman_constrained_mpc}.

%     \State Set
%     $\bu_r^{\mathrm{MPC}}
%     \gets\bu_{0|r}^*$.

%     \State Append
%     $(\bx_r^K,\bu_r^{\mathrm{MPC}})$
%     to $\mathcal D_{\mathrm{MPC}}$.
% \EndFor

% \State Construct
% $\bm{\mathsf A}_K$
% and $\bm{\mathsf y}_K$
% using \eqref{eq:koopman_mpc_policy_target}.

% \State Compute $\hat{\bW}(0)$
% using \eqref{eq:koopman_policy_initial_weights}.

% \State \Return $\hat{\bW}(0)$.
% \end{algorithmic}
% \end{algorithm}

\subsubsection{Practical Fixed-Time Critic Learning With Experience Replay}

During learning, a probing signal
\(
\be_p(t)\in\mathbb R^m
\)
is introduced to provide sufficient excitation
for critic-weight adaptation.
The behavior input is defined as
\(
\bu_b(t)=\hat{\bu}(t)+\be_p(t)
\),
with
\(
\be_p(t)
\)
selected such that
\(
\bu_b(t)\in\mathbb U
\).
Thus,
\(
\bu_b-\bu^*
=
(\bu_b-\hat{\bu})
+
(\hat{\bu}-\bu^*)
=
\be_p+(\hat{\bu}-\bu^*)
\).

Along the measured dynamics
\(
\dot{\bx}=\bfv(\bx)+\bg(\bx)\bu_b
\),
the optimal HJB equation is
\(
0
=
\ell(\bx,\bu^*)
+
\nabla V^{*\top}(\bx)
[\bfv(\bx)+\bg(\bx)\bu^*]
\).
Define the policy-suboptimality residual
\(
\delta_{\hat u}(\bx,t)
:=
\ell(\bx,\hat{\bu})
-
\ell(\bx,\bu^*)
+
\nabla V^{*\top}(\bx)\bg(\bx)
(\hat{\bu}-\bu^*)
\).
Consequently,
\(
\nabla V^{*\top}
[\bfv+\bg\hat{\bu}]
=
-\ell(\bx,\hat{\bu})
+
\delta_{\hat u}(\bx,t)
\).

Using
\(
\bfv+\bg\bu_b
=
\bfv+\bg\hat{\bu}
+
\bg(\bu_b-\hat{\bu})
\),
one obtains
\(
\dot V^*
=
-\ell(\bx,\hat{\bu})
+
\delta_{\hat u}(\bx,t)
+
\nabla V^{*\top}\bg(\bu_b-\hat{\bu})
\).
Integrating over
\(
[t-T,t]
\)
gives
\begin{equation}
\begin{aligned}
0
={}&
V^*(\bx(t))
-
V^*(\bx(t-T))
+
\int_{t-T}^{t}
\ell(\bx(\tau),\hat{\bu}(\tau))
\,d\tau
\\
&-
\int_{t-T}^{t}
\nabla V^{*\top}(\bx(\tau))
\bg(\bx(\tau))
[\bu_b(\tau)-\hat{\bu}(\tau)]
\,d\tau
\\
&-
\int_{t-T}^{t}
\delta_{\hat u}(\bx(\tau),\tau)
\,d\tau .
\end{aligned}
\label{eq:constrained_integral_bellman}
\end{equation}

Using the value-function approximation
\(
V^*(\bx)
=
V_B(\bx)
+
\bW^{*\top}\bphi(\bx)
+
\varepsilon(\bx)
\),
define
\(
\Delta\bphi(t)
:=
\bphi(\bx(t))-\bphi(\bx(t-T))
\)
and
\(
\Delta V_B(t)
:=
V_B(\bx(t))-V_B(\bx(t-T))
\).
Substituting into
\eqref{eq:constrained_integral_bellman}
yields
\(
\bW^{*\top}\boldsymbol\Omega(t)
+
\rho(t)
+
\varepsilon_B(t)=0
\),
where
\begin{equation}
\begin{aligned}
\boldsymbol\Omega(t)
={}&
\Delta\bphi(t)
-
\int_{t-T}^{t}
\nabla\bphi(\bx(\tau))
\bg(\bx(\tau))
[\bu_b(\tau)-\hat{\bu}(\tau)]
\,d\tau,
\\
\rho(t)
={}&
\Delta V_B(t)
+
\int_{t-T}^{t}
\ell(\bx(\tau),\hat{\bu}(\tau))
\,d\tau
\\
&-
\int_{t-T}^{t}
\nabla V_B^\top(\bx(\tau))
\bg(\bx(\tau))
[\bu_b(\tau)-\hat{\bu}(\tau)]
\,d\tau .
\end{aligned}
\label{eq:constrained_integral_regression}
\end{equation}

The aggregate Bellman residual is
\begin{equation}
\begin{aligned}
\varepsilon_B(t)
={}&
\varepsilon(\bx(t))
-
\varepsilon(\bx(t-T))-
\int_{t-T}^{t}
\nabla\varepsilon^\top(\bx(\tau))
\bg(\bx(\tau))
\\
&\qquad\times
[\bu_b(\tau)-\hat{\bu}(\tau)]
\,d\tau
-
\int_{t-T}^{t}
\delta_{\hat u}(\bx(\tau),\tau)
\,d\tau .
\end{aligned}
\label{eq:aggregate_bellman_residual}
\end{equation}
Hence,
\(
\varepsilon_B(t)
\)
accounts for both the value-function approximation
error and the policy-suboptimality residual.

Motivated by the analysis in Proposition 1
of~\cite{Gong2025}, we have:

\begin{proposition}[Bounded Bellman approximation residual]
\label{pr2}
Under Proposition~\ref{prop:critic_approximation_residual},
suppose that the probing signal satisfies
$\|\be_p(t)\|\le e_M$.
If the learned policy is uniformly close
to the optimal policy over the considered
learning interval, then, for sufficiently
large $N_C$, there exists
$\mathscr B_B>0$ such that
$|\bar{\varepsilon}_B(t)|
\le\mathscr B_B$ and
$|\bar{\varepsilon}_{B,j}|
\le\mathscr B_B$,
$j=1,\ldots,M$.
\end{proposition}

\begin{proof}
By Proposition~\ref{prop:critic_approximation_residual},
for any $\eta>0$ and sufficiently large
$N_C$,
$|\varepsilon(\bx)|\le\eta$ and
$\|\nabla\varepsilon(\bx)\|
\le\bar\varepsilon_g\le\eta$.

From the optimal policy expression,
\begin{equation}
\begin{aligned}
\bu(\bx,\bW^*)-\bu^*(\bx)
=
\frac12\bR^{-1}\bg^\top(\bx)
\nabla\varepsilon(\bx),
\end{aligned}
\label{eq:ideal_policy_approximation_error}
\end{equation}
giving
$\|\bu(\bx,\bW^*)-\bu^*(\bx)\|
\le\frac12\|\bR^{-1}\|
G\bar\varepsilon_g$.
Hence, Proposition~\ref{prop:critic_approximation_residual}
ensures that the ideal-weight policy
approximation error can be made arbitrarily
small by increasing $N_C$, since the
gradient approximation error bound
$\bar\varepsilon_g$ can be reduced
accordingly as $N_C$ increases.
For the learned policy, under the stated
uniform policy-accuracy condition, let
$\|\hat{\bu}-\bu^*\|\le\hat\epsilon$.
Using the HJB stationary condition
$\partial H/\partial\bu
|_{\bu=\bu^*}
=\mathbf0$,
one has
$2\bR\bu^*
+\bm L_1^\top(\bx)
+\bg^\top(\bx)\nabla V^*(\bx)
=\mathbf0$.

Let
$\Delta\bu=\hat{\bu}-\bu^*$.
Then,
$\hat{\bu}=\bu^*+\Delta\bu$,
yielding
\begin{equation*}
\begin{aligned}
&\ell(\bx,\hat{\bu})
-\ell(\bx,\bu^*)
=
\bm L_1(\bx)\Delta\bu
+2\bu^{*\top}\bR\Delta\bu
+\Delta\bu^\top\bR\Delta\bu.
\end{aligned}
\end{equation*}

Consequently,
\begin{equation*}
\begin{aligned}
\delta_{\hat u}
={}&
\ell(\bx,\hat{\bu})
-\ell(\bx,\bu^*)
+\nabla V^{*\top}(\bx)\bg(\bx)
\Delta\bu
\\
={}&
\big(
\bm L_1^\top(\bx)
+2\bR\bu^*
+\bg^\top(\bx)\nabla V^*(\bx)
\big)^\top
\Delta\bu
\\
&+
\Delta\bu^\top\bR\Delta\bu.
\end{aligned}
\end{equation*}

By the HJB stationary condition and
$\bR\succ0$, it follows that
\begin{equation}
\begin{aligned}
0\le\delta_{\hat u}
&=
\Delta\bu^\top\bR\Delta\bu
\le
\lambda_{\max}(\bR)
\|\Delta\bu\|^2
\le
\lambda_{\max}(\bR)
\hat\epsilon^2.
\end{aligned}
\label{eq:policy_suboptimality_residual_bound}
\end{equation}

We have
$\bu_b-\hat{\bu}=\be_p$.
By the triangle and Cauchy--Schwarz
inequalities,
\begin{equation}
\begin{aligned}
|\varepsilon_B(t)|
&\le
2\eta+
\int_{t-T}^{t}
\|\nabla\varepsilon\|
\|\bg\|
\|\be_p\|\,d\tau
\\
&\quad+
\int_{t-T}^{t}
|\delta_{\hat u}|\,d\tau
\le
2\eta+
TG e_M\bar\varepsilon_g
+
T\lambda_{\max}(\bR)
\hat\epsilon^2\\
&=:\mathscr B_B<\infty.
\end{aligned}
\label{eq:aggregate_residual_bound}
\end{equation}

The same bound applies to the historical
samples collected over the compact operating
set. Since the normalization denominators
are not smaller than one, the normalized
approximation residuals are also uniformly
bounded. Consequently, there exists a
finite constant $\mathscr B_B>0$
such that
$|\bar{\varepsilon}_B(t)|
\le\mathscr B_B$ and
$|\bar{\varepsilon}_{B,j}|
\le\mathscr B_B$,
for $j=1,\ldots,M$.
Moreover, increasing $N_C$ reduces
the $C^1$ approximation residual,
whereas the online policy error
additionally depends on critic learning.
\end{proof}

For the critic estimate, define
\(
\xi(t)
:=
\hat{\bW}^{\top}(t)\boldsymbol\Omega(t)
+
\rho(t)
\)
and
\(
\tilde{\bW}
:=
\hat{\bW}-\bW^*
\).
Then
\(
\xi(t)
=
\tilde{\bW}^{\top}\boldsymbol\Omega(t)
-
\varepsilon_B(t)
\).

Let
\(
m(t)
:=
1+
\boldsymbol\Omega^\top(t)\boldsymbol\Omega(t)
\),
\(
\bar{\boldsymbol\Omega}(t)
:=
\boldsymbol\Omega(t)/m(t)
\),
and
\(
e(t)
:=
\xi(t)/m(t)
\).
Thus,
\(
e(t)
=
\tilde{\bW}^{\top}
\bar{\boldsymbol\Omega}(t)
-
\bar\varepsilon_B(t)
\),
where
\(
\bar\varepsilon_B(t)
:=
\varepsilon_B(t)/m(t)
\).

Let
\(
\{
\boldsymbol\Omega_j,\rho_j
\}_{j=1}^{M}
\)
denote the recorded integral samples. For each stored sample, define
\(
m_j
:=
1+
\boldsymbol\Omega_j^\top\boldsymbol\Omega_j
\),
\(
\bar{\boldsymbol\Omega}_j
:=
\boldsymbol\Omega_j/m_j
\),
and
\(
e_j
:=
[
\hat{\bW}^{\top}\boldsymbol\Omega_j+\rho_j
]/m_j
\).
Then
\(
e_j
=
\tilde{\bW}^{\top}\bar{\boldsymbol\Omega}_j
-
\bar\varepsilon_{B,j}
\).

\begin{assumption}[Finite data informativity]
\label{ass:constrained_critic_ie}
The stored integral regressors satisfy
\begin{equation}
\sum_{j=1}^{M}
\bar{\boldsymbol\Omega}_j
\bar{\boldsymbol\Omega}_j^\top
\succeq
\alpha_{\mathrm{IE}}\bm I
\end{equation}
for some \(\alpha_{\mathrm{IE}}>0\).

\end{assumption}

Let \(0<\gamma_1<1<\gamma_2\) and define
\(\operatorname{sig}^{\gamma}(s)
:=|s|^\gamma\operatorname{sgn}(s)\).
Consider the two-power penalty
\(\Psi(e):=
\frac{k_1}{\gamma_1+1}|e|^{\gamma_1+1}
+\frac{k_2}{\gamma_2+1}|e|^{\gamma_2+1}\),
where \(k_1,k_2>0\).

Since
\(\partial e/\partial\hat{\bW}
=\bar{\boldsymbol\Omega}\)
and
\(\partial e_j/\partial\hat{\bW}
=\bar{\boldsymbol\Omega}_j\),
gradient descent on
\(\mathcal J_W
:=\Psi(e(t))+\sum_{j=1}^{M}\Psi(e_j)\)
gives
\begin{equation}
\begin{aligned}
\dot{\hat{\bW}}
={}&
-\bm\Gamma\bar{\boldsymbol\Omega}(t)
\big[
k_1\operatorname{sig}^{\gamma_1}(e(t))
+k_2\operatorname{sig}^{\gamma_2}(e(t))
\big]
\\
&-
\bm\Gamma\sum_{j=1}^{M}
\bar{\boldsymbol\Omega}_j
\big[
k_1\operatorname{sig}^{\gamma_1}(e_j)
+k_2\operatorname{sig}^{\gamma_2}(e_j)
\big],
\end{aligned}
\label{eq:constrained_fixed_time_critic_update}
\end{equation}
where \(\bm\Gamma=\bm\Gamma^\top\succ0\).

\begin{remark}[Distinction from existing learning schemes]
Fixed-time RL was investigated in
\cite{Wang2024FixedTime} without experience
replay, whereas \cite{Gong2025,Kokolakis2025Predefined}
incorporated replay-based learning under
known system dynamics. Safe fixed-time RL
with experience replay was further studied
in \cite{Wang2026SafeFixedTime} for zero-sum
games. In contrast, the present work explicitly
constructs an inverse-optimal running cost
combining a state-barrier penalty and a
non-quadratic input penalty to accommodate
safety and actuator constraints while
prescribing nominal fixed-time Lyapunov decay.
For unknown drift dynamics, the proposed
critic update employs an integral Bellman
residual constructed from finite-window
off-policy data and experience replay,
rather than relying on pointwise model-based
Bellman regression. The two-power update
retains the fixed-time learning principle
used in \cite{Gong2025,Kokolakis2025Predefined},
but differs in its integral regression
structure and its coupling with the
barrier-augmented inverse-optimal cost.
\end{remark}

\begin{theorem}[Practical fixed-time convergence under approximation error]
\label{thm:critic_practical_fixed_time_convergence}
Suppose that Assumption~\ref{ass:constrained_critic_ie}
holds and the critic weights are updated according to
\eqref{eq:constrained_fixed_time_critic_update}.
Then there exist constants
\(\bar C_{W1}>0\), \(\bar C_{W2}>0\),
and \(\Delta_W\geq0\) such that
\begin{equation}
\dot V_W
\leq
-\bar C_{W1}V_W^\mu
-\bar C_{W2}V_W^\nu
+\Delta_W.
\label{eq:critic_practical_fixed_time_inequality}
\end{equation}
Consequently, the critic-weight error converges in fixed
time to a bounded residual set.
\end{theorem}

\begin{proof}
For the nonzero approximation error,
\(e(t)=\tilde{\bm W}^{\top}
\bar{\boldsymbol\Omega}(t)-\bar\varepsilon_B(t)\)
and
\(e_j=\tilde{\bm W}^{\top}
\bar{\boldsymbol\Omega}_j-\bar\varepsilon_{B,j}\).
Hence,
\(\tilde{\bm W}^{\top}\bar{\boldsymbol\Omega}(t)
=e(t)+\bar\varepsilon_B(t)\)
and
\(\tilde{\bm W}^{\top}\bar{\boldsymbol\Omega}_j
=e_j+\bar\varepsilon_{B,j}\).

Using
\(V_W=\frac12\tilde{\bm W}^{\top}
\bm\Gamma^{-1}\tilde{\bm W}\)
and
\(\dot{\tilde{\bm W}}=\dot{\hat{\bm W}}\),
one again has
\(\dot V_W
=\tilde{\bm W}^{\top}
\bm\Gamma^{-1}\dot{\hat{\bm W}}\).
Substitution of
\eqref{eq:constrained_fixed_time_critic_update}
gives
\begin{equation}
\begin{aligned}
\dot V_W
={}&
-[e(t)+\bar\varepsilon_B(t)]
\big[
k_1\operatorname{sig}^{\gamma_1}(e(t))
+k_2\operatorname{sig}^{\gamma_2}(e(t))
\big]
\\
&-
\sum_{j=1}^{M}
[e_j+\bar\varepsilon_{B,j}]
\big[
k_1\operatorname{sig}^{\gamma_1}(e_j)
+k_2\operatorname{sig}^{\gamma_2}(e_j)
\big].
\end{aligned}
\end{equation}

Using
\(s\operatorname{sig}^{\gamma}(s)
=|s|^{\gamma+1}\),
one obtains
\begin{equation}
\begin{aligned}
\dot V_W
={}&
-k_1|e(t)|^{\gamma_1+1}
-k_2|e(t)|^{\gamma_2+1}
\\
&-
k_1\sum_{j=1}^{M}|e_j|^{\gamma_1+1}
-k_2\sum_{j=1}^{M}|e_j|^{\gamma_2+1}
\\
&-
\bar\varepsilon_B(t)
\big[
k_1\operatorname{sig}^{\gamma_1}(e(t))
+k_2\operatorname{sig}^{\gamma_2}(e(t))
\big]
\\
&-
\sum_{j=1}^{M}\bar\varepsilon_{B,j}
\big[
k_1\operatorname{sig}^{\gamma_1}(e_j)
+k_2\operatorname{sig}^{\gamma_2}(e_j)
\big].
\end{aligned}
\end{equation}

For any \(\gamma>0\), Young's inequality gives
\(|\bar\varepsilon||e|^\gamma
\leq\frac12|e|^{\gamma+1}
+c_\gamma|\bar\varepsilon|^{\gamma+1}\),
where
\(c_\gamma:=
\frac{2^\gamma\gamma^\gamma}
{(\gamma+1)^{\gamma+1}}\).

Applying this inequality to the current and recorded
approximation terms, together with
Assumption~\ref{ass:constrained_critic_ie}, gives
\begin{equation}
\begin{aligned}
\dot V_W
\leq{}&
-\frac{k_1}{2}|e(t)|^{\gamma_1+1}
-\frac{k_2}{2}|e(t)|^{\gamma_2+1}
\\
&-
\frac{k_1}{2}\sum_{j=1}^{M}|e_j|^{\gamma_1+1}
-\frac{k_2}{2}\sum_{j=1}^{M}|e_j|^{\gamma_2+1}
+\Delta_{\varepsilon W},
\end{aligned}
\end{equation}
where
\(\Delta_{\varepsilon W}
:=(M+1)
[k_1c_{\gamma_1}\mathscr E_B^{\gamma_1+1}
+k_2c_{\gamma_2}\mathscr E_B^{\gamma_2+1}]\).

Discarding the two nonpositive current-data terms yields
\begin{equation}
\begin{aligned}
\dot V_W
\leq{}&
-\frac{k_1}{2}
\sum_{j=1}^{M}|e_j|^{\gamma_1+1}
-
\frac{k_2}{2}
\sum_{j=1}^{M}|e_j|^{\gamma_2+1}
+\Delta_{\varepsilon W}.
\end{aligned}
\end{equation}

Let \(p:=\gamma_1+1=2\mu\) and
\(q:=\gamma_2+1=2\nu\),
so that \(1<p<2<q\).
Define \(z_j:=\bar{\boldsymbol\Omega}_j^\top
\tilde{\bm W}\),
\(\bm B:=
[\bar{\boldsymbol\Omega}_1,\ldots,
\bar{\boldsymbol\Omega}_M]^\top
\in\mathbb R^{M\times N_c}\),
and
\(\bm z:=[z_1,\ldots,z_M]^\top
=\bm B\tilde{\bm W}\).

Since \(e_j=z_j-\bar\varepsilon_{B,j}\),
the inequality
\(|a-b|^r\geq2^{1-r}|a|^r-|b|^r\),
valid for \(r\geq1\), gives
\(\sum_{j=1}^{M}|e_j|^p
\geq2^{1-p}\sum_{j=1}^{M}|z_j|^p
-\sum_{j=1}^{M}|\bar\varepsilon_{B,j}|^p\)
and
\(\sum_{j=1}^{M}|e_j|^q
\geq2^{1-q}\sum_{j=1}^{M}|z_j|^q
-\sum_{j=1}^{M}|\bar\varepsilon_{B,j}|^q\).

From the IE condition,
\(\|\bm z\|_2^2
=\tilde{\bm W}^{\top}\bm B^\top\bm B
\tilde{\bm W}
\geq\alpha_{\mathrm{IE}}
\|\tilde{\bm W}\|^2\).

For \(1<p<2\), one has
\(\|\bm z\|_p\geq\|\bm z\|_2\),
and therefore
\(\sum_{j=1}^{M}|z_j|^p
\geq\alpha_{\mathrm{IE}}^{p/2}
\|\tilde{\bm W}\|^p\).
For \(q>2\), the inequality
\(\|\bm z\|_q
\geq M^{1/q-1/2}\|\bm z\|_2\)
yields
\(\sum_{j=1}^{M}|z_j|^q
\geq M^{1-q/2}
\alpha_{\mathrm{IE}}^{q/2}
\|\tilde{\bm W}\|^q\).

Using
\(|\bar\varepsilon_{B,j}|
\leq\mathscr E_B\),
one obtains
\begin{equation}
\begin{aligned}
\sum_{j=1}^{M}|e_j|^p
\geq{}&
2^{1-p}\alpha_{\mathrm{IE}}^{p/2}
\|\tilde{\bm W}\|^p
-M\mathscr E_B^p,
\\
\sum_{j=1}^{M}|e_j|^q
\geq{}&
2^{1-q}M^{1-q/2}
\alpha_{\mathrm{IE}}^{q/2}
\|\tilde{\bm W}\|^q
-M\mathscr E_B^q.
\end{aligned}
\end{equation}

Substitution gives
\begin{equation}
\begin{aligned}
\dot V_W
\leq{}&
-k_1 2^{-p}\alpha_{\mathrm{IE}}^{p/2}
\|\tilde{\bm W}\|^p
-
k_2 2^{-q}M^{1-q/2}
\alpha_{\mathrm{IE}}^{q/2}
\|\tilde{\bm W}\|^q
+\Delta_W,
\end{aligned}
\end{equation}
where
\(\Delta_W:=
\Delta_{\varepsilon W}
+\frac{M}{2}
[k_1\mathscr E_B^p
+k_2\mathscr E_B^q]\).

Since
\(V_W\leq
[2\lambda_{\min}(\bm\Gamma)]^{-1}
\|\tilde{\bm W}\|^2\),
one has
\(\|\tilde{\bm W}\|^p
\geq[2\lambda_{\min}(\bm\Gamma)]^{p/2}
V_W^{p/2}\)
and
\(\|\tilde{\bm W}\|^q
\geq[2\lambda_{\min}(\bm\Gamma)]^{q/2}
V_W^{q/2}\).

Therefore,
\(\dot V_W
\leq-\bar C_{W1}V_W^{p/2}
-\bar C_{W2}V_W^{q/2}
+\Delta_W\),
where
\(\bar C_{W1}:=
k_1 2^{-p}\alpha_{\mathrm{IE}}^{p/2}
[2\lambda_{\min}(\bm\Gamma)]^{p/2}\)
and
\(\bar C_{W2}:=
k_2 2^{-q}M^{1-q/2}
\alpha_{\mathrm{IE}}^{q/2}
[2\lambda_{\min}(\bm\Gamma)]^{q/2}\).

Finally, by the prescribed choice
\(\gamma_1=2\mu-1\) and
\(\gamma_2=2\nu-1\),
one has
\(p/2=\mu\) and \(q/2=\nu\),
with \(\frac12<\mu<1<\nu\).
Hence,
\(\dot V_W
\leq-\bar C_{W1}V_W^\mu
-\bar C_{W2}V_W^\nu
+\Delta_W\),
which proves
\eqref{eq:critic_practical_fixed_time_inequality}.

For any \(0<\chi_W<1\), define
\(V_{W,\Delta}>0\) as the unique positive
solution of
\(\bar C_{W1}V_{W,\Delta}^{\mu}
+\bar C_{W2}V_{W,\Delta}^{\nu}
=\Delta_W/\chi_W\).

For every \(V_W\geq V_{W,\Delta}\), one has
\(\Delta_W\leq
\chi_W[
\bar C_{W1}V_W^\mu
+\bar C_{W2}V_W^\nu]\),
and therefore
\(\dot V_W
\leq-(1-\chi_W)\bar C_{W1}V_W^\mu
-(1-\chi_W)\bar C_{W2}V_W^\nu\).

Thus, the critic-weight error enters the residual set
\(V_W\leq V_{W,\Delta}\)
within the uniformly bounded time
\begin{equation}
\begin{aligned}
T_{W,\Delta}
\leq{}&
\frac{1}{
(1-\chi_W)\bar C_{W1}(1-\mu)}
+
\frac{1}{
(1-\chi_W)\bar C_{W2}(\nu-1)}.
\end{aligned}
\end{equation}

When \(\mathscr E_B=0\),
one has \(\Delta_W=0\),
and the exact fixed-time convergence result
is recovered. This completes the proof.
\end{proof}

\section{Closed-Loop Practical Fixed-Time Stability}
\label{sec:closed_loop_stability}

The preceding results establish nominal fixed-time stability
under the constrained inverse-optimal policy and practical
fixed-time convergence of the critic weights.
To analyze the complete learning closed loop, the following
technical lemma provides the integral comparison needed
for the subsequent composite Lyapunov analysis.

\begin{lemma}[Derivative-limited low-order window bound]
\label{lem:derivative_limited_window_bound}
Let
\(
y:[t-T,t]\rightarrow\mathbb R_{\geq0}
\)
be absolutely continuous and satisfy
\(
|\dot y(\tau)|\leq L_y
\)
for almost every
\(
\tau\in[t-T,t]
\),
where
\(
L_y>0
\).
Assume that
\(
\int_{t-T}^{t}y(\tau)\,d\tau\leq\bar I
\)
for some finite
\(
\bar I>0
\).
For
\(
0<\gamma_1<1
\),
define
\(
\mu=(\gamma_1+1)/2
\).
Then
\begin{equation}
\int_{t-T}^{t}
y^{\gamma_1}(\tau)\,d\tau
\geq
\rho_{\gamma_1,T}
\left(
\int_{t-T}^{t}
y(\tau)\,d\tau
\right)^\mu ,
\label{eq:derivative_limited_window_bound}
\end{equation}
where
\begin{equation}
\rho_{\gamma_1,T}
:=
\min
\left\{
2^{\gamma_1-1}
L_y^{\frac{\gamma_1-1}{2}},
\;
4^{\gamma_1-1}
T^{1-\gamma_1}
\bar I^{\frac{\gamma_1-1}{2}}
\right\}.
\label{eq:rho_gamma_window}
\end{equation}
\end{lemma}

\begin{proof}
See Appendix~\ref{app:derivative_limited_window_bound}
\end{proof}

For the IRL state trajectory, let
\(
y(\tau)=V_d(\bx(\tau))
\).
On the compact operating set
\(
\Omega
\),
continuity of the closed-loop dynamics and
\(
\nabla V_d
\)
implies the existence of
\(
L_{V_d}>0
\)
such that
\(
|\dot V_d(\bx(\tau))|\leq L_{V_d}
\).
Moreover, there exists a finite
\(
\bar I_V>0
\)
such that
\(
\int_{t-T}^{t}V_d(\bx(\tau))\,d\tau\leq\bar I_V
\).
Hence,
Lemma~\ref{lem:derivative_limited_window_bound}
gives
\begin{equation}
\int_{t-T}^{t}
V_d^{\gamma_1}(\bx(\tau))\,d\tau
\geq
\rho_{\gamma_1,T}^{V_d}
\left(
\int_{t-T}^{t}
V_d(\bx(\tau))\,d\tau
\right)^\mu ,
\label{eq:Vd_window_low_order_bound}
\end{equation}
where
\begin{equation}
\rho_{\gamma_1,T}^{V_d}
:=
\min
\left\{
2^{\gamma_1-1}
L_{V_d}^{\frac{\gamma_1-1}{2}},
\;
4^{\gamma_1-1}
T^{1-\gamma_1}
\bar I_V^{\frac{\gamma_1-1}{2}}
\right\}.
\label{eq:Vd_window_rho}
\end{equation}
This estimate will be used below in the Leibniz derivative of the finite-window composite Lyapunov functional.

% ============================================================
% FINITE-WINDOW COMPOSITE ANALYSIS
% ============================================================

Define the finite-window state energy
\(I_V(t):=\int_{t-T}^{t}V_d(\bx(\tau))\,d\tau\)
and the composite Lyapunov functional
\begin{equation}
\mathcal J(t)
:=V_d(\bx(t))+\eta I_V(t)+\lambda_W V_W(t),
\qquad \eta,\lambda_W>0.
\label{eq:constrained_integral_composite_energy}
\end{equation}

\begin{theorem}[Practical fixed-time stability and constraint satisfaction]
\label{thm:composite_constrained_stability}
Suppose that Assumption~\ref{ass:constrained_critic_ie}
and Lemma~\ref{lem:identification_error} hold,
with the critic update law
\eqref{eq:constrained_fixed_time_critic_update}
and the continuously implemented policy
\eqref{eq:constrained_critic_policy}.
Let $\gamma_1=2\mu-1$ with
$\frac34<\mu<1<\nu$.
For $t\geq t_\star\geq T$, assume that
the nominal fixed-time identity holds and
all sampling windows remain in an admissible
region where
$V_d\leq\bar V_d$,
$V_W\leq\bar V_W$,
$|\dot V_d|\leq L_{V_d}$, and
$I_V\leq\bar I_V$, with uniform finite
bounds over the considered initial-condition class.

Then there exist $C_1,C_2>0$ and
$\Delta_J\geq0$ such that the composite
functional \eqref{eq:constrained_integral_composite_energy}
satisfies
\begin{equation}
\dot{\mathcal J}
\leq -C_1\mathcal J^\mu
-C_2\mathcal J^\nu+\Delta_J.
\label{eq:composite_constrained_FxT_bound}
\end{equation}
Consequently, the state and critic-weight
errors are practically fixed-time stable
relative to this class from $t_\star$,
while $|u_j(t)|<\bar u_j$ for all
$j=1,\ldots,m$.
Forward invariance follows under
Lemma~\ref{lem:constrained_forward_invariance}.
If $\Delta_J=0$, the exact fixed-time
convergence comparison is recovered.
\end{theorem}

\begin{proof}
Under the nominal identified dynamics \(\dot{\bx}=\hat{\bfv}(\bx,t)+\bg(\bx)\bu_c(\bx)\), Theorem~\ref{thm:fixed_time_inverse_optimal} gives \(\nabla V_d^\top(\bx)[\hat{\bfv}(\bx,t)+\bg(\bx)\bu_c(\bx)]=-aV_d^{\gamma_1}(\bx)-bV_d^\nu(\bx)\). For the actual plant under the behavior policy during learning, \(\dot{\bx}=\bfv(\bx)+\bg(\bx)\bu_b(t)\), where \(\bu_b(t)=\hat{\bu}(\bx,t)+\be_p(t)\). Adding and subtracting \(\hat{\bfv}\) and \(\bg\bu_c\) gives \(\dot{\bx}=\hat{\bfv}+\bg\bu_c+(\bfv-\hat{\bfv})+\bg(\hat{\bu}-\bu_c)+\bg\be_p\). Hence,
\[
\dot V_d
=
\nabla V_d^\top[\hat{\bfv}+\bg\bu_c]
+
\nabla V_d^\top(\bfv-\hat{\bfv})
+
\nabla V_d^\top\bg(\hat{\bu}-\bu_c)
+
\nabla V_d^\top\bg\be_p,
\]
and therefore
\[
\dot V_d
=
-aV_d^{\gamma_1}
-bV_d^\nu
+
\nabla V_d^\top(\bfv-\hat{\bfv})
+
\nabla V_d^\top\bg(\hat{\bu}-\bu_c)
+
\nabla V_d^\top\bg\be_p.
\]

Since \(\Omega\) is compact, \(\nabla V_d\) is continuous, and \(\|\bfv-\hat{\bfv}\|\leq\bar\Delta_f\), there exists \(\bar V_g>0\) such that \(\|\nabla V_d(\bx)\|\leq\bar V_g\) for every \(\bx\in\Omega\). Consequently, \(\nabla V_d^\top(\bfv-\hat{\bfv})\leq|\nabla V_d^\top(\bfv-\hat{\bfv})|\leq\|\nabla V_d\|\|\bfv-\hat{\bfv}\|\leq\bar V_g\bar\Delta_f\). Define \(\Delta_{fV}:=\bar V_g\bar\Delta_f\). In addition, suppose that the probing signal is chosen such that \(\|\be_p(t)\|\leq\bar e_p\) and the applied behavior input remains admissible, \(\bu_b(t)\in\bu_b(t)\in\mathbb{U}\). Since \(\bg\) is continuous on \(\Omega\), let \(\bar g:=\sup_{\bx\in\Omega}\|\bg(\bx)\|<\infty\). On the compact region where \(\|\nabla V_d\|\leq\bar V_g\) holds, \(\nabla V_d^\top\bg\be_p\leq |\nabla V_d^\top\bg\be_p|\leq\bar V_g\bar g\bar e_p\). Define \(\Delta_{pV}:=\bar V_g\bar g\bar e_p\). Thus,
\[
\dot V_d
\leq
-aV_d^{\gamma_1}
-bV_d^\nu
+
\Delta_{fV}
+
\Delta_{pV}
+
\nabla V_d^\top\bg(\hat{\bu}-\bu_c).
\]

By \eqref{eq:general_constrained_optimal_policy}, the nominal constrained policy can be written as \(\bu_c=-\bm U\tanh(\bm\zeta_d)\), where \(\bm\zeta_d:=\frac{1}{2}\bm U^{-1}\bR^{-1}[\bm L_1^\top+\bg^\top\nabla V_d]\). Similarly, \eqref{eq:constrained_critic_policy} gives \(\hat{\bu}=-\bm U\tanh(\hat{\bm\zeta})\), where \(\hat{\bm\zeta}:=\frac{1}{2}\bm U^{-1}\bR^{-1}[\bm L_1^\top+\bg^\top\nabla\hat V]\). Since \(V_d=V_B+\bW^{*\top}\bphi+\varepsilon\) and \(\hat V=V_B+\hat{\bW}^\top\bphi\), one has \(\nabla V_d=\nabla V_B+(\nabla\bphi)^\top\bW^*+\nabla\varepsilon\) and \(\nabla\hat V=\nabla V_B+(\nabla\bphi)^\top\hat{\bW}\). Hence,
\[
\nabla\hat V-\nabla V_d
=
(\nabla\bphi)^\top(\hat{\bW}-\bW^*)
-\nabla\varepsilon
=
(\nabla\bphi)^\top\tilde{\bW}
-\nabla\varepsilon.
\]
It follows that
\[
\hat{\bm\zeta}-\bm\zeta_d
=
\frac{1}{2}
\bm U^{-1}\bR^{-1}\bg^\top
\left[
(\nabla\bphi)^\top\tilde{\bW}
-
\nabla\varepsilon
\right].
\]

Since \(|d\tanh(s)/ds|=\operatorname{sech}^2(s)\leq1\), the componentwise mapping \(\tanh(\cdot)\) is globally Lipschitz with unit constant. Therefore,
\[
\begin{aligned}
\|\hat{\bu}-\bu_c\|
&=
\left\|
-\bm U
[
\tanh(\hat{\bm\zeta})
-
\tanh(\bm\zeta_d)
]
\right\|
\\
&\leq
\|\bm U\|
\|
\tanh(\hat{\bm\zeta})
-
\tanh(\bm\zeta_d)
\|
\\
&\leq
\|\bm U\|
\|
\hat{\bm\zeta}-\bm\zeta_d
\|.
\end{aligned}
\]
Using the preceding expression for \(\hat{\bm\zeta}-\bm\zeta_d\), one obtains
\[
\begin{aligned}
\|\hat{\bu}-\bu_c\|
\leq{}&
\frac{1}{2}
\|\bm U\|
\left\|
\bm U^{-1}\bR^{-1}
\bg^\top
(\nabla\bphi)^\top
\right\|
\|\tilde{\bW}\|
\\
&+
\frac{1}{2}
\|\bm U\|
\left\|
\bm U^{-1}\bR^{-1}
\bg^\top
\nabla\varepsilon
\right\|.
\end{aligned}
\]

Because \(\Omega\) is compact and the involved mappings are continuous, there exist finite constants \(c_\phi>0\) and \(c_\varepsilon\geq0\) such that \(\frac{1}{2}\|\bm U\|\|\bm U^{-1}\bR^{-1}\bg^\top(\bx)(\nabla\bphi(\bx))^\top\|\leq c_\phi\) and \(\frac{1}{2}\|\bm U\|\|\bm U^{-1}\bR^{-1}\bg^\top(\bx)\nabla\varepsilon(\bx)\|\leq c_\varepsilon\). Hence, \(\|\hat{\bu}-\bu_c\|\leq c_\phi\|\tilde{\bW}\|+c_\varepsilon\).

Moreover, compactness of \(\Omega\) and continuity of \(\bg^\top\nabla V_d\) imply the existence of \(c_V>0\) such that \(\|\bg^\top\nabla V_d\|\leq c_V\). Thus,
\[
\begin{aligned}
\nabla V_d^\top\bg(\hat{\bu}-\bu_c)
&=
(\bg^\top\nabla V_d)^\top(\hat{\bu}-\bu_c)
\leq
\|\bg^\top\nabla V_d\|
\|\hat{\bu}-\bu_c\|
\\
&\leq
c_V
\left(
c_\phi\|\tilde{\bW}\|
+
c_\varepsilon
\right)
=
c_Vc_\phi\|\tilde{\bW}\|
+
c_Vc_\varepsilon.
\end{aligned}
\]

Recall that \(V_W=\frac{1}{2}\tilde{\bW}^\top\bm\Gamma^{-1}\tilde{\bW}\). Since \(\bm\Gamma=\bm\Gamma^\top\succ0\), one has \(\tilde{\bW}^\top\bm\Gamma^{-1}\tilde{\bW}\geq\lambda_{\min}(\bm\Gamma^{-1})\|\tilde{\bW}\|^2\). Since \(\lambda_{\min}(\bm\Gamma^{-1})=1/\lambda_{\max}(\bm\Gamma)\), it follows that \(V_W\geq[2\lambda_{\max}(\bm\Gamma)]^{-1}\|\tilde{\bW}\|^2\), and hence \(\|\tilde{\bW}\|^2\leq2\lambda_{\max}(\bm\Gamma)V_W\) and \(\|\tilde{\bW}\|\leq[2\lambda_{\max}(\bm\Gamma)]^{1/2}V_W^{1/2}\). Define \(c_W:=c_Vc_\phi[2\lambda_{\max}(\bm\Gamma)]^{1/2}\). Therefore, \(\nabla V_d^\top\bg(\hat{\bu}-\bu_c)\leq c_WV_W^{1/2}+c_Vc_\varepsilon\), and hence
\[
\dot V_d
\leq
-aV_d^{\gamma_1}
-bV_d^\nu
+
c_WV_W^{1/2}
+
\Delta_{fV}
+
\Delta_{pV}
+
c_Vc_\varepsilon.
\]

For every \(s\geq0\) and \(0<\mu<1<\nu\), one has \(s^{1/2}\leq1+s^\mu+s^\nu\). Indeed, for \(0\leq s\leq1\), \(s^{1/2}\leq1\), whereas for \(s\geq1\), \(s^{1/2}\leq s^\nu\). Thus, \(c_WV_W^{1/2}\leq c_W+c_WV_W^\mu+c_WV_W^\nu\). Define \(\varrho_1:=c_W\), \(\varrho_2:=c_W\), and \(\Delta_x:=\Delta_{fV}+\Delta_{pV}+c_Vc_\varepsilon+c_W\). The state-value derivative therefore satisfies
\begin{equation}
\dot V_d
\leq
-aV_d^{\gamma_1}
-bV_d^\nu
+
\varrho_1V_W^\mu
+
\varrho_2V_W^\nu
+
\Delta_x.
\label{eq:state_critic_coupling_bound}
\end{equation}

From Theorem~\ref{thm:critic_practical_fixed_time_convergence},
there exist \(h_1,h_2>0\) and \(\Delta_W\geq0\)
such that
\(
\dot V_W
\leq
-h_1V_W^\mu-h_2V_W^\nu+\Delta_W
\),
where
\(h_1:=\bar C_{W1}\)
and
\(h_2:=\bar C_{W2}\).
Consider
\(\mathcal J=V_d+\eta I_V+\lambda_WV_W\),
with \(\eta,\lambda_W>0\).
By the Leibniz integral rule,

\begin{equation}
\dot I_V(t)
=V_d(\bx(t))-V_d(\bx(t-T))
=\int_{t-T}^{t}\dot V_d(\bx(\tau))\,d\tau.
\label{eq:leibniz_window_state}
\end{equation}
Integrating \eqref{eq:state_critic_coupling_bound}
over \([t-T,t]\) therefore gives
\begin{equation}
\begin{aligned}
\dot I_V
\leq{}&
-a\int_{t-T}^{t}V_d^{\gamma_1}(\tau)\,d\tau
-b\int_{t-T}^{t}V_d^\nu(\tau)\,d\tau\\
&+\varrho_1\int_{t-T}^{t}V_W^\mu(\tau)\,d\tau
+\varrho_2\int_{t-T}^{t}V_W^\nu(\tau)\,d\tau
+T\Delta_x.
\end{aligned}
\label{eq:window_state_derivative_bound}
\end{equation}
Here and below, arguments along the trajectories are
suppressed whenever unambiguous.

Since \(\gamma_1=2\mu-1\),
Lemma~\ref{lem:derivative_limited_window_bound}
provides the low-order estimate
\eqref{eq:Vd_window_low_order_bound}, namely
\(\int_{t-T}^{t}V_d^{\gamma_1}(\tau)\,d\tau
\geq \rho_{\gamma_1,T}^{V_d} I_V^\mu\).
For \(\nu>1\), Jensen's inequality gives
\(\int_{t-T}^{t}V_d^\nu(\tau)\,d\tau
\geq T^{1-\nu}I_V^\nu\).
Furthermore, the uniformly bounded critic energy implies
\(\int_{t-T}^{t}V_W^\mu(\tau)\,d\tau
\leq T\bar V_W^\mu\) and
\(\int_{t-T}^{t}V_W^\nu(\tau)\,d\tau
\leq T\bar V_W^\nu\).
Thus,
\begin{equation}
\begin{aligned}
\dot I_V
\leq{}&-a\rho_{\gamma_1,T}^{V_d}I_V^\mu
-bT^{1-\nu}I_V^\nu\\
&+T\bigl(\varrho_1\bar V_W^\mu
+\varrho_2\bar V_W^\nu+\Delta_x\bigr).
\end{aligned}
\label{eq:window_energy_fxt_bound}
\end{equation}
This is the point at which the derivative-limited
window bound enters the closed-loop proof.

For the pointwise state term, since
\(0\leq V_d\leq\bar V_d\) and
\(\gamma_1<\mu\), one has
\(V_d^{\gamma_1}\geq\bar V_d^{\gamma_1-\mu}V_d^\mu\).
Define \(a_0:=a\bar V_d^{\gamma_1-\mu}>0\).
Choosing
\(\lambda_W>\max\{\varrho_1/h_1,\varrho_2/h_2\}\)
and setting
\(d_1:=\lambda_Wh_1-\varrho_1>0\) and
\(d_2:=\lambda_Wh_2-\varrho_2>0\),
we obtain from the pointwise state inequality,
\eqref{eq:window_energy_fxt_bound}, and the critic inequality
\begin{equation}
\begin{aligned}
\dot{\mathcal J}
\leq{}&-a_0V_d^\mu-bV_d^\nu
-\eta a\rho_{\gamma_1,T}^{V_d}I_V^\mu\\
&-\eta bT^{1-\nu}I_V^\nu
-d_1V_W^\mu-d_2V_W^\nu+\Delta_J,
\end{aligned}
\label{eq:window_composite_intermediate}
\end{equation}
where
\(\Delta_J:=(1+\eta T)\Delta_x
+\eta T(\varrho_1\bar V_W^\mu
+\varrho_2\bar V_W^\nu)+\lambda_W\Delta_W\).

For the low-order terms, define
\(c_1:=\min\{a_0,
 a\rho_{\gamma_1,T}^{V_d}\eta^{1-\mu},
 d_1/\lambda_W^\mu\}>0\).
Then

\begin{equation}
\begin{aligned}
&a_0V_d^\mu
+\eta a\rho_{\gamma_1,T}^{V_d}I_V^\mu
+d_1V_W^\mu
\\
&\quad\geq
c_1\bigl[
V_d^\mu
+(\eta I_V)^\mu
+(\lambda_WV_W)^\mu
\bigr]
\\
&\quad\geq
c_1\mathcal J^\mu.
\end{aligned}
\end{equation}

where the last step uses
\(r^\mu+s^\mu+u^\mu\geq(r+s+u)^\mu\)
for \(0<\mu<1\).

For the high-order terms, define
\(c_2:=\min\{b,bT^{1-\nu}\eta^{1-\nu},
 d_2/\lambda_W^\nu\}>0\).
Then
\begin{equation}
\begin{aligned}
bV_d^\nu+\eta bT^{1-\nu}I_V^\nu+d_2V_W^\nu
&\geq c_2\bigl[V_d^\nu+(\eta I_V)^\nu
+(\lambda_WV_W)^\nu\bigr]\\
&\geq 3^{1-\nu}c_2\mathcal J^\nu,
\end{aligned}
\end{equation}
where convexity gives
\(r^\nu+s^\nu+u^\nu\geq
3^{1-\nu}(r+s+u)^\nu\) for \(\nu>1\).
Define \(C_1:=c_1\) and \(C_2:=3^{1-\nu}c_2\).
Combining these estimates yields
\begin{equation}
\dot{\mathcal J}
\leq-C_1\mathcal J^\mu-C_2\mathcal J^\nu+\Delta_J,
\end{equation}
which proves \eqref{eq:composite_constrained_FxT_bound}.

Choose \(0<\chi<1\). When \(\Delta_J>0\), define
\(\mathcal J_\Delta>0\) as the unique positive solution of
\begin{equation}
C_1\mathcal J_\Delta^\mu+C_2\mathcal J_\Delta^\nu
=\frac{\Delta_J}{\chi}.
\label{eq:constrained_practical_residual_radius}
\end{equation}
When \(\Delta_J=0\), set \(\mathcal J_\Delta:=0\).
Since \(\frac{d}{ds}(C_1s^\mu+C_2s^\nu)
=\mu C_1s^{\mu-1}+\nu C_2s^{\nu-1}>0\)
for every \(s>0\), the positive solution is unique.
For \(\mathcal J\geq\mathcal J_\Delta\),
\(\Delta_J\leq\chi[C_1\mathcal J^\mu+C_2\mathcal J^\nu]\),
and therefore
\(\dot{\mathcal J}\leq-(1-\chi)C_1\mathcal J^\mu
-(1-\chi)C_2\mathcal J^\nu\).

Let \(\mathcal J_0:=\mathcal J(t_\star)\).
Whenever \(\mathcal J>\mathcal J_\Delta\),
\(\frac{d\mathcal J}{dt}\leq-(1-\chi)
[C_1\mathcal J^\mu+C_2\mathcal J^\nu]\).
If \(\mathcal J_0>\mathcal J_\Delta\), integration gives
\begin{equation}
\begin{aligned}
T_\Delta
&\leq\int_{\mathcal J_\Delta}^{\mathcal J_0}
\frac{ds}{(1-\chi)(C_1s^\mu+C_2s^\nu)}
\leq\frac{1}{1-\chi}\int_0^\infty
\frac{ds}{C_1s^\mu+C_2s^\nu}.
\end{aligned}
\end{equation}
Following the fixed-time comparison argument
of Lemma~\ref{lem:fixed_time}, splitting the integral
at \(s=1\) gives
\begin{equation}
\begin{aligned}
T_\Delta
&\leq\frac{1}{1-\chi}
\left[\int_0^1\frac{ds}{C_1s^\mu}
+\int_1^\infty\frac{ds}{C_2s^\nu}\right]\\
&=\frac{1}{(1-\chi)C_1(1-\mu)}
+\frac{1}{(1-\chi)C_2(\nu-1)}.
\end{aligned}
\end{equation}
If \(\mathcal J_0\leq\mathcal J_\Delta\), then
\(T_\Delta=0\). Moreover, the derivative of
\(\mathcal J\) is strictly negative at
\(\mathcal J=\mathcal J_\Delta>0\), so the residual
set \(\{\mathcal J\leq\mathcal J_\Delta\}\)
is forward invariant. Consequently, the composite
Lyapunov functional enters the residual set within
\begin{equation}
T_\Delta\leq
\frac{1}{(1-\chi)C_1(1-\mu)}
+\frac{1}{(1-\chi)C_2(\nu-1)},
\label{eq:constrained_practical_fixed_time}
\end{equation}
independently of \(\mathcal J(t_\star)\), over the
class for which the foregoing constants are uniform.
This establishes practical fixed-time convergence.
\end{proof}

\begin{lemma}[Forward Invariance of the Safe Set]
\label{lem:constrained_forward_invariance}
Consider the closed-loop system under the conditions
of Theorem~\ref{thm:composite_constrained_stability}.
Suppose that
\(
\bx(t_\star)\in\operatorname{Int}\mathcal C
\),
the closed-loop solution is forward complete
as long as it remains in the safe region,
and the composite Lyapunov inequality
\eqref{eq:composite_constrained_FxT_bound}
holds uniformly throughout
\(
\operatorname{Int}\mathcal C
\).
Assume also that every finite-time exit from
the admissible operating region must pass through
a state-constraint boundary characterized by
\(
h_\ell(\bx)=0
\)
for some
\(
\ell\in\{1,\ldots,n_h\}
\).
Then,
\(
\operatorname{Int}\mathcal C
\)
is forward invariant, that is,
\(
\bx(t)\in\operatorname{Int}\mathcal C
\)
for all
\(
t\geq t_\star
\).
\end{lemma}

\begin{proof}
From
\eqref{eq:composite_constrained_FxT_bound}
and
\eqref{eq:constrained_practical_residual_radius},
one has
\(
\dot{\mathcal J}
\leq
-(1-\chi)C_1\mathcal J^\mu
-(1-\chi)C_2\mathcal J^\nu<0
\)
whenever
\(
\mathcal J\geq\mathcal J_\Delta
\).
Consequently, the sublevel set
\(
\{\mathcal J\leq\mathcal J_\Delta\}
\)
is forward invariant, and
\(
\mathcal J(t)\leq M_J
\),
where
\(
M_J:=
\max\{\mathcal J(t_\star),\mathcal J_\Delta\}
<\infty
\).
Since
\(
\mathcal J(t)
=
V_d(\bx(t))+\eta I_V(t)+\lambda_WV_W(t)
\),
with
\(
V_W(t)\geq0
\),
and
\(
V_d(\bx)=V_0(\bx)+\kappa_BB(\bx)
\),
where
\(
V_0(\bx)\geq0
\)
and
\(
\kappa_B>0
\),
it follows that
\(
\kappa_BB(\bx(t))
\leq
V_d(\bx(t))
\leq
\mathcal J(t)
\leq M_J
\).
Hence,
\(
B(\bx(t))
\leq M_J/\kappa_B
<\infty
\).

Suppose, by contradiction, that a trajectory
starting at \(t_\star\) in
\(
\operatorname{Int}\mathcal C
\)
reaches the state-constraint boundary at
some finite first-exit time
\(
t_e>0
\).
By continuity of
\(
\bx(t)
\),
there exists
\(
\ell\in\{1,\ldots,n_h\}
\)
such that
\(
h_\ell(\bx(t))\to0^+
\)
as
\(
t\to t_e^-
\).
From the logarithmic barrier definition,
\(
b_\ell(\bx(t))
=
\log\left(
1+
[\gamma h_\ell(\bx(t))]^{-1}
\right)
\),
one obtains
\(
b_\ell(\bx(t))\to+\infty
\).
Since
\(
B(\bx(t))
=
\sum_{j=1}^{n_h}
\beta_j
[b_j(\bx(t))-b_j(\bm0)]^2
\)
with
\(
\beta_j>0
\),
it follows that
\(
B(\bx(t))
\geq
\beta_\ell
[b_\ell(\bx(t))-b_\ell(\bm0)]^2
\to+\infty
\)
as
\(
t\to t_e^-
\).
This contradicts
\(
B(\bx(t))
\leq M_J/\kappa_B<\infty
\).
Thus, no trajectory starting in
\(
\operatorname{Int}\mathcal C
\)
can reach the state-constraint boundary
in finite time. Together with the
forward-completeness and boundary assumptions,
this establishes
\(
\bx(t)\in\operatorname{Int}\mathcal C
\)
for all
\(
t\geq t_\star
\).
\end{proof}

\subsection{Simulation Setup}
\label{subsec:simulation_setup}

The simulated plant is \(\dot{\bx}=\bfv(\bx)+\bu\), 
where \(\bfv(\bx)=[-\tanh(x_1+x_2)-0.4x_1^3,\,
-\sin(x_2+x_3)-0.3x_1x_2,\,
-x_3/(1+x_3^2)-0.2\tanh(x_1x_3)]^\top\), 
\(\bg(\bx)=\bm I_3\), \(\bm U=8\bm I_3\), and 
\(\bR=0.4\bm I_3\), with \(|u_j|\leq8\), \(j=1,2,3\).
The safe domain is defined by 
\(h_0(\bx)=2.4^2-\bx^\top\bx>0\) and 
\(h_j(\bx)=\|\bx-\bm c_j\|_2^2-(r_j+d_{\mathrm s})^2>0\), 
where \(\bm c_1=[0.55,0.25,0.10]^\top\), 
\(\bm c_2=[-0.50,0.30,0.15]^\top\), 
\(\bm c_3=[0.10,-0.45,0.25]^\top\), 
\([r_1,r_2,r_3]=[0.13,0.14,0.13]\), and 
\(d_{\mathrm s}=0.055\).
The barrier parameters are \(\gamma_\ell=\beta_\ell=1\), 
\(\ell=0,\ldots,3\), and \(\kappa_B=0.015\).
The desired value function is 
\(V_d(\bx)=0.5\bx^\top\bx+\kappa_B B(\bx)\), 
with decay parameters \(a=0.8\), \(b=0.1\), 
\(\mu=3/4\), and \(\nu=2\).

Integral concurrent learning uses the basis
\(\bm\psi_f(\bx)=[\tanh(x_1+x_2),x_1^3,
\sin(x_2+x_3),x_1x_2,x_3/(1+x_3^2),
\tanh(x_1x_3)]^\top\),
with \(1400\) data intervals of length \(0.04\,\mathrm{s}\),
quadrature step \(0.004\,\mathrm{s}\), stored-data gain 
\(k_f=2\), and \(4000\) updates of step size \(0.005\).
The identification states are sampled from 
\([-1.25,1.25]^3\), with inputs 
\(\bu=-1.5\bx+\bm\xi\), 
\(\bm\xi\sim\mathcal U([-2,2]^3)\), subject to safety 
and input constraints.
The identification weights are initialized with 
\(18\) equally spaced entries in \([-0.06,0.06]\), 
arranged column-wise, and 
\(\bm\Gamma_f=0.5(\bm G_f+10^{-10}\bm I_6)^{-1}\),
where \(\bm G_f\) is the normalized identification Gramian.
The random seed is \(7\).

The Koopman lifting is 
\(\bm z=[\bx^\top,\bm\psi_f^\top(\bx)]^\top
\in\mathbb R^9\), with regression regularization \(10^{-7}\)
using the same identification dataset.
The initialization MPC uses horizon \(N_p=35\),
sampling period \(0.04\,\mathrm{s}\),
\(\bm Q_K=\operatorname{diag}(3\bm I_3,\bm 0_{6\times6})\),
\(\bm R_K=0.15\bm I_3\), input-tightening parameter
\(\epsilon_u=10^{-5}\), and prediction safety margin
\(2\times10^{-4}\).
The constrained MPC problem is solved using MATLAB's
\texttt{fmincon} with sequential quadratic programming.
The terminal matrix \(\bP_K\) is obtained by solving
\eqref{eq:koopman_terminal_sdp} in YALMIP with
\(\bm Q_f=\operatorname{diag}(\bm I_3,10^{-3}\bm I_6)\)
and \(\bm R_f=0.2\bm I_3\).
The initialization-only robust CBF uses
\(\alpha_\ell(h_\ell)=3h_\ell\) and
\(\bar{\Delta}_f=0.05\).
A total of \(159\) state-input pairs are collected 
from five teacher trajectories for transformed policy 
matching with ridge regularization \(\lambda_K=10^{-2}\).

The critic basis is 
\(\bphi(\bx)=[x_1^2,x_2^2,x_3^2,x_1x_2,x_1x_3,
x_2x_3,x_1^4,x_2^4,x_3^4]^\top\), with evaluation 
reference \(\bW^*=[0.5,0.5,0.5,0,0,0,0,0,0]^\top\).
The MATLAB-computed Koopman--MPC initialization is
\[
\bW_K \simeq
\left[
\begin{gathered}
1.2028,\;1.1371,\;1.3395,\\
-0.0299,\;-0.1261,\;-0.0112,\\
-0.1436,\;-0.2910,\;-1.1938
\end{gathered}
\right]^{\top}.
\]
Learning uses \(144\) fixed replay windows of length 
\(0.16\,\mathrm{s}\), shifted by \(0.04\,\mathrm{s}\), 
with residuals and their gradients reevaluated under 
the current critic.
The parameters are \(k_1=1\), \(k_2=0.1\), 
\(\gamma_1=0.5\), and \(\gamma_2=3\), with 
\(\bm\Gamma=8(\bm G_K+
10^{-5}\lambda_{\max}(\bm G_K)\bm I_9)^{-1}\),
where \(\bm G_K\) is the normalized replay Gramian 
evaluated at \(\bW_K\), satisfying 
\(\lambda_{\min}(\bm G_K)=1.632\times10^{-3}\).
The lower-order decay and critic update terms are 
numerically regularized with \(\epsilon_V=10^{-6}\) 
and \(\epsilon_e=10^{-5}\), respectively.
The Bellman residual RMS stopping threshold is \(10^{-5}\).
A sinusoidal probing signal with amplitude \(0.02\),
angular frequencies \([1.6,3.7,6.1]^\top\), and phases
\([0.2,1.1,2.2]^\top\) is introduced inside the 
hyperbolic tangent during the first second.

The five initial states are 
\(\bx_0^{(1)}=[1.45,0.48,0.39]^\top\), 
\(\bx_0^{(2)}=[1.25,0.75,0.10]^\top\), 
\(\bx_0^{(3)}=[-1.25,0.60,0.60]^\top\), 
\(\bx_0^{(4)}=[-1.05,0.80,0.18]^\top\), and 
\(\bx_0^{(5)}=[0.42,-1.18,0.50]^\top\).
The baseline uses \(\hat{\bW}(0)=\bW_K\), and the 
comparison considers 
\(\hat{\bW}(0)\in
\{\bW_K,\bm0_9,0.15\bm1_9,-\bW_K,2\bW_K\}\)
for each initial state, with identical replay data 
and learning gains.
Simulations run for up to \(15\,\mathrm{s}\), using
\texttt{ode15s} for critic adaptation and adaptive RK4 
for the plant dynamics, with maximum integration steps 
of \(0.02\,\mathrm{s}\) and \(0.005\,\mathrm{s}\), 
respectively.
The state settling threshold is \(10^{-3}\).

\begin{figure}[H]
    \centering
    \scalebox{1}[1.2]{%
        \includegraphics[width=\columnwidth]{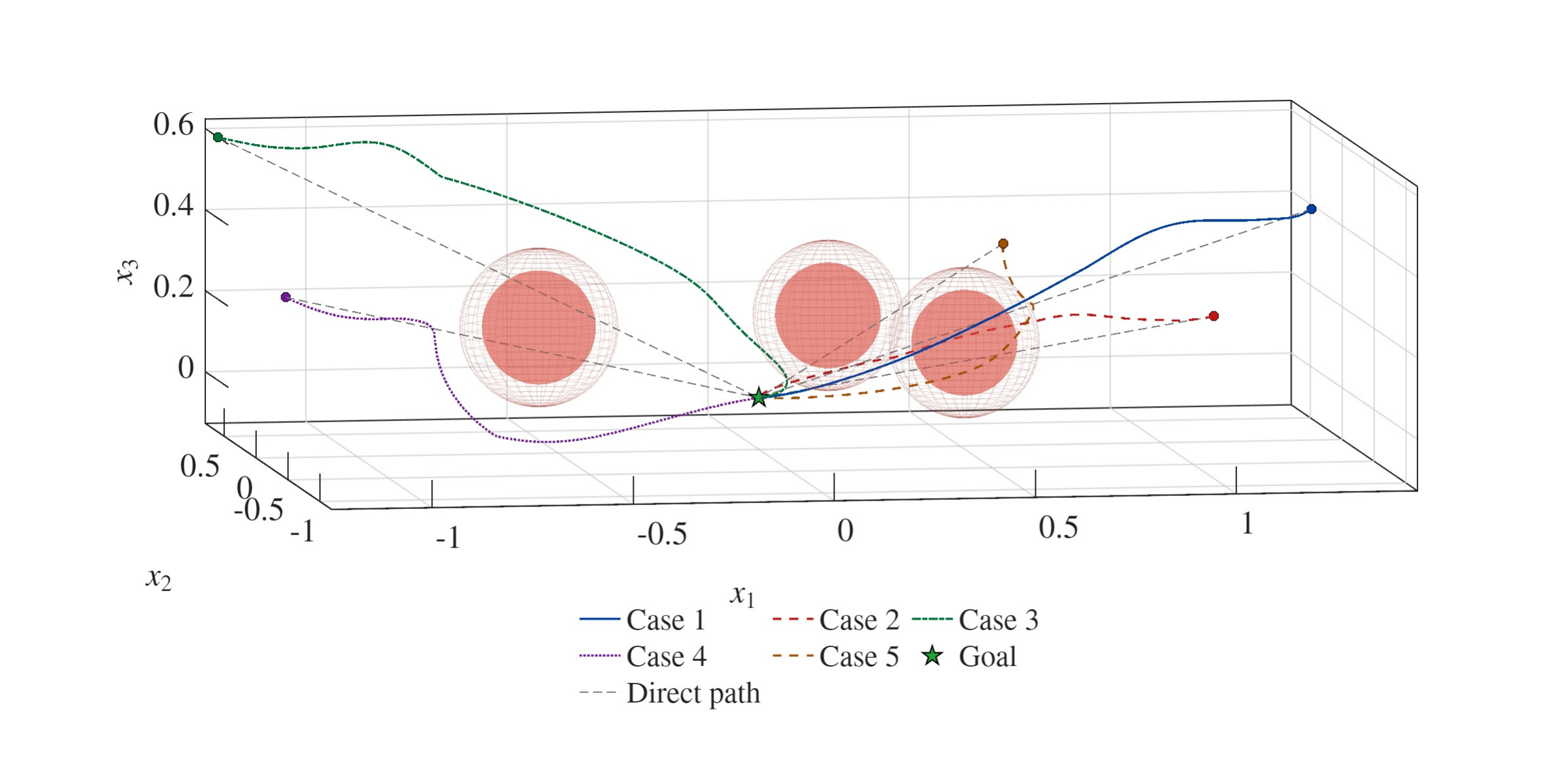}%
    }
    \caption{Three-dimensional state trajectories with direct-path references.}
    \label{fig:trajectories_3d_direct}
\end{figure}

\begin{figure}[H]
    \centering
    \scalebox{1}[1.0]{%
        \includegraphics[width=\columnwidth]{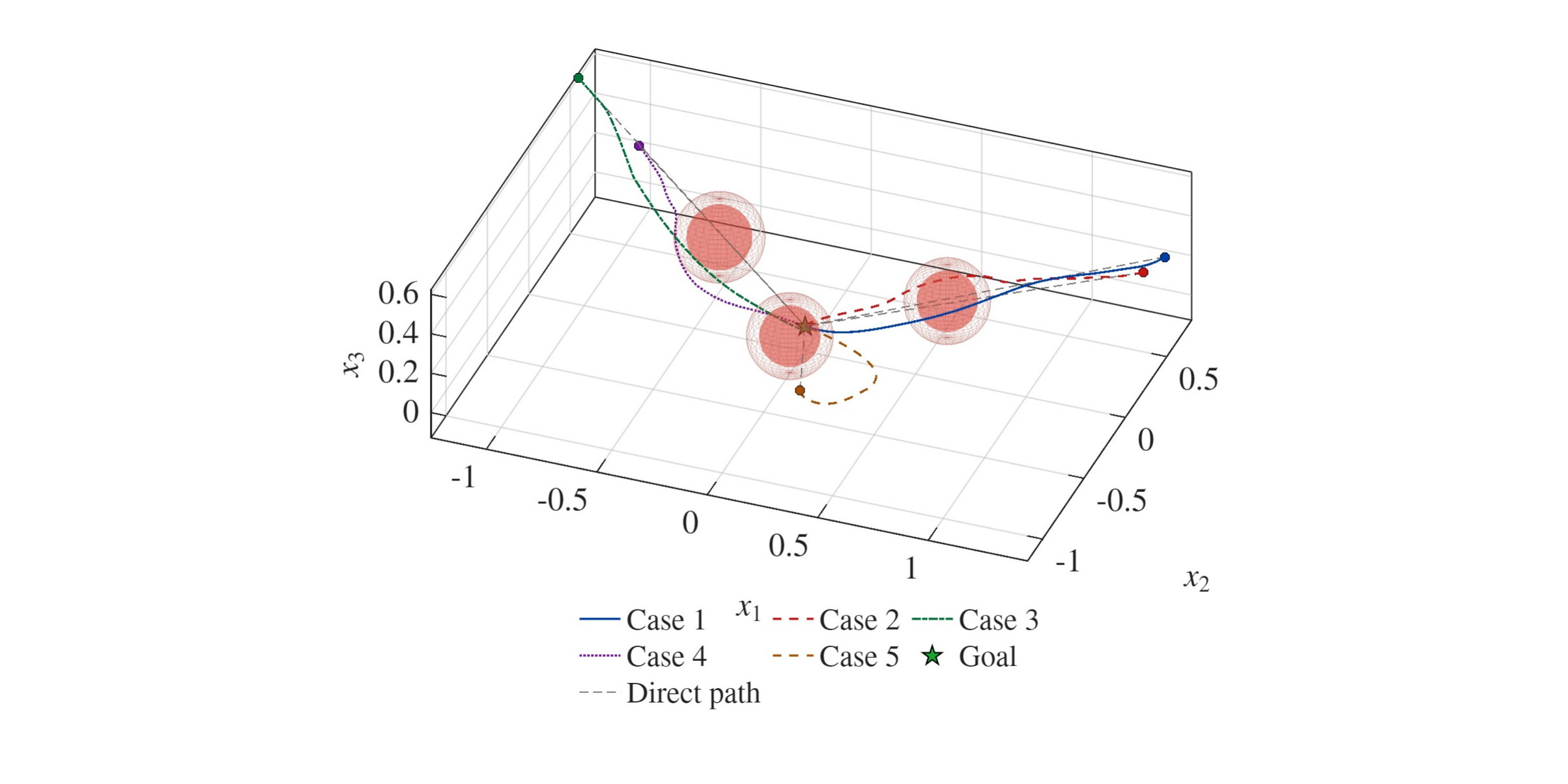}%
    }
    \caption{Three-dimensional state trajectories under five initial conditions.}
    \label{fig:trajectories_3d}
\end{figure}

\begin{figure}[H]
    \centering
    \scalebox{1}[1.0]{%
        \includegraphics[width=\columnwidth]{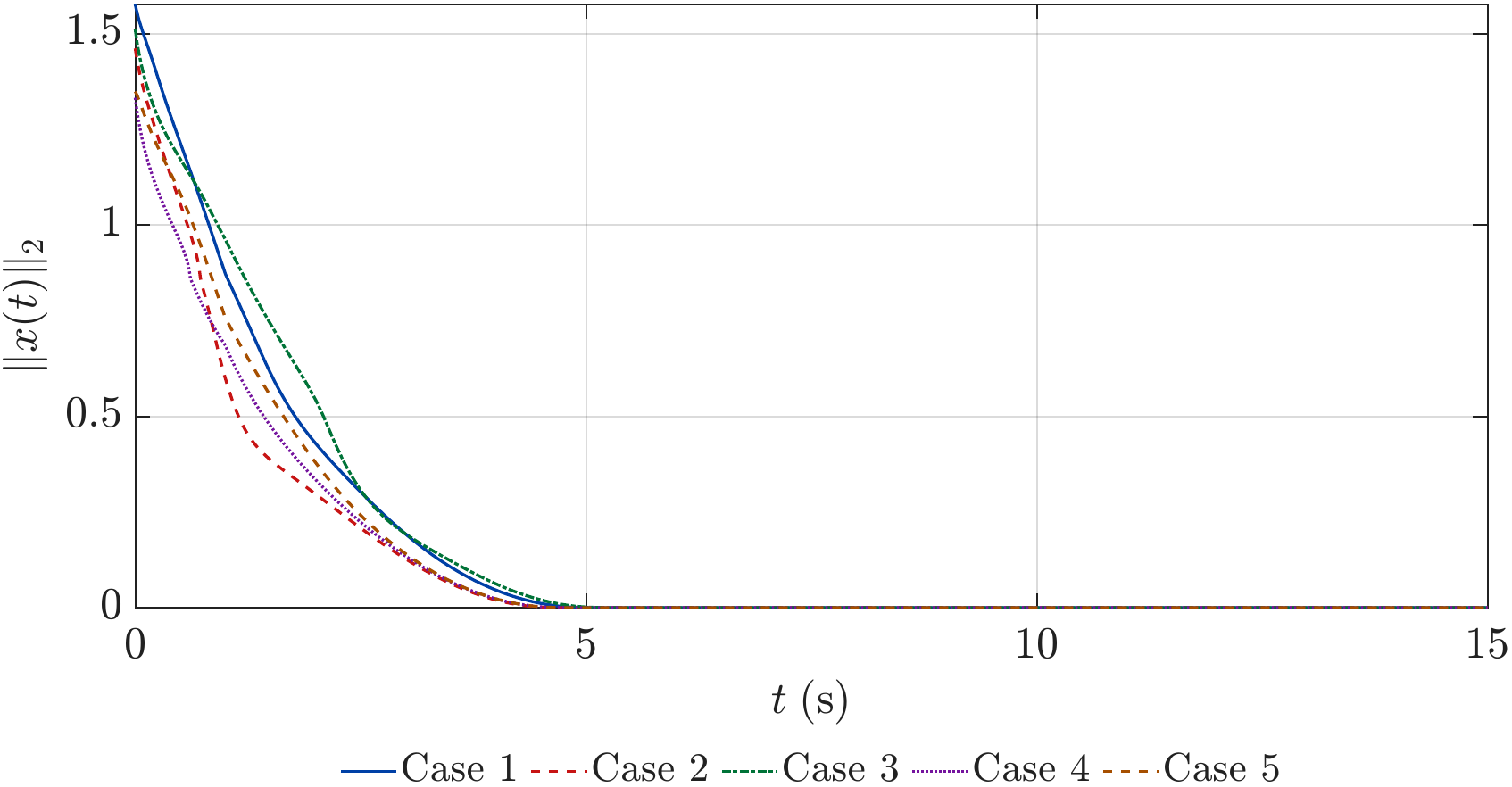}%
    }
    \caption{Evolution of the state norm under five initial conditions.}
    \label{fig:state_norm}
\end{figure}

\begin{figure}[H]
    \centering
    \scalebox{1}[1.0]{%
        \includegraphics[width=\columnwidth]{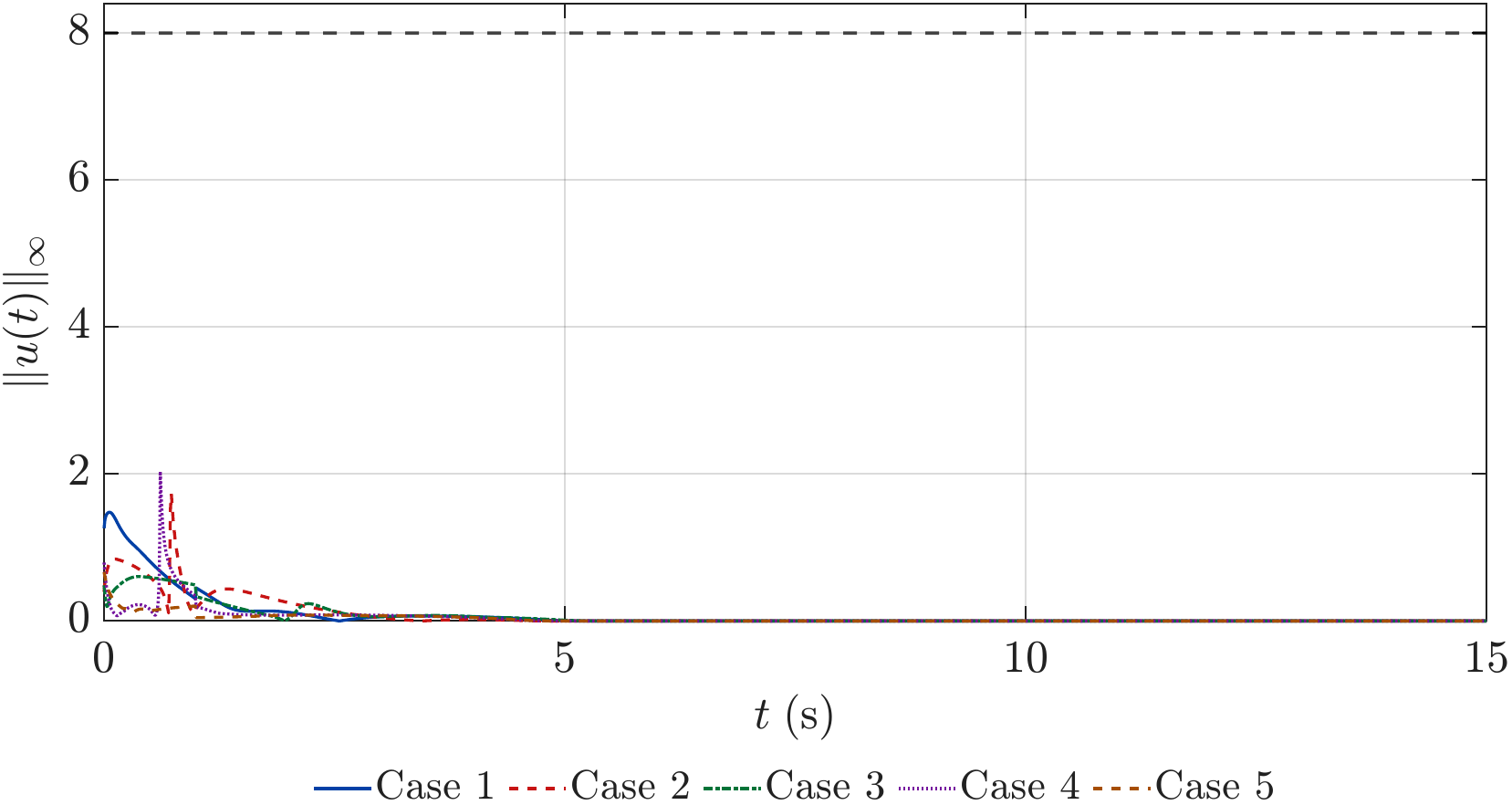}%
    }
    \caption{Evolution of the control-input infinity norm and its admissible bound.}
    \label{fig:input_norm}
\end{figure}

\begin{figure}[H]
    \centering
    \scalebox{1}[1.0]{%
        \includegraphics[width=\columnwidth]{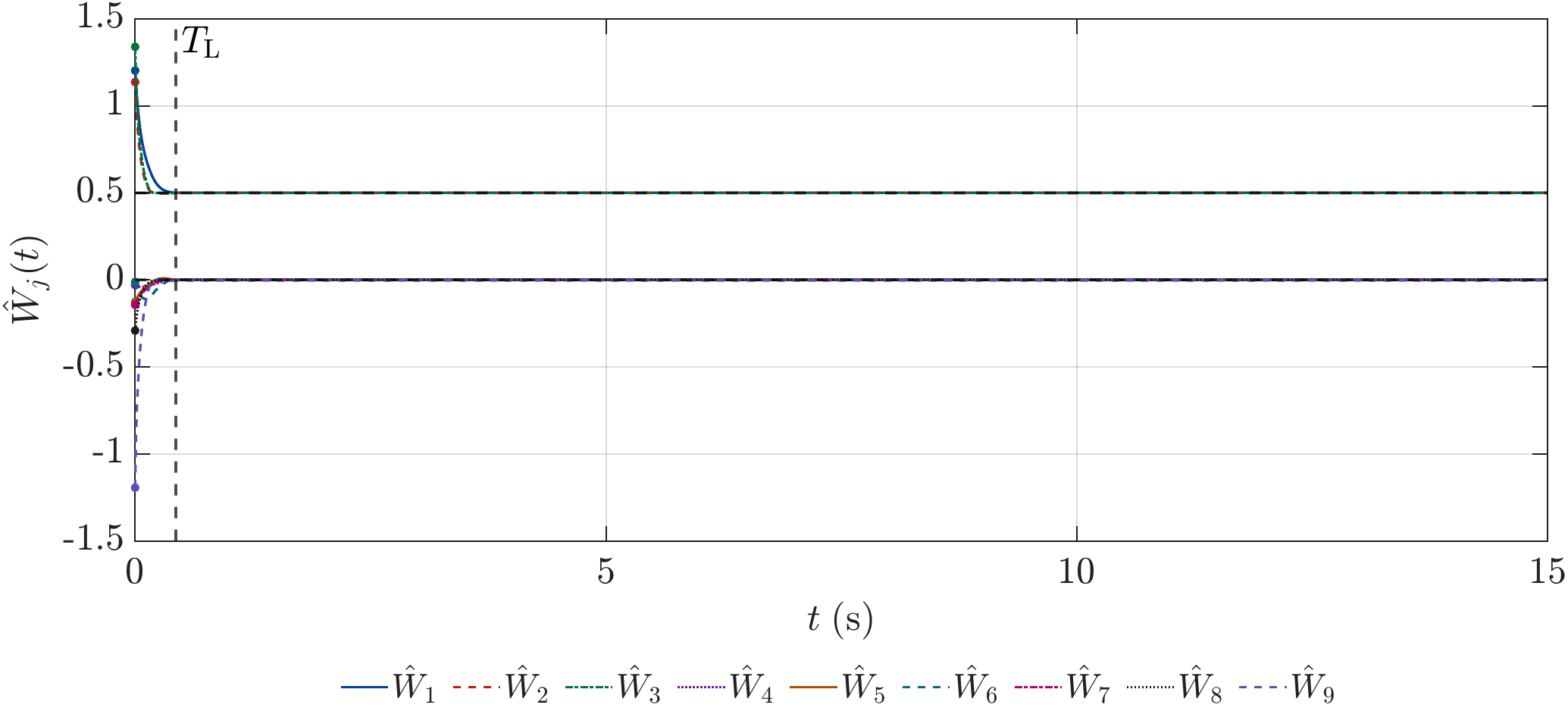}%
    }
    \caption{Evolution of the critic weights with Koopman--MPC initialization.}
    \label{fig:critic_weights}
\end{figure}

\begin{figure}[H]
    \centering
    \scalebox{1}[1.0]{%
        \includegraphics[width=\columnwidth]{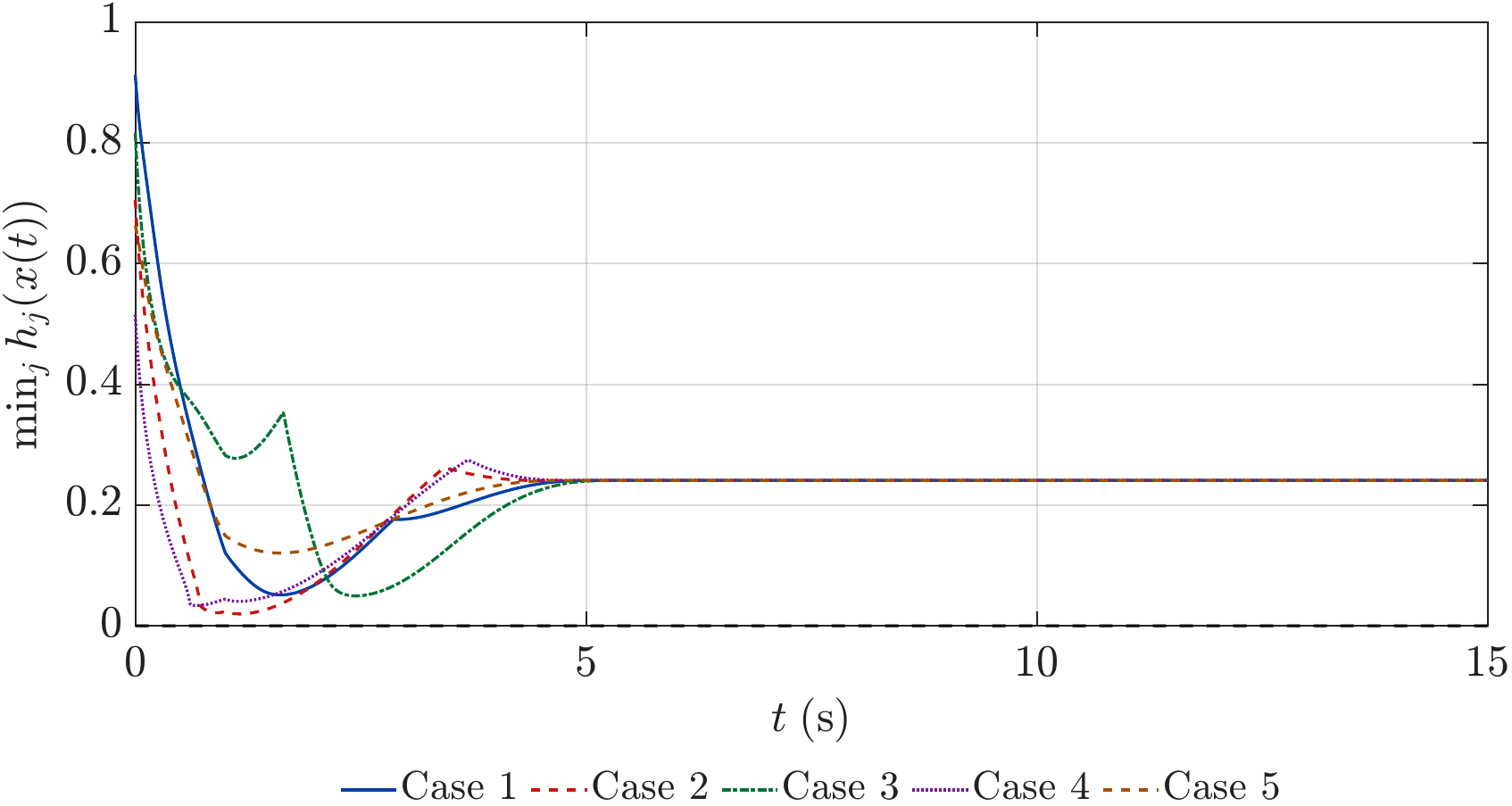}%
    }
    \caption{Evolution of the minimum safety-function value under five initial conditions.}
    \label{fig:minimum_safety}
\end{figure}

\begin{figure}[H]
    \centering
    \scalebox{1}[1.0]{%
        \includegraphics[width=\columnwidth]{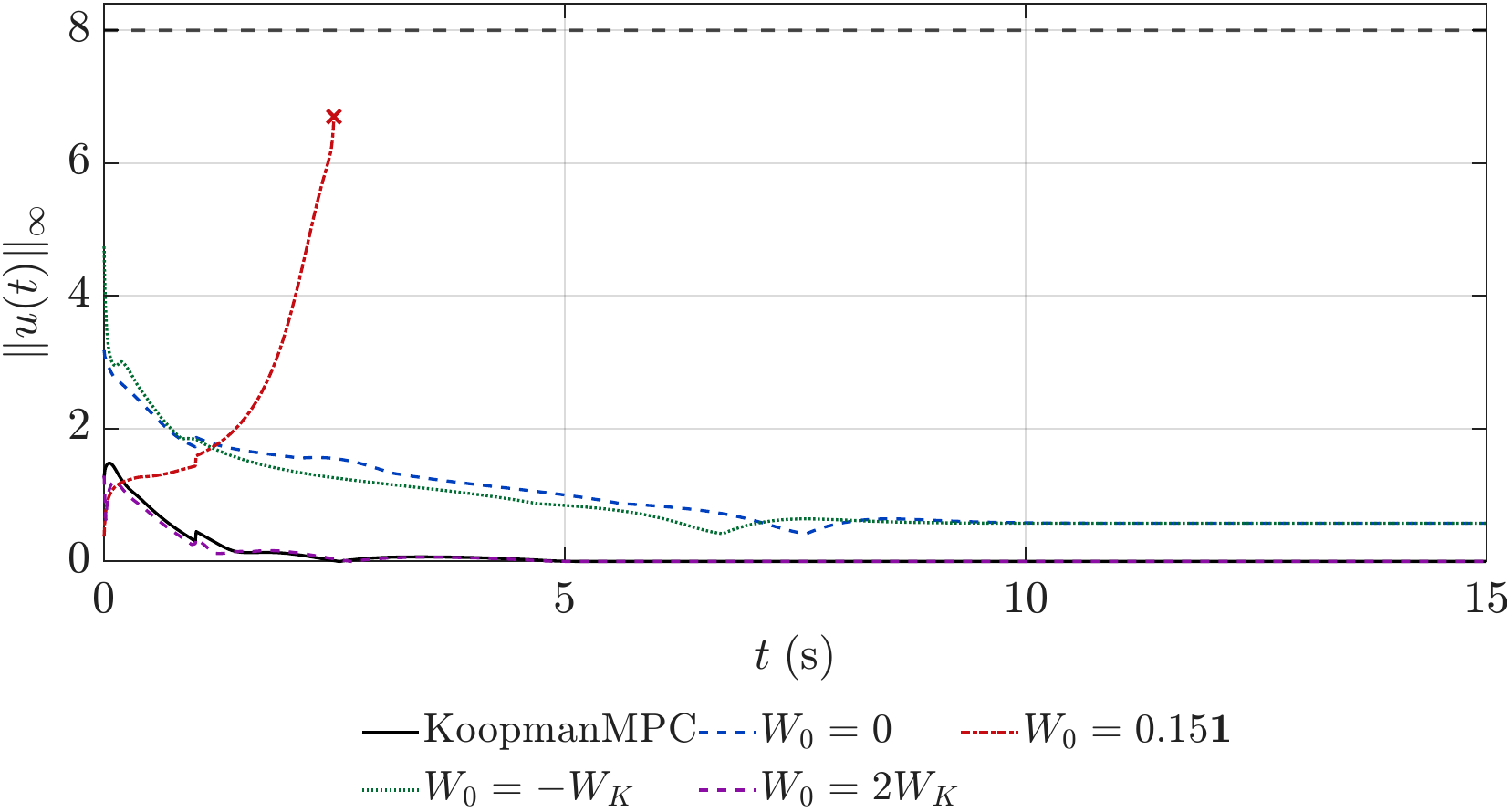}%
    }
    \caption{Control-input infinity norms for five critic initializations under the first initial state.}
    \label{fig:initialization_input_comparison}
\end{figure}

\begin{figure}[H]
    \centering
    \scalebox{1}[1.0]{%
        \includegraphics[width=\columnwidth]{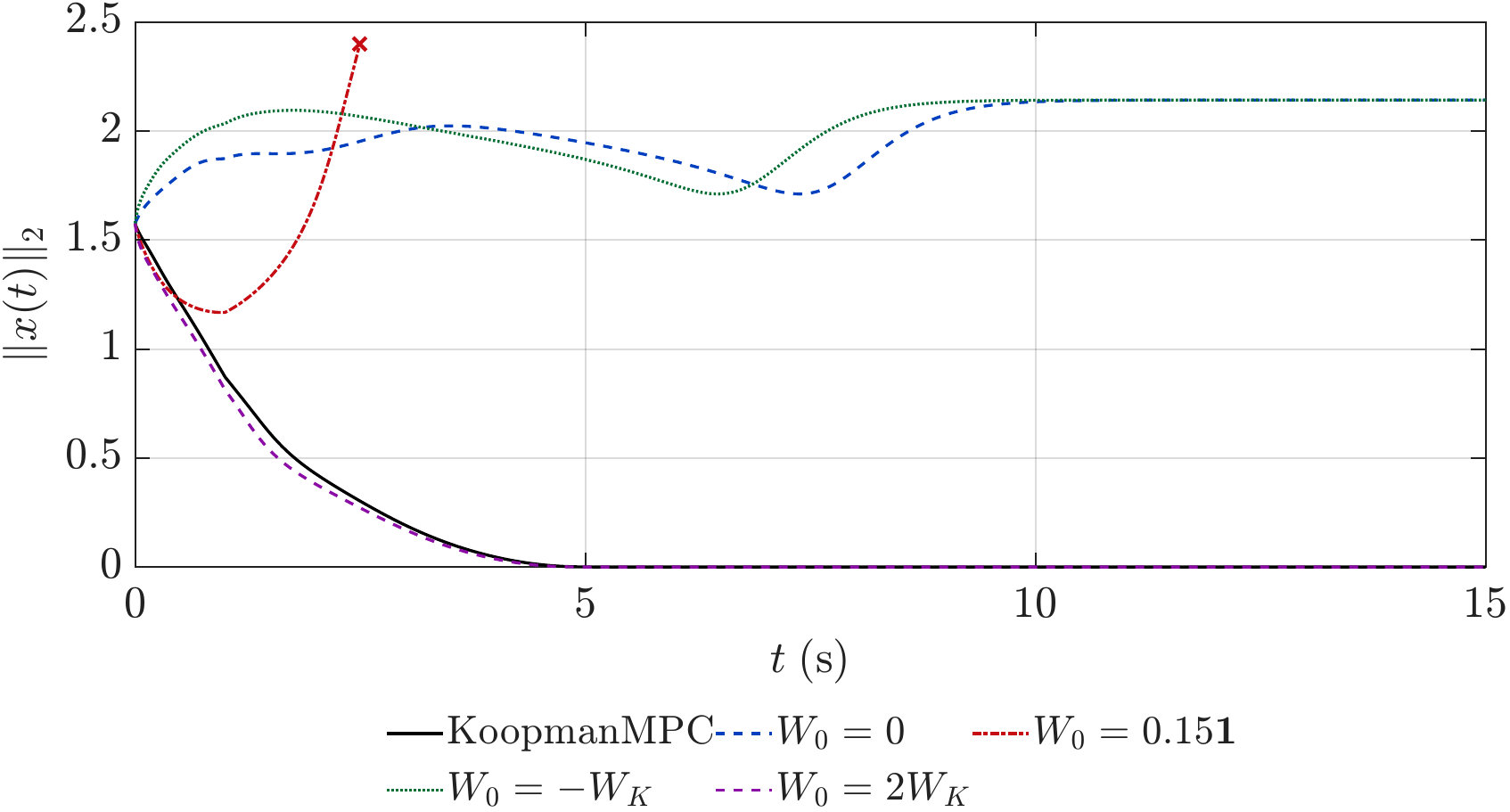}%
    }
    \caption{State norms for five critic initializations under the first initial state.}
    \label{fig:initialization_state_comparison}
\end{figure}

Figs.~\ref{fig:trajectories_3d_direct} and 
\ref{fig:trajectories_3d} illustrate the effectiveness 
of the proposed controller in generating collision-free 
trajectories, despite all direct paths toward the origin 
being obstructed. As shown in Fig.~\ref{fig:state_norm}, 
the state trajectories reach the prescribed settling 
threshold of \(10^{-3}\) within \(4.55\)--\(5.07\,\mathrm{s}\), 
which is consistent with the desired practical fixed-time 
convergence. Fig.~\ref{fig:input_norm} confirms input 
constraint satisfaction, with a maximum control-input 
magnitude of \(2.0345\), substantially below the 
admissible bound of \(8\).

Fig.~\ref{fig:critic_weights} demonstrates rapid critic 
adaptation, with the Bellman residual reaching the 
prescribed threshold of \(10^{-5}\) within 
\(0.43152\,\mathrm{s}\) and a final weight-reference error 
of \(2.9864\times10^{-3}\). Moreover, 
Fig.~\ref{fig:minimum_safety} shows that the minimum 
safety-function value remains positive across all five 
cases, with an overall minimum of \(0.019589\).
These results numerically demonstrate simultaneous 
state convergence, safety, and input constraint 
satisfaction without an additional online safety filter.

Figs.~\ref{fig:initialization_input_comparison} and 
\ref{fig:initialization_state_comparison} highlight 
the importance of critic initialization. Under the first 
initial state, the zero and negative initializations 
fail to stabilize the origin, whereas the uniform 
initialization \(0.15\bm 1_9\) reaches the domain or 
decay-feasibility limit at \(2.4944\,\mathrm{s}\).
Across all five initial states, the proposed Koopman--MPC 
initialization achieves a \(100\%\) convergence rate, 
compared with \(0\%\), \(20\%\), and \(0\%\) for 
\(\bm 0_9\), \(0.15\bm 1_9\), and \(-\bW_K\), respectively.
Although \(2\bW_K\) also achieves a \(100\%\) convergence 
rate, the proposed initialization reduces the worst-case 
peak control input by \(32.75\%\) and the critic adaptation 
time by \(9.87\%\). These results demonstrate the benefits 
of Koopman--MPC initialization in improving convergence 
reliability, limiting control transients, and accelerating 
critic adaptation.

% ============================================================
% CONCLUSION
% ============================================================

\section{Conclusion}
\label{sec:conclusion}

This paper developed a safety-critical Koopman-initialized
practical fixed-time integral reinforcement learning framework
for optimal control of unknown nonlinear systems.
The unknown drift was reconstructed from integral data and
used to construct a barrier-augmented inverse-optimal
fixed-time control problem.
A Koopman-based initialization scheme provided data-informed
initial critic weights, while a critic-only learning law
with experience replay achieved practical fixed-time
convergence without requiring persistent excitation.
A composite Lyapunov analysis established practical
fixed-time closed-loop stability, and the barrier-augmented
construction ensured forward invariance of the prescribed
safe set under the stated assumptions.
Simulation results demonstrated the effectiveness of
the proposed framework.

\section*{Acknowledgment}
The authors would like to thank Ho Chi Minh City University of Technology (HCMUT) and Vietnam National University Ho Chi Minh City (VNU-HCM) for supporting this research.

\appendices

\section{Proof of Lemma~\ref{lem:derivative_limited_window_bound}}
\label{app:derivative_limited_window_bound}

\begin{proof}
    Let
\(
M_y
:=
\max_{\tau\in[t-T,t]}y(\tau)
\)
and let
\(
\tau_m\in[t-T,t]
\)
satisfy
\(
y(\tau_m)=M_y
\).
At least one of
\(
[t-T,\tau_m]
\)
and
\(
[\tau_m,t]
\)
has length no smaller than
\(
T/2
\).
Since
\(
|\dot y(\tau)|\leq L_y
\),
the function
\(
y
\)
is
\(
L_y
\)-Lipschitz, and hence
\(
y(\tau)
\geq
M_y-L_y|\tau-\tau_m|
\).
Therefore,
\(
y(\tau)\geq M_y/2
\)
whenever
\(
|\tau-\tau_m|\leq M_y/(2L_y)
\).
There is consequently a one-sided interval contained in
\(
[t-T,t]
\)
of length at least
\(
\min\{T/2,M_y/(2L_y)\}
\)
on which
\(
y(\tau)\geq M_y/2
\).
It follows that
\(
\int_{t-T}^{t}y(\tau)\,d\tau
\geq
(M_y/2)
\min\{T/2,M_y/(2L_y)\}
\).

Suppose first that
\(
M_y\leq L_yT
\).
Then
\(
M_y/(2L_y)\leq T/2
\),
and therefore
\(
\int_{t-T}^{t}y(\tau)\,d\tau
\geq
M_y^2/(4L_y)
\).
Hence,
\(
M_y
\leq
2
\left[
L_y
\int_{t-T}^{t}y(\tau)\,d\tau
\right]^{1/2}
\).
Since
\(
\gamma_1-1<0
\)
and
\(
0\leq y(\tau)\leq M_y
\),
one has
\(
y^{\gamma_1}(\tau)
=
y(\tau)y^{\gamma_1-1}(\tau)
\geq
y(\tau)M_y^{\gamma_1-1}
\).
Consequently,
\(
\int_{t-T}^{t}y^{\gamma_1}(\tau)\,d\tau
\geq
M_y^{\gamma_1-1}
\int_{t-T}^{t}y(\tau)\,d\tau
\).
Because
\(
\gamma_1-1<0
\),
the preceding upper bound on
\(
M_y
\)
implies
\(
M_y^{\gamma_1-1}
\geq
2^{\gamma_1-1}
L_y^{\frac{\gamma_1-1}{2}}
\left(
\int_{t-T}^{t}y(\tau)\,d\tau
\right)^{\frac{\gamma_1-1}{2}}
\).
Thus,
\(
\int_{t-T}^{t}y^{\gamma_1}(\tau)\,d\tau
\geq
2^{\gamma_1-1}
L_y^{\frac{\gamma_1-1}{2}}
\left(
\int_{t-T}^{t}y(\tau)\,d\tau
\right)^{\frac{\gamma_1+1}{2}}
\).
Using
\(
\mu=(\gamma_1+1)/2
\),
one obtains
\(
\int_{t-T}^{t}y^{\gamma_1}(\tau)\,d\tau
\geq
2^{\gamma_1-1}
L_y^{\frac{\gamma_1-1}{2}}
\left(
\int_{t-T}^{t}y(\tau)\,d\tau
\right)^\mu
\).

Suppose next that
\(
M_y>L_yT
\).
Then
\(
M_y/(2L_y)>T/2
\),
and the same window estimate gives
\(
\int_{t-T}^{t}y(\tau)\,d\tau
\geq
M_yT/4
\).
Therefore,
\(
M_y
\leq
(4/T)
\int_{t-T}^{t}y(\tau)\,d\tau
\).
Again,
\(
y^{\gamma_1}(\tau)
\geq
y(\tau)M_y^{\gamma_1-1}
\),
and hence
\(
\int_{t-T}^{t}y^{\gamma_1}(\tau)\,d\tau
\geq
M_y^{\gamma_1-1}
\int_{t-T}^{t}y(\tau)\,d\tau
\).
Since
\(
\gamma_1-1<0
\),
one obtains
\(
M_y^{\gamma_1-1}
\geq
4^{\gamma_1-1}
T^{1-\gamma_1}
\left(
\int_{t-T}^{t}y(\tau)\,d\tau
\right)^{\gamma_1-1}
\),
and therefore
\(
\int_{t-T}^{t}y^{\gamma_1}(\tau)\,d\tau
\geq
4^{\gamma_1-1}
T^{1-\gamma_1}
\left(
\int_{t-T}^{t}y(\tau)\,d\tau
\right)^{\gamma_1}
\).

Moreover,
\(
\gamma_1-\mu
=
(\gamma_1-1)/2<0
\).
Since
\(
\int_{t-T}^{t}y(\tau)\,d\tau
\leq
\bar I
\),
one has
\(
\left(
\int_{t-T}^{t}y(\tau)\,d\tau
\right)^{\gamma_1-\mu}
\geq
\bar I^{\gamma_1-\mu}
=
\bar I^{\frac{\gamma_1-1}{2}}
\).
Hence,
\(
\left(
\int_{t-T}^{t}y(\tau)\,d\tau
\right)^{\gamma_1}
\geq
\bar I^{\frac{\gamma_1-1}{2}}
\left(
\int_{t-T}^{t}y(\tau)\,d\tau
\right)^\mu
\).
Thus,
\(
\int_{t-T}^{t}y^{\gamma_1}(\tau)\,d\tau
\geq
4^{\gamma_1-1}
T^{1-\gamma_1}
\bar I^{\frac{\gamma_1-1}{2}}
\left(
\int_{t-T}^{t}y(\tau)\,d\tau
\right)^\mu
\).

Combining the two cases gives
\(
\int_{t-T}^{t}
y^{\gamma_1}(\tau)\,d\tau
\geq
\rho_{\gamma_1,T}
\left(
\int_{t-T}^{t}y(\tau)\,d\tau
\right)^\mu
\),
where
\(
\rho_{\gamma_1,T}
\)
is defined by
\eqref{eq:rho_gamma_window}.
This proves
\eqref{eq:derivative_limited_window_bound}.
\hfill\IEEEQED
\end{proof}

% ============================================================
% REFERENCES
% ============================================================

\bibliographystyle{IEEEtran}
\bibliography{references}

\end{document}